\documentclass[10pt]{article}
\usepackage[a4paper,textwidth=43pc,hcentering,top=20mm,bottom=20mm]{geometry}   %
\usepackage[T1]{fontenc}
\usepackage[utf8]{inputenc}
\usepackage{lmodern}
\usepackage{textcomp}
\usepackage{amsmath,amssymb}
\usepackage{graphicx}
\usepackage{setspace}
\usepackage{array,booktabs,xltabular}
\usepackage{needspace}   %
\usepackage{enumitem}
\usepackage[ruled,vlined,linesnumbered]{algorithm2e}   %
\SetAlgorithmName{Algorithm}{Algorithm}{List of Algorithms}
\SetKwInput{KwIn}{Input}
\SetKwInput{KwOut}{Output}
\SetKwProg{Op}{Operation}{:}{}   %
\SetKwProg{Fn}{Function}{:}{}   %
\DontPrintSemicolon

\SetCommentSty{algcomsty}
\setlist{itemsep=1pt,topsep=3pt,parsep=0pt}
\usepackage[dvipsnames]{xcolor}
\usepackage[most]{tcolorbox}
\usepackage[backend=biber,style=ieee,sorting=none,doi=false,url=false,isbn=false,dashed=false]{biblatex}
\DeclareFieldFormat{urldate}{\mkbibparens{Accessed\addcolon\space#1}}
\DeclareCapitalPunctuation{.}
\DeclareRangeChars{:~,;-+/}   %
\usepackage{xpatch}
\xpatchbibdriver{online}{\printtext[parens]{\usebibmacro{date}}}{\iffieldundef{year}{}{\printtext[parens]{\usebibmacro{date}}}}{}{}
\usepackage[hidelinks]{hyperref}
\usepackage{xurl}
\newcolumntype{Y}{>{\raggedright\arraybackslash}X}
\makeatletter
\def\@seccntformat#1{\csname pre@#1\endcsname\csname the#1\endcsname\csname post@#1\endcsname\ }
\def\post@section{.}
\g@addto@macro\appendix{\def\pre@section{Appendix~}}
\makeatother
\newcommand{\IEEEfigcap}[1]{\setbox0\hbox{#1}\ifdim\wd0>\linewidth\parbox[t]{\linewidth}{\noindent #1}\else\hbox to\linewidth{\hfil\box0\hfil}\fi}
\newenvironment{thmblock}[1]{\par\addvspace{4pt}\noindent #1\ \itshape\ignorespaces}{\par\addvspace{4pt}}
\newtcolorbox{reviewbox}{breakable,colback=gray!6,colframe=gray!45,boxrule=0.4pt,arc=1pt,left=4pt,right=4pt,top=2pt,bottom=2pt,fontupper=\small\setstretch{1.15}}
\newtcolorbox{figbox}{colback=blue!3,colframe=blue!40,boxrule=0.5pt,arc=1pt,left=4pt,right=4pt,top=2pt,bottom=2pt,fontupper=\small\setstretch{1.15}}
\begin{document}

\begin{center}
{\LARGE\bfseries Per-Query Recall Guarantees for Approximate Nearest Neighbor Search in Dynamic Vector Databases\par}
\vspace{8pt}
{\normalsize \begin{tabular}{c@{\hspace{3em}}c}Woong-Kee Loh & Suan Lee\\ Gachon University & Semyung University\\ Seongnam, Republic of Korea & Jecheon, Republic of Korea\\ \texttt{wkloh2@gachon.ac.kr} & \texttt{suanlee@semyung.ac.kr}\end{tabular}\par}
\vspace{8pt}
\parbox{0.9\linewidth}{\small\itshape This is the extended version of the paper submitted to ICDE 2027. It contains the complete proofs, the experiments omitted from the conference version, and supplementary results. This work has been submitted to the IEEE for possible publication. Copyright may be transferred without notice, after which this version may no longer be accessible.}
\end{center}

\section*{\centering Abstract}

Vector search now runs as an operator inside database systems, where a missed neighbor makes a query answer wrong, yet approximate nearest neighbor (ANN) indexes are tuned only for a target \textit{average} recall. Tuned to an average recall of 0.99, indexes left 5–19\% of in-distribution queries and, without a query log, up to 99.6\% of out-of-distribution queries below it (more under updates) and changed a top-5-tag aggregate over the results for up to 27\% of queries. In the spirit of approximate query processing, we present PQR-Search, a $k$-ANN operator that guarantees $\Pr[\mathrm{Recall}(q) < \tau] \le \delta$ for every query and, for $\tau < 1$, $\mathbb E[\mathrm{Recall}(q)] \ge \tau$. The probability is over a random rotation inside the index alone, so no distributional assumption is needed, only that queries and updates do not depend on it, and the guarantee holds after insertions and deletions without retuning. PQR-Search prunes 1-bit codes of rotated IVF residuals with exact error quantiles and predicts its cost without data-fitted constants. On 20 datasets of up to 100M vectors, failure rates stayed at most $\delta$ in every audit, including the NeurIPS'23 streaming runbook, and the aggregate changed in at most 0.04\% of trials. The guarantee costs 2.5$\times$ the single-query latency of HNSW tuned as above and 1.3$\times$ lower batch throughput (geometric means). PQR-Search was 26$\times$ faster than FAISS exhaustive search in batch, faster than IVF with $n_{\rm probe}$ chosen afterwards to meet the target on every test query, and 27–76$\times$ faster than PUFFINN, which also guarantees per-query recall.

\bigskip

\section{Introduction}\label{sec:intro}

With the advance of deep neural networks, it has become common to represent unstructured data such as text, images, and audio as high-dimensional embedding vectors~\cite{r38,r158}. Accordingly, $k$-nearest neighbor ($k$-NN) search, which finds the vectors close to a query vector in a vector database, has become a core operation of many applications such as information retrieval~\cite{r156}, recommendation~\cite{r154}, and retrieval-augmented generation~\cite{r157}. However, exact $k$-NN search on large-scale high-dimensional data is very expensive, and even classical indexes such as pivot-based indexes and metric trees are known to face lower bounds stating that they must examine most of the data in high dimensions~\cite{r70,r71}. For this reason, practice widely relies on approximate nearest neighbor (ANN) search, which gives up some accuracy in exchange for a large reduction in latency~\cite{r23,r158}. Graph-based indexes~\cite{r55,r56}, partition-based indexes such as IVF~\cite{r23,r24}, and quantization-based methods~\cite{r19,r25,r62} have achieved both high recall and low latency on real data. The search accuracy of these methods is usually set by tuning parameters such as $n_{\rm probe}$ or efSearch until the average recall reaches a target, and the chosen value is applied equally to every query~\cite{r3,r13,r88}.

\begin{figure}[t]
\centering
\includegraphics{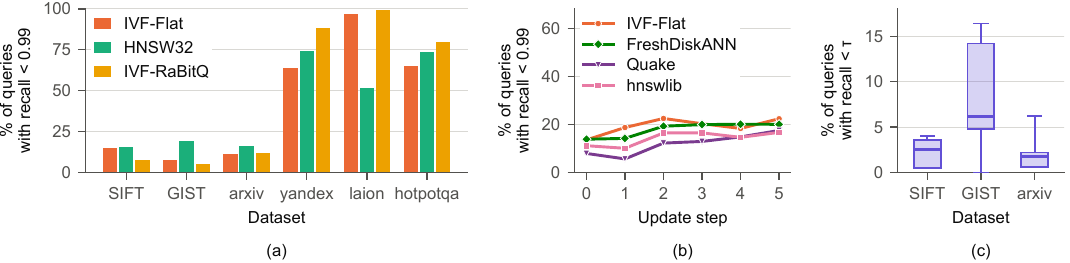}
\refstepcounter{figure}\label{fig:motivation}
\par\vspace{3pt}{\small\setstretch{1.15}\IEEEfigcap{Figure \thefigure. Per-query failures of average-recall tuning and distribution-level guarantees. (a) Indexes tuned to average recall 0.99 ($k = 100$; the left three datasets have in-distribution queries and the right three datasets have real out-of-distribution queries). (b) An index updated after tuning (yandex, $k = 100$, 10\% insertion and 10\% deletion per step). (c) CTR (one box contains the 9 settings of one dataset). The datasets are described in Table~\ref{tab:datasets}.}\par}
\end{figure}

A high average recall does not mean that the result of an individual query is accurate. Benchmark queries tend to be easier than real workloads~\cite{r47}, and since the average is determined by such queries, the target can be met even if some hard queries have low recall. In our measurements, IVF-Flat, HNSW, and IVF-RaBitQ indexes tuned to an average recall of 0.99 met the target average recall on in-distribution queries, yet at $k = 100$, 5.0–19.2\% of the queries had recall below 0.99 (Fig.~\ref{fig:motivation}(a), Section~\ref{sec:motivation}). It has also been reported that 4.8\% of the queries had recall 0 on DiskANN with an average Recall@10 of 0.9~\cite{r79}. When the distribution of queries differs from that of the data, the performance of ANN indexes drops sharply~\cite{r38,r122,r123}. In the same measurements, when queries that search for images by text or real user questions were sent to indexes tuned with data vectors in the absence of a query log, the fraction of queries with recall below 0.99 was 51.4–99.6\% at $k = 100$ (Fig.~\ref{fig:motivation}(a), Section~\ref{sec:motivation}).

How much the recall of an individual query matters depends on the application. In applications where the returned results are interchangeable, such as retrieval-augmented generation and recommendation, missing some neighbors does not significantly degrade downstream quality~\cite{r77,r144,r153,r154}. In other applications, however, a single missed neighbor is itself an error, and the average recall is not sufficient. For example, in image copy detection, a missed copy keeps circulating. Copy detection is performed by matching learned image descriptors~\cite{r39,r148} with ANN search, and a large-scale service processing billions of images per day~\cite{r41} explicitly aims to increase recall while maintaining precision~\cite{r142}. The same holds for vector databases and vector extensions of relational DBMSs. These systems provide vector search as an operation of query processing~\cite{r119,r134,r135,r152}, and its result set becomes the input of filters, joins, and aggregates, so a missed neighbor leads to an error in the final query result. In our measurements, tag aggregates over the results of indexes tuned to an average recall of 0.99 differed from those over exact results for up to 27\% of queries (Section~\ref{sec:scale}). However, these systems only let users specify a target accuracy or search parameters and give no guarantee on the result of an individual query~\cite{r119,r134,r135}. Moreover, in searches where the cost of a missed document is high, such as legal document review and systematic literature reviews, the goal is recall as close to 100\% as possible~\cite{r45}. Approximate query processing (AQP) has met the same need for aggregates by attaching to each answer an error bound that holds with a stated probability over sampling~\cite{r100,r95,r96}. Vector operators lack such a contract: methods that give per-query guarantees in vector search are rare, and even these are limited to per-neighbor probabilities, that is, the expected recall, or to a single nearest neighbor (Section~\ref{sec:related}).

The problem grows in environments where data are continuously updated. The data are not fixed in the applications mentioned above either. Copy detection systems continuously inspect newly uploaded images, and misinformation newly verified by fact-checking organizations as well as new variants found by the system again become subjects of matching and review~\cite{r142}. In a DBMS, vector columns are inserted, modified, and deleted like any other column, and Oracle advises checking the accuracy and rebuilding the index because such updates can degrade the accuracy of IVF indexes over time~\cite{r134}. In applications where documents are constantly added, modified, and deleted, such as Wikipedia search, the distributions of queries and updates also change over time, and such dynamic workloads increase the latency of the index and degrade recall~\cite{r50}. Accordingly, methods that maintain index quality under updates have been studied, such as graph repair~\cite{r93}, splitting and merging of partitions~\cite{r48,r50}, and local reclustering according to reconstruction error~\cite{r49}. However, as Quake~\cite{r50} pointed out, once the index structure changes, the initially chosen parameters cannot maintain the same recall. In our measurements as well, when five steps of 10\% insertions and deletions were applied to an index initially tuned to an average recall of 0.99 on yandex, the fraction of queries with recall below 0.99 rose from 13.5\% to 22.3\% for IVF-Flat, and from 7.9\% to 17.5\% even for Quake~\cite{r50}, which takes a recall target directly ($k = 100$, Fig.~\ref{fig:motivation}(b), Section~\ref{sec:motivation}). Retuning the parameters is costly because it again requires calibration queries with ground truth~\cite{r3,r13,r88}.

Research on guaranteeing the accuracy of search results has also been conducted. RaBitQ~\cite{r62} used a random transformation to provide probabilistic error bounds for each individual distance estimate, and this paper starts from this result. ADSampling~\cite{r64} turns error bounds of random projections into a high-probability guarantee of returning the $k$ nearest candidates, which becomes a guarantee of exact search when all vectors are candidates. Its upper bound, however, contains a constant of unknown value, and it targets neither recall below exactness nor updates. ConANN~\cite{r1} and CTR~\cite{r11} presented distribution-level guarantees on recall using conformal prediction. Their guarantees, however, concern a risk averaged over the query distribution, and, as the authors of ConANN noted, per-query conditional guarantees cannot be obtained by conformal methods~\cite{r1,r6}. In our measurements, in 27 settings in which CTR was run with its own loss target, up to 16.4\% of the queries had recall below the target $\tau$, which its averaged guarantee permits (Fig.~\ref{fig:motivation}(c), Section~\ref{sec:cost}). Per-query probabilistic guarantees have been given by PUFFINN~\cite{r179}, which scans an LSH index and stops once each true neighbor has been found with probability at least $1-\delta$; by Confirmation Sampling~\cite{r169}, which repeatedly queries LSH data structures to find the exact nearest neighbor of each query with probability $1-\delta$; and by TPU-KNN~\cite{r168}, which computes all distances and then meets a target expected recall under an assumption on the placement of the neighbors. Among LSH methods from the database community, SRS~\cite{r186}, VHP~\cite{r189}, and R2LSH~\cite{r191} also return the exact nearest neighbor of each query with any given probability over data-independent random projections. However, PUFFINN and TPU-KNN bound each true neighbor's miss probability, that is, the per-query expected recall, from which Markov's inequality gives tail bounds on recall. PUFFINN is limited to cosine and Jaccard similarity and does not support deletion, and its public implementation, in the setting covered by its proof, was 27–76$\times$ slower than our method at one thread (Appendix~\ref{app:puffinn}). Confirmation Sampling, SRS, VHP, and R2LSH address a single nearest neighbor ($k = 1$), and TPU-KNN presupposes computing all distances. Moreover, to the best of our knowledge, none of these guarantees was analyzed under updates (SRS, VHP, and R2LSH support updates through tree indexes but state no guarantee under them), and the dynamic index studies mentioned above~\cite{r48,r49,r50,r93} were evaluated by average recall (Section~\ref{sec:related}).

This paper addresses the problem of $k$-ANN search that guarantees per-query recall in dynamic vector databases. Given a target recall $\tau \in (0, 1]$ and a failure probability $\delta$, the goal is to guarantee $\Pr[\mathrm{Recall}(q) < \tau] \le \delta$ for every given query without assumptions on the data or the query distribution, provided only that the queries and updates do not depend on the randomness inside the index, whose seed is kept secret from whoever issues them; we call this the \textit{$(\tau, \delta)$-recall guarantee}. Exact search ($\tau = 1$) is a special case of it. We also guarantee $\mathbb E[\mathrm{Recall}(q)] \ge \tau$ for every query, which we call the \textit{expected-recall guarantee}. In both guarantees, the probability and the expectation are taken not over multiple queries but, for a given single query, over the randomness inside the index. Hence the expected-recall guarantee ensures for each individual query the goal of conventional tuning, which brought the average recall over many queries to at least $\tau$. In applications where missing some neighbors is acceptable, such as retrieval-augmented generation and recommendation, it suffices that the recall of each query reaches the target on average, so the expected-recall guarantee meets that requirement at lower cost than the $(\tau, \delta)$-recall guarantee. These guarantees must hold without retuning while insertions and deletions continue. To give a per-query guarantee, the probability of missing true neighbors remaining in unvisited regions must be bounded, but methods that do not inspect unvisited regions, such as graph traversal, cannot directly count the neighbors remaining in those regions. Re-ranking an HNSW result with exact distances removes false neighbors but cannot reveal missed ones, since finding them requires examining the unvisited vectors.

To solve this problem, we propose PQR-Search (per-query recall search), an index and search method that defines the probability of the guarantee not over the query distribution but over randomness inside the index. Its index applies a random orthogonal transformation to vectors and stores them as 1-bit binary codes. Given a query, it first computes the exact distances to some vectors and uses the $k$-th of these distances as the pruning radius, progressively tests with codes that the remaining vectors cannot lie within this radius and prunes them, and then re-ranks only the remaining vectors by exact distances. Whenever an exact distance is computed during the test, the radius is shrunk to prune more vectors. Under a random orthogonal transformation, the test error follows a data-independent distribution~\cite{r62}, so an upper bound on the probability of missing a true neighbor can be computed for the given data and query. The premise needed for this guarantee is that the queries, the updates, and the construction of the index are independent of the orthogonal transformation, so the same guarantee holds after updates without retuning. What is encoded here is not the vector itself but its residual with respect to the centroid of an IVF partition, and the vectors whose exact distances are computed first are also chosen from partitions whose centroids are close to the query. Since the error of a distance estimated from a 1-bit code is proportional to the magnitude of the residual~\cite{r62}, the smaller the residual, the more accurate the test, and residual norms give deterministic bounds via the triangle inequality~\cite{r15,r16}. Why this design provides the same guarantee at lower cost than exhaustive search is intuitive. To determine the result, most vectors need not the value of their distance but only a test showing that they cannot lie within the pruning radius; for vectors far from the radius this test is settled by a binary code of tens of bytes, so the original vectors are read to compute distances only for the few vectors that pass the test (Section~\ref{sec:costlaw}). This design also favors high-dimensional data. The cost of testing unvisited vectors, as with scanning compressed codes such as PQ~\cite{r19}, is determined not by the dimension of the data but by the number of bits of the binary code, so for data of higher dimension, such as embedding vectors~\cite{r38}, the gain over exhaustive search, which computes all distances, when queries are processed one at a time actually becomes larger. In our measurements, with the first-stage code length fixed, this speedup grew from 19.9–24.6$\times$ when the same data were reduced to 128 dimensions to 40.4–62.6$\times$ at the original dimensions (768–1024) (Section~\ref{sec:scale}). The design also favors updates. As updates continue, vectors drift away from the centroids to which they are assigned and the search cost increases, so a dynamic index maintains an efficient index, that is, an index with low search cost, by moving centroids or splitting partitions. Since insertion, deletion, and maintenance all take place per partition, an update affects only part of the index, and the conditions under which the guarantee holds during updates are simple.

The main contributions of this paper are as follows.

\begin{itemize}
\item \textbf{A per-query contract for vector operators (Sections~\ref{sec:problem}, \ref{sec:devices}, \ref{sec:dynamic}).} We show by measurement that average-recall tuning and distribution-level guarantees based on conformal prediction cannot protect individual queries, and, like an AQP error bound, we define as the problem a $(\tau, \delta)$-recall guarantee that is stated before execution and holds for every fixed dataset and query. The guarantee covers true neighbors in unvisited regions as well, and since its only randomness is the random orthogonal transformation $P$ inside the index, it holds for out-of-distribution queries too, and any workload chosen without looking at $P$ expects at most a $\delta$ fraction of failing queries. The principle that keeps this guarantee cheap is to let the guarantee come from $P$ alone, to make the decisions that affect the guarantee (the stage configuration, maintenance operations, and their timing) by predictions that do not read $P$, and to confine the choices that may depend on $P$ (the test order and early exact computation) to what Theorem 5 permits. Decisions based on values that depend on $P$, such as measured pass counts, introduce selection bias; this principle is therefore what keeps the guarantee intact without retuning, even under updates.
\item \textbf{Per-query recall guarantees with exact quantiles (Sections~\ref{sec:rule}, \ref{sec:guarantee}, Appendices~\ref{app:proofpass}, \ref{app:lowerbound}).} For search that tests unvisited vectors with prefix-readable binary codes of rotated residuals, we show that the $(\tau, \delta)$-recall guarantee holds for every given query without assumptions on the data or query distribution, for queries and updates that do not depend on the index's internal randomness, and that for $\tau < 1$ the same run also satisfies the expected-recall guarantee under a mild condition. The guarantee theorems themselves are short, because they build on the error identity of RaBitQ~\cite{r62} and the union bound over the $k$ true neighbors of ADSampling~\cite{r64}. What is new is the condition under which a radius set from exact distances of visited vectors also applies to unvisited true neighbors; the exact error distributions of prefix codes and 2-bit codes, stated for any code that is a function of $P^{-1}\boldsymbol{o}$, which give \textit{error-bound parameters} from exact Beta quantiles instead of an unspecified constant (with these quantiles alone, even a RaBitQ full scan becomes a guaranteed search, 12.1–20.5$\times$ faster than FAISS exhaustive search in geometric mean; Section~\ref{sec:cost}); the \textit{adaptive radius} (Theorem 5), with which we make explicit that the guarantee survives a radius that shrinks during testing although it depends on $P$, also with per-thread radii and stale radii; and the soundness of the implementation including floating-point error (Section~\ref{sec:process}, Appendix~\ref{app:proofpass}). ADSampling's per-comparison argument uses only that the distance is within the threshold, so it already tolerates such a dependence, but it does not discuss it, and its verification avoided it with a fixed threshold~\cite{r64}. Moreover, for analyses that use only per-neighbor probabilities, the Markov step of the bound is tight; it is attained exactly only by a placement of exact duplicates with $k = m + 1$ (Proposition 1). Separately from the Markov step, we give lower bounds on cost: in a probe model, a per-query guarantee must probe at least a $1 - \delta^{1/(m+1)}$ fraction of the hidden cells in expectation, and a deterministic rule all but $m$ of them (Theorems 7 and 8), and under randomized SETH no data structure with polynomial preprocessing can give such guarantees in query time $n^{1-\epsilon}$ on worst-case inputs with $D = \omega(\log n)$ (Theorem 9 in Appendix~\ref{app:lowerbound}).
\item \textbf{Guarantees under updates (Section~\ref{sec:dynamic}).} We extend the index and search algorithm of Section~\ref{sec:search} to dynamic environments by adding only a delta segment that holds new vectors separately, a bitmap that marks deletions, and merges. An insertion consists of encoding one vector and rearranging the delta region of its partition, and a deletion of marking one bit; unlike graph indexes, no connections need to be repaired. We show that if the updates and the decisions for maintaining an efficient index are independent of the orthogonal transformation, the static guarantee holds for a query at any time without retuning and without splitting the failure probability $\delta$ among segments. We also show that if the same orthogonal transformation keeps being used, the failure of a query can persist to later times, and that a fresh orthogonal transformation at merge time breaks this persistence, and we analyze the risk of adaptive updates. The same frame covers filtered $k$-NN: marking the vectors that fail a predicate in the deletion bitmap keeps the guarantee against the true neighbors among the qualifying vectors for any selectivity (Corollary 2). As a result, the decisions for maintaining an efficient index do not affect the guarantee and change only the search cost, and we determine them with the formula of Section~\ref{sec:costlaw}.
\item \textbf{Residual cost law (Section~\ref{sec:costlaw}).} We derive a formula that predicts the pass rate of the first-stage test, which sets the cost of the same guarantee, without data-fitted constants and without reading the orthogonal transformation. The formula shows that the cost is determined by the residual scale and the test radius; it thus answers quantitatively the normalization (centroid) choice that RaBitQ~\cite{r62} left as future work, and it grounds the residual codes and the index design. Following the principle above, the formula sets the parameters of the index (Section~\ref{sec:devices}) and, in dynamic settings, the operations that keep the index efficient and their timing (Section~\ref{sec:dynamic}) without retuning.
\item \textbf{Index and search (Section~\ref{sec:search}).} Following the theory, we design an IVF index and a query processing algorithm and implement them with AVX-512. Unlike standard IVF, vectors in unvisited partitions are also tested, so the guarantee holds regardless of $n_{\rm probe}$, and queries are processed fast with SIMD tests and multithreading within what the theorems of Section~\ref{sec:theory} permit. The parameter selection rule based on the cost law changes only the query processing cost, so the guarantee is unchanged even if the distributions of calibration queries and actual queries differ.
\item \textbf{Evaluation (Section~\ref{sec:experiments}).} We validated each contribution on 20 datasets (up to 100M points) reflecting the applications above (copy detection, cross-modal search, document retrieval, and database operators) and on the official NeurIPS'23 streaming runbook~\cite{r40}. Failure rates of the per-query guarantees stayed at most $\delta$ in every audit. In 168 static settings the $(\tau, \delta)$-recall guarantee failed 45 times out of 4,176,000 (at most 0.5\% of $\delta$ in each configuration), and, unlike CTR, which gives a distribution-level guarantee at the same target, no query had an average recall over orthogonal transformations significantly below the target. The price of the guarantee was modest. Where HNSW32 tuned to an average recall of 0.99 met the target on calibration queries, PQR-Search had 2.5$\times$ its 1-thread latency and HNSW32 had 1.3$\times$ higher batch throughput (geometric means), yet HNSW32 still left 3.3–29.2\% of test queries below recall 0.99. Tuned instead to the tail target (at most 1\% of calibration queries below recall 0.99), HNSW32 reached it within its search grid in only 10 of 18 settings and was there 1.1$\times$ (1-thread) and 1.9$\times$ (batch) slower than PQR-Search, whereas at a target of 0.9 HNSW32 tuned post hoc on the test queries was 2.0$\times$ faster at 1 thread. Widening the tail-tuned baselines on out-of-distribution queries to HNSW with $M$ up to 64 and efSearch up to 16,384, DiskANN, RoarGraph, and multi-bit RaBitQ kept PQR-Search 2.9$\times$ (1-thread) and 5.3$\times$ (batch) faster than the fastest baseline reaching 0.99, whereas on two 100M datasets with in-distribution queries tail-tuned HNSW32 was 4.8$\times$ faster at 1 thread. In tag aggregates over the search results, indexes tuned to an average recall of 0.99 answered differently from exact search for up to 27\% of queries, and PQR-Search for at most 0.04\% of trials. The pass-rate prediction of the cost law agreed well with the measurements across partitioning methods, scales, and updates without data-fitted constants (log correlation of at least 0.999 for static indexes). Our index and search were faster than FAISS exhaustive search, which gives the same guarantee, in all 52 main settings (geometric means of 22.9$\times$ when queries are processed one at a time and 26.4$\times$ when 1,000 queries are processed at once) and in every setting on the three 100M-scale datasets, and in all 52 main settings also faster than IVF whose $n_{\rm probe}$ was chosen afterwards. They were also 18.3$\times$ faster at one thread than PDX-BOND~\cite{r194}, an exact scan that prunes with partial distances. The parameter selection rule was 1.18$\times$ faster than the default configuration in geometric mean without slowing down any setting. The RaBitQ-equivalent configuration (all partitions tested with the full code), which gives the same guarantee, was also 12.1–20.5$\times$ faster than FAISS exhaustive search in geometric mean per metric, and our method was faster than it by another 1.64$\times$ (1-thread), 2.33$\times$ (32-thread), and 1.24$\times$ (batch), and 3.07$\times$ faster than the official RaBitQ implementation run with the same partitions and error-bound parameter (1-thread). It was 27–76$\times$ faster at one thread than the public implementation of PUFFINN, which also gives per-query guarantees, in the setting covered by its proof. On the streaming runbook, the failure rate stayed at most $\delta$ in all 33 search steps without retuning, whereas IVF-Flat tuned at the first time point exceeded its target failure rate by up to 4.2$\times$, and residual-based maintenance made search 1.29–8.97$\times$ faster than an index without maintenance.
\end{itemize}

The rest of this paper is organized as follows. Section~\ref{sec:background} shows by measurement the per-query failures of indexes tuned by average recall, defines the problem, and reviews related work and the prior theory on which our method is based. Section~\ref{sec:theory} presents the theory of per-query guarantees and the pass-rate formula that determines the cost. Section~\ref{sec:search} describes the index and search algorithm designed from these results, their implementation, and parameter selection based on the cost law. Section~\ref{sec:dynamic} addresses guarantees under updates and how to maintain an efficient index based on the formula of Section~\ref{sec:costlaw}. Section~\ref{sec:experiments} presents the experimental results, and Section~\ref{sec:conclusion} concludes the paper. Appendix~\ref{app:proofs} contains the proofs, Appendix~\ref{app:lowerbound} the lower bound for per-query guarantees, and Appendix~\ref{app:extra} supplementary experiments. The code, the scripts for all experiments and audits, and the result files from which all tables and figures are generated are available at \url{https://github.com/suanlab/pqr-search}.

\bigskip

\section{Background and Related Work}\label{sec:background}

In this section, we first show by measurement that indexes tuned by average recall miss neighbors on some queries, that such queries increase sharply when the query distribution differs from the distribution used for tuning, and that the initial tuning cannot maintain recall when the data are updated (Section~\ref{sec:motivation}). We then summarize the definitions of recall and of guarantees and formally define the problem addressed in this paper (Section~\ref{sec:problem}), and classify related work by the form of its guarantee (Section~\ref{sec:related}). Finally, we summarize the prior theory that our method inherits (Section~\ref{sec:prelim}).

\subsection{Motivating Experiments}\label{sec:motivation}

Widely used ANN indexes set their parameters so that the average recall measured on calibration queries reaches a target~\cite{r3,r13,r88}. To examine what this practice guarantees for individual queries, we repeated 20 times an experiment that splits the queries in half into calibration and test queries on three 1M-scale datasets (SIFT, GIST~\cite{r19}, arxiv~\cite{r38}) (see Section~\ref{sec:setup} for the data and the measurement method). When the parameters of IVF-Flat, HNSW32, and IVF-RaBitQ were set to an average recall of 0.99 on the calibration queries, the average recall of the test queries met the target, but the fraction of queries with recall below 0.99 was 5.8–7.3\% at $k = 10$ and 5.0–19.2\% at $k = 100$ (Fig.~\ref{fig:motivation}(a)). IVF-RaBitQ at $k = 10$ did not reach an average of 0.99 even on the calibration queries on SIFT and GIST and was excluded from these ranges (see Section~\ref{sec:setup} for the implementation).

This problem becomes much more severe when the distribution of real queries differs from the distribution used for tuning. In benchmarks, it is customary to split the data randomly and use the smaller part as queries~\cite{r47}, and without a query log, data points must be used for tuning as well. In our measurements, when real queries were sent to indexes tuned in this way to an average recall of 0.99, the fraction of queries with recall below 0.99 rose to 19.9–98.4\% at $k = 10$ and 51.4–99.6\% at $k = 100$, and 8.1–95.8\% of queries fell even below recall 0.9 at $k = 100$ (yandex and laion, which search for images by text, and hotpotqa~\cite{r38}, which consists of real user questions; Fig.~\ref{fig:motivation}(a)). For IVF-Flat on laion, the average recall dropped to 0.37 at $k = 10$, and 60\% of the queries had recall below 0.5. In contrast, when tuning used, instead of data vectors, a separate set of 1,000 real queries drawn from the same distribution as the test queries (corresponding to a query log), in the cases that reached an average recall of 0.99, the fraction of queries with recall below 0.99 at $k = 100$ was 4.5–18.7\%, on the same level as the 5.0–19.2\% for in-distribution queries. Hence the cause of this failure can be attributed to the distribution mismatch between the queries used for tuning and the real queries. Even when tuned on a query log, HNSW32 on laion ($k = 100$) and IVF-RaBitQ on yandex and laion did not reach an average recall of 0.99 even at the largest search parameter, and the HNSW32 result agrees with prior reports that graph-based ANN struggles with out-of-distribution (OOD) queries~\cite{r122,r123}.

When the data are updated, the initially tuned parameters cannot maintain the same recall. We built an index on 50\% of the yandex data and tuned it to an average recall of 0.99, and then, without retuning, applied five consecutive steps, each inserting 10\% of the data arriving in cluster order and deleting the oldest 10\%. The fraction of queries with recall below 0.99 at $k = 100$ rose from 13.5\% to 22.3\% for IVF-Flat, from 13.8\% to 20.0\% for FreshDiskANN, and from 11.1\% to 16.7\% for hnswlib, and also rose from 7.9\% to 17.5\% for Quake~\cite{r50}, which takes a recall target directly (Fig.~\ref{fig:motivation}(b)). The per-query shortfalls of CTR~\cite{r11}, which gives a distribution-level guarantee, are also shown in Fig.~\ref{fig:motivation}(c), and the settings are given in Section~\ref{sec:cost}.

\subsection{Problem Definition}\label{sec:problem}

Let $\mathcal D$ ($\lvert\mathcal D\rvert = n$) be the set of data vectors $\boldsymbol{o}_r \in \mathbb R^D$, and let $\boldsymbol{q}_r$ be the vector of query $q$. Each data vector has a distinct id, and distances are Euclidean. The $k$ vectors closest to the query form the true $k$-nearest neighbor set $\mathcal N_k^*(q)$ (ties are broken by a fixed rule), and $d_i$ denotes the distance from the query to the $i$-th closest vector ($d_1 \le \dots \le d_n$). Letting $\mathcal N_k(q)$ be the set of $k$ vectors returned by the search, we define the recall@$k$ of query $q$ as \begin{equation}\mathrm{Recall}(q) = \frac{\lvert\mathcal N_k(q) \cap \mathcal N_k^*(q)\rvert}{k}\label{eq:recall}\end{equation} (we omit $k$, which is determined by the context)~\cite{r11}. As in SPANN~\cite{r24}, however, a vector at the same distance as the true $k$-th neighbor counts as a correct answer that takes the corresponding place in $\mathcal N_k^*(q)$ (Appendix~\ref{app:proofstatic}). A search result is exact if $\mathrm{Recall}(q) = 1$. For a target recall $\tau$, letting $m = \lfloor(1-\tau)k\rfloor$ be the number of allowed misses, $\mathrm{Recall}(q) < \tau$ is equivalent to missing at least $m + 1$ neighbors.

The term recall guarantee differs greatly in meaning depending on what the probability is defined over. Table~\ref{tab:levels} divides recall guarantees into five levels by placing, between empirical tuning without a guarantee (L0) and deterministic exhaustive search (L4), the marginal guarantee~\cite{r4} and the high-probability guarantee over the calibration set~\cite{r99}, which the conformal prediction literature distinguishes, and the per-query guarantee over the randomness inside the algorithm~\cite{r66}. The input-conditional guarantee of the conformal prediction literature is not included as a level, because it cannot be obtained without distributional assumptions~\cite{r6}. Methods that tune by average recall, like the indexes of Section~\ref{sec:motivation}, give no guarantee for new queries (L0). Methods based on conformal prediction give an L1 guarantee~\cite{r1,r4}, which bounds the risk averaged over the calibration set and the query distribution, or an L2 guarantee that holds with high probability over the calibration set~\cite{r99}. However, since the L1 guarantee is an average over calibration sets, the risk can exceed the target for a single drawn calibration set~\cite{r99}, and above all, such guarantees do not prevent failures of individual queries. This paper targets the L3 guarantee for each given query, provides it at lower cost than exhaustive search (L4) without assumptions on the query distribution, and maintains it under updates. Prior work that guarantees only per-neighbor probabilities or a single exact nearest neighbor per query is listed separately as L3 (part) (Section~\ref{sec:related}).

\Needspace{16\baselineskip}
{\centering \textbf{Table \refstepcounter{table}\label{tab:levels}\thetable{}. Levels of recall guarantees.}\par}\nopagebreak

{\small\setstretch{1.1}
\begin{xltabular}{\linewidth}{YYYY}
\toprule
\textbf{Level} & \textbf{Form} & \textbf{Probability space} & \textbf{Examples} \\
\midrule
\endfirsthead
\multicolumn{4}{l}{\emph{Table~\thetable\ (continued)}} \\
\toprule
\textbf{Level} & \textbf{Form} & \textbf{Probability space} & \textbf{Examples} \\
\midrule
\endhead
L4 Deterministic & $\mathrm{Recall}(q) = 1$ & None & Exhaustive search \\
\textbf{L3 Per-query} & $\Pr[\mathrm{Recall}(q) < \tau] \le \delta$ for every given $q$ (also satisfies $\mathbb E[\mathrm{Recall}(q)] \ge \tau$ if $\delta(m+1) \le (1-\tau)k$) & Randomness inside the index & \textbf{This work} \\
L3 (part) & Per-neighbor probability, or one exact nearest neighbor, for every given $q$ & LSH, random projections & PUFFINN~\cite{r179}, TPU-KNN~\cite{r168}, Confirmation Sampling~\cite{r169}, SRS~\cite{r186}, VHP~\cite{r189}, R2LSH~\cite{r191} \\
L2 High probability over the calibration set & $\Pr_{\text{cal}}[\text{risk} \le \alpha] \ge 1 - \delta$ & Calibration set & LTT certification~\cite{r99} of CTR~\cite{r11} \\
L1 Marginal expectation & $\mathbb E_{\text{cal},q}[\mathrm{loss}] \le \alpha$ & Calibration set and query distribution & ConANN~\cite{r1}, $\alpha$ target of CTR~\cite{r11} \\
L0 Empirical average (no guarantee) & Average $\mathrm{Recall} \ge \tau$ on calibration queries (measured) & None & HNSW/IVF tuning~\cite{r55,r23}, DARTH~\cite{r13}, LAET~\cite{r3}, Quake~\cite{r50} \\
\bottomrule
\end{xltabular}}

In this paper, the only randomness over which the probability of the guarantees is defined is the random orthogonal transformation $P$ used by the index, where $P$ is a rotation drawn uniformly (from the Haar measure) from $O(D)$, the set of all $D \times D$ orthogonal matrices; we generate it by QR decomposition of a $D \times D$ matrix with independent standard normal entries, correcting the signs of the columns of $Q$ by the signs of the diagonal of $R$. $P$ is drawn without looking at the data or the queries when the index is built, and is replaced according to the schedule of Section~\ref{sec:dynguarantee}. Since the error of the binary code test depends on $P$, it is random even for a given query, and its distribution is determined regardless of the data and the query (Section~\ref{sec:rule}). The failure probability in this paper is the probability of missing a true neighbor because of this error. This framework, which fixes the query and defines the probability over the randomness inside the algorithm, is the same as that of LSH theory~\cite{r66}, and we apply it to a target on $k$-NN recall rather than an approximation ratio. All probabilities $\Pr$ and expectations $\mathbb E$ in this paper are over $P$, and the data, the query, and everything determined without looking at $P$ (partition centroids and assignments, the partitions to visit, the initial radius (determined in Section~\ref{sec:guarantee}), the error-bound parameters, the decisions of the \textit{parameter selection rule} made from the prediction of Eq.~\eqref{eq:passprob}, and so on) are regarded as fixed. Decisions to enable techniques that are made by measuring times that depend on the orthogonal transformation also do not change the upper bound on the failure probability (Section~\ref{sec:devices}). Therefore, all guarantees in this paper hold for fixed data and for each fixed (given) query. Within this framework, the problem addressed in this paper is as follows. Given a target $\tau \in (0, 1]$ and $\delta \in (0, 1)$, the problem is to perform, as fast as possible, a $k$-ANN search that guarantees for every query \begin{equation}\Pr[\mathrm{Recall}(q) < \tau] \le \delta\label{eq:goal}\end{equation} Eq.~\eqref{eq:goal} is the $(\tau, \delta)$-recall guarantee mentioned in Section~\ref{sec:intro}, and for $\tau = 1$ it is the exact-search guarantee (T1 in the experiment tables). The expected-recall guarantee, which guarantees $\mathbb E[\mathrm{Recall}(q)] \ge \tau$ for every query, is written E in the experiment tables. In general, neither guarantee implies the other, but the $(\tau, \delta)$-recall guarantee of our method also satisfies the expected-recall guarantee for $\tau < 1$ under a mild condition, and if only the expected-recall guarantee is needed, a lower-cost mode can be used (Section~\ref{sec:guarantee}). With the same method, we also handle a form that bounds, per query, the loss $\mathbb E[\max(0, \tau - \mathrm{Recall}(q))]$ that CTR bounds over the query distribution (Ectr), and range queries, which bound the probability of missing even one vector within radius $r$ (Section~\ref{sec:guarantee}).

We note two points about these guarantees. First, the queries and the updated points must be independent of $P$ (in practice, the seed from which $P$ is drawn is kept secret from whoever issues them), because an adversary who has seen the orthogonal transformation or the stored codes can make the failure probability 1 (Section~\ref{sec:dynguarantee}). Second, since all queries use the same orthogonal transformation, the fraction of failing queries in any workload chosen without looking at $P$ is at most $\delta$ in expectation, but the failures of multiple queries within the same index are not independent of one another, and a failing query may keep failing until the transformation is replaced (Section~\ref{sec:dynguarantee}). Meanwhile, a ratio guarantee, which states that the distance of the returned $k$-th point is within $1 + \varepsilon$ times the true $k$-th distance, is logically independent of recall guarantees, and in high dimensions it becomes nearly unrelated to recall because of distance concentration~\cite{r125}. For example, in our measurements on SSNPP~\cite{r39}, for 92\% of the queries the top 100 all lie within 1.1 times the nearest-neighbor distance, so even a result that misses the nearest neighbor (recall@1 = 0) satisfies the ratio guarantee with $\varepsilon = 0.1$. The guarantees of PAC-NN~\cite{r10}, LSH~\cite{r66}, and most LSH methods from the database community~\cite{r184,r185,r187,r188,r190,r192} are per-query, but since they concern such ratio guarantees, we exclude them from our comparison. Per-query guarantees based on LSH or random projections that allow ratio 1 and return the exact nearest neighbor~\cite{r169,r186,r189,r191} are discussed in Section~\ref{sec:related}.

Table~\ref{tab:notation} summarizes the notation of this paper. Vectors are written in bold italic, and quantities that depend on the test stage $\ell$ are marked with the superscript $(\ell)$. Each definition is explained again where it first appears.

\Needspace{16\baselineskip}
{\centering \textbf{Table \refstepcounter{table}\label{tab:notation}\thetable{}. Summary of notation.}\par}\nopagebreak

{\small\setstretch{1.1}
\begin{xltabular}{\linewidth}{YY}
\toprule
\textbf{Symbol} & \textbf{Meaning} \\
\midrule
\endfirsthead
\multicolumn{2}{l}{\emph{Table~\thetable\ (continued)}} \\
\toprule
\textbf{Symbol} & \textbf{Meaning} \\
\midrule
\endhead
$D$, $n$ & Dimension of the vectors, number of data vectors \\
$\mathcal D$, $q$, $\boldsymbol{q}_r$, $\boldsymbol{o}_r$ & Set of data vectors, query, query vector, data vector \\
$k$, $\mathcal N_k^*(q)$, $\mathcal N_k(q)$, $d_k$ & Number of neighbors to find, true $k$-nearest neighbor set, set of $k$ vectors returned by the search, distance to the true $k$-th neighbor \\
$\tau$, $\delta$, $m$ & Target recall, failure probability, number of allowed misses $\lfloor(1-\tau)k\rfloor$ \\
$P$ & Random orthogonal transformation (the probabilities of the guarantees are over $P$) \\
$\boldsymbol{c}$, $x$, $\Phi$ & Partition centroid, residual norms of the data vector and the query $\lVert\boldsymbol{o}_r - \boldsymbol{c}\rVert$, $\lVert\boldsymbol{q}_r - \boldsymbol{c}\rVert$ \\
$\boldsymbol{o}$, $\boldsymbol{q}$, $\rho$ & Unit vectors normalized with respect to the centroid, their inner product $\langle\boldsymbol{o}, \boldsymbol{q}\rangle$ \\
$r$, $\theta(r)$ & Pruning radius, pruning threshold (Eq.~\eqref{eq:theta}) \\
$L$, $m_\ell$ & Number of test stages, number of code coordinates read at stage $\ell$ \\
$\bar{\boldsymbol{x}}^{(\ell)}$, $\bar{\boldsymbol{o}}^{(\ell)}$, $\hat\rho^{(\ell)}$ & Binary code, code vector, and estimate of $\rho$ at stage $\ell$ (Eq.~\eqref{eq:est}) \\
$\kappa^{(\ell)}$, $\zeta^{(\ell)}$, $G_D$ & Magnitude of the estimation error (Eq.~\eqref{eq:kappa}), normalized error, its cumulative distribution function (Eq.~\eqref{eq:gd}) \\
$\varepsilon_{0,\ell}$, $f$ & Error-bound parameter, slack factor (Eq.~\eqref{eq:rule}) \\
$\sigma_{\det}$ & Deterministic slack added to the test in the implementation (Section~\ref{sec:process}) \\
$\boldsymbol{c}_0$ & Mean of all data (global centering, Section~\ref{sec:devices}) \\
$\beta$, $\delta_\ell$ & Failure probability budget, its per-stage share ($\beta = \sum_\ell\delta_\ell$) \\
$N$ & Number of partitions \\
$C$, $V$, $M$ & Candidate set, set of unvisited true neighbors, number of missed true neighbors \\
$g$, $R_{\mathrm{vis}}$ & Margin (Eq.~\eqref{eq:gap}), relative radius (the initial radius divided by the residual scale $\sqrt{x\Phi}$, Appendix~\ref{app:relradius}) \\
$L_t$, $\mathcal N_{k,t}^*(q)$, $\mathrm{Recall}_t(q)$ & Set of vectors live at time $t$, true $k$-nearest neighbor set over it, recall with respect to it (Section~\ref{sec:dynguarantee}) \\
$Q_H$, $C_{\rm m}(\alpha, p)$, $B_p(\alpha)$, $\alpha_p^*$ & Number of queries until the next merge, cost and benefit of applying maintenance operation $\alpha$ to partition $p$ (Eq.~\eqref{eq:maint}), operation with the largest benefit (Section~\ref{sec:maintenance}) \\
\bottomrule
\end{xltabular}}

\subsection{Related Work}\label{sec:related}

Table~\ref{tab:related} classifies related work by the form of its guarantee. Below, we first examine error-bounded distance estimation (RaBitQ~\cite{r62}, ADSampling~\cite{r64}) and conformal prediction (CTR~\cite{r11}, ConANN~\cite{r1}), which are closest to our work (summarizing in between the weaknesses under updates of RaBitQ-family methods that are fast on static data), and then discuss in turn recall prediction and dynamic indexes, prior work that gave per-query probabilistic guarantees (including LSH methods from the database community), approximate query processing, and lower bounds.

\Needspace{16\baselineskip}
{\centering \textbf{Table \refstepcounter{table}\label{tab:related}\thetable{}. Classification by the form of guarantee.}\par}\nopagebreak

{\small\setstretch{1.1}
\begin{xltabular}{\linewidth}{YYYYY}
\toprule
\textbf{Family} & \textbf{Examples} & \textbf{Guarantee} & \textbf{Per-query} & \textbf{Remarks} \\
\midrule
\endfirsthead
\multicolumn{5}{l}{\emph{Table~\thetable\ (continued)}} \\
\toprule
\textbf{Family} & \textbf{Examples} & \textbf{Guarantee} & \textbf{Per-query} & \textbf{Remarks} \\
\midrule
\endhead
Graph, IVF, quantization & HNSW~\cite{r55}, DiskANN~\cite{r56}, IVF-PQ~\cite{r19}, ScaNN~\cite{r25} & None & — & Tuned to average recall \\
Error-bounded distance estimation & RaBitQ~\cite{r62}, ADSampling~\cite{r64} & Probabilistic bound on each distance estimate (ADSampling: the $k$ nearest candidates) & Only within the explored or candidate range & Unexplored regions only as candidates; unknown constant \\
Target recall prediction & DARTH~\cite{r13}, LAET~\cite{r3}, Ada-ef~\cite{r88} & Prediction & No & — \\
Dynamic indexes & FreshDiskANN~\cite{r93}, SPFresh~\cite{r48}, Quake~\cite{r50}, Ada-IVF~\cite{r49} & None (evaluated by average recall) & No & Tuning breaks down after updates (Sections~\ref{sec:motivation}, \ref{sec:expdynamic}) \\
Conformal prediction & ConANN~\cite{r1}, CTR~\cite{r11} & Risk over the query distribution $\le \alpha$ & No & Does not prevent per-query shortfalls (Section~\ref{sec:cost}) \\
Distributional assumption & Auncel~\cite{r2} & Error bound under a local uniformity assumption & Under the assumption & Assumption rejected on real embeddings~\cite{r1} \\
Probabilistic approximation & PAC-NN~\cite{r10}, LSH~\cite{r66}, LSB-forest~\cite{r184}, C2LSH~\cite{r185}, QALSH~\cite{r187}, PM-LSH~\cite{r188}, LCCS-LSH~\cite{r190}, DB-LSH~\cite{r192} & $(1 + \varepsilon)$- or $c^2$-approximation (mostly with constant probability) & Per-query in form & Independent of recall guarantees (Section~\ref{sec:problem}); theorems mostly for $k = 1$ \\
LSH per-neighbor probability & PUFFINN~\cite{r179} & Finds each true neighbor with probability $1-\delta$ (expected recall $\ge 1-\delta$) & Per-query & Tail bound via Markov's inequality; cosine and Jaccard similarity; no deletion; not analyzed under updates \\
LSH and random-projection exact search & Confirmation Sampling~\cite{r169}, SRS~\cite{r186}, VHP~\cite{r189}, R2LSH~\cite{r191} & Exact nearest neighbor with a given probability $1-\delta$ & Per-query & $k = 1$ (mainly, for Confirmation Sampling); target $\tau < 1$ and guarantees under updates not addressed (SRS, VHP, R2LSH support updates) \\
Expected recall & TPU-KNN~\cite{r168} & Expected recall $\ge$ target & Per-query (under an assumption) & Computes all distances; assumes uniform placement of the top $k$ \\
Approximate query processing & Online Aggregation~\cite{r100}, BlinkDB~\cite{r95}, VerdictDB~\cite{r96} & Confidence intervals of aggregate answers (probability over samples) & Per answer & Error reported after execution; a pre-specified error bound cannot be guaranteed for every query~\cite{r193} \\
Deterministic & Exhaustive search & recall = 1 & Per-query & Cost linear in the data size \\
\textbf{This work} & — & $\Pr[\mathrm{Recall} < \tau] \le \delta$ (also satisfies $\mathbb E[\mathrm{Recall}] \ge \tau$ if $\delta(m+1) \le (1-\tau)k$) & \textbf{Per-query} & No assumptions; maintained under updates without retuning \\
\bottomrule
\end{xltabular}}

RaBitQ~\cite{r62} gives probabilistic error bounds for distance estimation with a random orthogonal transformation and 1-bit binary codes, and inside IVF it discards vectors in the probed partitions whose lower bound on the distance is larger than the distance of the nearest neighbor found so far, arguing that the true nearest neighbor in the probed partitions is sent to re-ranking with high probability (with an unspecified constant). Vectors in unprobed partitions are not covered and no per-query recall target is stated, and the choice of centroids was left as future work. Concurrent work, RaBitQ-SSD~\cite{r181}, gives an unbiased estimate and a probabilistic error bound from a prefix of the code alone, and uses it to discard, within the probed lists, candidates whose lower bound is at least the current $k$-th distance, reducing SSD reads. Its bound is a concentration inequality with an unspecified constant and its recall is reported empirically, whereas we show that the error of a prefix code follows an exact distribution and is independent of all of $P^{-1}\boldsymbol{o}$ (Lemma 1), which yields tests with exact quantiles and the pass-rate formula, and we guarantee per-query recall including unvisited partitions. ADSampling~\cite{r64} compares distances while progressively sampling coordinates after a random orthogonal transformation, and gives a per-query guarantee by arguing that if all distance comparisons for the true $k$-NN among the candidates are correct, those $k$ vectors are returned. This guarantee states that the $k$ closest among the candidates produced by the index are returned, which becomes a guarantee of exact search when all vectors are candidates (Corollary 4.2 of~\cite{r64}). However, since the upper bound on the failure probability contains a constant of unknown value, parameters matching a target failure probability cannot be computed (the implementation uses empirically determined values), and its analysis of a threshold that depends on the random transformation, the running $k$-th distance, is implicit: the per-comparison argument uses only that the distance is within the threshold, and its verification fixes the threshold (see the discussion after Theorem 5). This paper states the guarantee for the recall of the query by testing unvisited vectors with binary codes of IVF residuals, determines the error-bound parameters by exact quantiles of the Beta distribution (Section~\ref{sec:rule}), and shows by a formula how centroids affect the test cost (Section~\ref{sec:costlaw}).

Recent methods that push the time-accuracy trade-off of the RaBitQ family further on static data have weaknesses under updates (Table~\ref{tab:staticfast}). SymphonyQG~\cite{r65} computes, for each graph vertex, the RaBitQ codes of its neighbors with the vertex's own vector as the centroid, stores these duplicated codes next to the vertex, and sets out-degrees to fit the 32-vector batches of FastScan. Inserting or deleting a vector therefore requires recomputing the neighbor codes and rebuilding the batches of every vertex whose edges change, and the authors leave handling deletions by rebuilding the edges of the vertices pointing to the deleted vertex as future work. MRQ~\cite{r182} and SAQ~\cite{r183} apply a PCA projection learned from the data and concentrate bits on the leading high-variance dimensions, choosing the number of leading dimensions (MRQ) or the dimension segmentation and bit allocation (SAQ) from the per-dimension variances of the data. The error bound that MRQ uses for the discarded tail dimensions is likewise a Chebyshev bound based on the per-dimension variances of the training data and on the assumption that the data are centered. Both methods were evaluated only on fixed data; when vectors from a different distribution accumulate, the learned projection and bit allocation drift from the data, and retraining requires re-encoding every vector. Multi-bit RaBitQ (Extended RaBitQ~\cite{r63}) encodes each vector independently without training, but its encoding time grows as $O(2^B D\log D)$ in the number of bits $B$, which SAQ reduces (by more than 80$\times$ at $B = 9$). In our index, the only data-dependent components are the IVF centroids, and the codes are 1-bit prefix codes and a 2-bit final stage, so an insertion only encodes the new vector with the current centroid and orthogonal transformation and changes no other code (Section~\ref{sec:dynops}). IVF partitions confine the effect of an update to one partition, so the FastScan layout is rebuilt only for that partition of the delta segment, and entirely at a merge. Moreover, since our guarantee requires testing unvisited vectors with their codes, IVF, which stores one code per vector contiguously by partition, suits this test. When the distribution shifts, the guarantee holds and only the cost changes (Sections~\ref{sec:dynguarantee} and \ref{sec:expdynamic}).

\Needspace{16\baselineskip}
{\centering \textbf{Table \refstepcounter{table}\label{tab:staticfast}\thetable{}. Methods of the RaBitQ family that lead in time-accuracy on static data, and their updates.}\par}\nopagebreak

{\small\setstretch{1.1}
\begin{xltabular}{\linewidth}{llYY}
\toprule
\textbf{Method} & \textbf{Index} & \textbf{Determined from data} & \textbf{Extra work per insertion or deletion} \\
\midrule
\endfirsthead
\multicolumn{4}{l}{\emph{Table~\thetable\ (continued)}} \\
\toprule
\textbf{Method} & \textbf{Index} & \textbf{Determined from data} & \textbf{Extra work per insertion or deletion} \\
\midrule
\endhead
SymphonyQG~\cite{r65} & Graph & Edges (neighbor codes duplicated with the vertex vector as centroid) & Recompute neighbor codes and rebuild 32-vector batches of every vertex whose edges change \\
MRQ~\cite{r182} & IVF & PCA projection, number of leading dimensions, tail-dimension variances (used in the error bound) & Under distribution shift, retrain and re-encode all vectors \\
SAQ~\cite{r183} & IVF & PCA projection, dimension segmentation and bit allocation & Under distribution shift, retrain and re-encode all vectors \\
Extended RaBitQ~\cite{r63} & IVF & Centroids & Encode only the new vector (time $O(2^B D\log D)$) \\
\textbf{Ours} & IVF + delta segment & Centroids & Encode only the new vector (1-bit prefix, 2-bit), re-lay out only its partition \\
\bottomrule
\end{xltabular}}

CTR~\cite{r11} first certifies the search result of HNSW and rectifies only the queries that fail certification with a costlier exact search. The certification step applies conformal risk control~\cite{r4} and Learn-then-Test~\cite{r99} to scores obtained during the HNSW search. It defines the loss of a query as the amount by which recall falls short of $\tau$ (recall shortfall), $\max(0, \tau - \mathrm{Recall}(q))$, and shows that the expected loss of a new query drawn from the query distribution is at most $\alpha(1-\tau)$ plus a certification error term of the rectification step. ConANN~\cite{r1} calibrates a per-query $n_{\rm probe}$ of IVF by conformal prediction, and its guarantee concerns the average over the query distribution of the fraction of missed true neighbors (false negative rate). Both methods rely on the assumption that calibration queries and new queries come from the same distribution~\cite{r4}, and cannot give (conditional) guarantees for each individual input without distributional assumptions~\cite{r6}. Indeed, in our measurements, some queries of CTR had recall below $\tau$ on SIFT and GIST, the evaluation datasets of the CTR paper, and also on arxiv (Fig.~\ref{fig:motivation}(c), Section~\ref{sec:cost}). In contrast, the probability in our work is over the random orthogonal transformation of the index, so the guarantee holds for every given query, including out-of-distribution queries, and calibration queries are used only to determine the cost. CTR's loss can also be bounded by $\alpha(1-\tau)$ per query with the same method (Sections~\ref{sec:guarantee}, \ref{sec:cost}).

Methods that learn early termination or predict recall (LAET~\cite{r3}, DARTH~\cite{r13}, Ada-ef~\cite{r88}, Quake's APS~\cite{r50}) only predict recall and do not guarantee it. Pruning based on the triangle inequality or the law of cosines~\cite{r14,r15,r16} accelerates search by skipping vectors that cannot be answers without computing their distances. CLIP~\cite{r14} and TRIM~\cite{r16} relax the bounds and allow a slight drop in recall, whereas Tribase~\cite{r15} uses only pruning that does not change the result, so it becomes deterministic exact search when all partitions of IVF are searched. However, when only some partitions are searched, it yields the same recall as IVF with the same $n_{\rm probe}$ and does not guarantee per-query recall, and the speedup Tribase reported at recall 1 was also measured at the $n_{\rm probe}$ chosen so that the recall of the test queries became 1. We compare against Tribase searching all partitions as a baseline for the exact-search guarantee (Section~\ref{sec:cost}). Dynamic ANN studies maintain recall under updates by graph repair (FreshDiskANN~\cite{r93}), size-based splitting and merging (LIRE of SPFresh~\cite{r48}), splitting and merging driven by a cost model together with an adaptive $n_{\rm probe}$ (Quake~\cite{r50}), local reclustering according to reconstruction error (Ada-IVF~\cite{r49}), and periodic recomputation of centroids (DeDrift~\cite{r53}). However, among these,~\cite{r48,r49,r50,r93} were all evaluated by average recall without guarantees, and the theory of deletions in graphs~\cite{r147} also only shows that the random-walk statistics of the graph are preserved after deletions and does not guarantee the recall of query results. Our guarantee, on the other hand, holds without retuning if the updates and the decisions for maintaining an efficient index are independent of the orthogonal transformation (Theorem 6 in Section~\ref{sec:dynguarantee}), so the maintenance policy determines only the cost (Section~\ref{sec:maintenance}).

Prior works that gave per-query probabilistic guarantees that also cover true neighbors outside the candidates are PUFFINN, Confirmation Sampling, and TPU-KNN, together with the LSH methods from the database community discussed in the next paragraph. PUFFINN~\cite{r179} scans multiple LSH tries starting from the longest hash matches and stops after examining a number of tries computed from the collision probability of the current $k$-th candidate, thereby finding each of the true $k$-nearest neighbors with probability at least $1-\delta$. This is a guarantee that the expected recall of every query is at least $1-\delta$, of the same form as our expected-recall guarantee. Since the probability is per neighbor, the probability of finding all $k$ neighbors is at least $1-k\delta$ by the union bound, and since the expected number $M$ of missed neighbors is at most $k\delta$, Markov's inequality gives the tail bound $\Pr[M \ge m+1] \le k\delta/(m+1)$. With a per-neighbor probability of $\delta(m+1)/k$ this bound equals the $(\tau, \delta)$-recall guarantee; our bound uses the same Markov step, which cannot be improved given only per-neighbor probabilities (Proposition 1). However, the implementation supports only cosine and Jaccard similarity, filters candidates before exact distance computation with a sketch filter that the proof does not cover, and does not support deletion, and the original paper validated the guarantee only by average recall. Our guarantees contain PUFFINN's per-neighbor problem (Corollary 3 in Appendix~\ref{app:puffinn}) and hold under Euclidean distance and under updates including deletions. In a comparison with the same public implementation, its default setting missed even the average-recall target because of a candidate filter that the proof does not cover, and with the filter disabled it was 27–76$\times$ slower at one thread than our method configured for the expected-recall guarantee at the same level (Appendix~\ref{app:puffinn}). Confirmation Sampling~\cite{r169} queries, one after another, LSH data structures whose success probabilities are unknown, and returns the exact nearest neighbor of each query with probability $1-\delta$ without assumptions on the data. This work addresses the exact nearest neighbor for $k = 1$ (only outlining an extension to $k$-NN in its conclusion) and does not address a target recall $\tau < 1$ or updates. TPU-KNN~\cite{r168} computes all distances and then divides the data into several bins and collects only the closest point in each bin; under the assumption that the top $k$ are placed in the bins independently and uniformly, it computes the expected recall and chooses the number of bins that meets the target. Our expected-recall guarantee holds over the random orthogonal transformation of the index without such placement assumptions, and does not compute all distances.

The database community has developed many LSH variants with probabilistic guarantees. Like ours, their probability is over data-independent hash functions or random projections and therefore holds for every fixed dataset and query, but most of them bound the distance ratio of one returned point with constant probability. LSB-forest~\cite{r184} returns a 4-approximate nearest neighbor, and C2LSH~\cite{r185}, QALSH~\cite{r187}, PM-LSH~\cite{r188}, and DB-LSH~\cite{r192} a $c^2$-approximate one (QALSH with probability at least $1/2-\delta'$ for its parameter $\delta'$, PM-LSH and DB-LSH at least $1/2-1/e$, DB-LSH in sublinear time); LCCS-LSH~\cite{r190} answers the $(R, c)$-near-neighbor problem with probability at least $1/4$. Their $k$-NN variants also bound distance ratios, with guarantees absent, conditional on early termination (SRS~\cite{r186}), or omitted (VHP~\cite{r189}). SRS~\cite{r186}, VHP~\cite{r189}, and R2LSH~\cite{r191} also allow ratio $c = 1$ and return the exact nearest neighbor with any given probability, a per-query exact-search guarantee for $k = 1$; for SRS, the worst-case expected cost in this case is linear in the data size. SRS, VHP, and R2LSH keep the projections fixed and support insertions and deletions through R-trees or B$^+$-trees, so their guarantees might be read as surviving updates independent of the projections, but none of them states a guarantee under updates. We found no method in this line that bounds $\Pr[\mathrm{Recall}(q) < \tau]$ for $k > 1$ or analyzes its guarantee under updates.

Approximate query processing (AQP) is close to our work in that it trades exactness for speed under a probabilistic error bound. Online aggregation~\cite{r100} progressively reports a confidence interval that contains the final answer with a stated probability, computed from records retrieved in random order, and BlinkDB~\cite{r95} and VerdictDB~\cite{r96} answer from precomputed samples and attach confidence intervals (e.g., a 1\% error at 95\% confidence). Their probability is over sampling and ours is over a rotation inside the index, but both hold for fixed data and queries; and just as precomputed samples are shared by queries, our rotation is shared by all queries, so failures may be correlated across queries (Section~\ref{sec:problem}). The difference lies in what is fixed in advance. Sampling-based AQP fixes the sample or the time and reports the error after execution, and cannot guarantee a pre-specified error bound for every query~\cite{r193}; we fix the target $(\tau, \delta)$ before execution and let the cost vary per query.

Meanwhile, on adversarially constructed data, the search time of graph indexes is linear in the data size; for example, DiskANN does not find any of the top 5 neighbors before examining 10\% or more of the data~\cite{r60}. Theoretically as well, if SETH (strong exponential time hypothesis)~\cite{r106,r107,r108}, a hypothesis widely used as the basis of lower bounds in computational complexity theory, is true, then nearest neighbor search with a sufficiently small approximation error in high dimensions requires, even with polynomial-time preprocessing, time nearly linear in the data size per query~\cite{r112}, and the inner product similarity join cannot be solved essentially faster than quadratic time, even approximately~\cite{r111}. By a reduction of the same kind, we show that this lower bound also applies to per-query recall guarantees. That is, if randomized SETH is true, then for $D = \omega(\log n)$ no data structure with polynomial-time preprocessing that gives a per-query $(\tau, \delta)$-recall guarantee or expected-recall guarantee can achieve query time $n^{1-\varepsilon}$ on worst-case inputs (Theorem 9 in Appendix~\ref{app:lowerbound}). Thus, if SETH is true, no index that gives per-query guarantees without assumptions on the data can avoid nearly linear cost in the worst case, so this paper targets lower cost than exhaustive search not in the worst case but on real data.

\subsection{Preliminaries}\label{sec:prelim}

This section summarizes the residual binary codes of RaBitQ~\cite{r62} and their error, the triangle-inequality bounds of residual norms~\cite{r15,r16}, and the test argument of ADSampling~\cite{r64}, on which our index and search method are based. RaBitQ divides the data vectors into IVF partitions and normalizes each vector with respect to the centroid $\boldsymbol{c}$ of its partition. For a data vector $\boldsymbol{o}_r$ and a query $\boldsymbol{q}_r$, letting $x = \lVert \boldsymbol{o}_r - \boldsymbol{c}\rVert$, $\Phi = \lVert \boldsymbol{q}_r - \boldsymbol{c}\rVert$, $\boldsymbol{o} = (\boldsymbol{o}_r - \boldsymbol{c})/x$, and $\boldsymbol{q} = (\boldsymbol{q}_r - \boldsymbol{c})/\Phi$, we have \begin{equation}\lVert \boldsymbol{o}_r - \boldsymbol{q}_r\rVert^2 = x^2 + \Phi^2 - 2x\Phi\,\langle \boldsymbol{o}, \boldsymbol{q}\rangle\label{eq:dist}\end{equation} Since $x$ and $\Phi$ need to be computed only once per vector or per partition, distance computation reduces to computing the inner product of unit vectors $\rho = \langle \boldsymbol{o}, \boldsymbol{q}\rangle$. Whether a vector lies within distance $r$ of the query can also be decided by the inner product alone. We define the \textit{pruning threshold} for distance $r$ as \begin{equation}\theta(r) = \frac{x^2 + \Phi^2 - r^2}{2x\Phi}\label{eq:theta}\end{equation} Then, by Eq.~\eqref{eq:dist}, for $x, \Phi > 0$, $\lVert \boldsymbol{o}_r - \boldsymbol{q}_r\rVert \le r$ and $\rho \ge \theta(r)$ are equivalent. Moreover, residual norms give a deterministic bound through the triangle inequality $\lvert x - \Phi\rvert \le \lVert \boldsymbol{o}_r - \boldsymbol{q}_r\rVert$~\cite{r15,r16}, so only vectors with $\lvert x - \Phi\rvert \le r$ need to be tested. Hereafter we call this condition the \textit{triangle-inequality range}.

RaBitQ stores, as the binary code $\bar{\boldsymbol{x}} = \mathrm{sign}(P^{-1}\boldsymbol{o})/\sqrt D$, only the sign of each coordinate of $P^{-1}\boldsymbol{o}$, the unit vector $\boldsymbol{o}$ transformed by the inverse of a random orthogonal matrix $P$, and estimates the inner product with the code vector $\bar{\boldsymbol{o}} = P\bar{\boldsymbol{x}}$ as $\hat\rho = \langle\bar{\boldsymbol{o}}, \boldsymbol{q}\rangle/\langle\bar{\boldsymbol{o}}, \boldsymbol{o}\rangle$. The denominator does not depend on the query and is precomputed at indexing time, and the numerator $\langle\bar{\boldsymbol{x}}, P^{-1}\boldsymbol{q}\rangle$ can be computed fast with lookup tables (LUTs). Under a random orthogonal transformation, the error of this estimate follows a data-independent distribution. The error is proportional to $\sqrt{1-\rho^2}$ and to the deviation of the code vector from the direction of $\boldsymbol{o}$, and its proportionality coefficient $\zeta$ is symmetric about 0, has variance 1, and is close to the standard normal distribution when $D$ is large~\cite{r76}. Since the squares of the coordinates of a point uniformly distributed on the sphere follow a Beta distribution, its cumulative distribution function $G_D(s) = \Pr[\zeta < s]$ is given, for $s \ge 0$, by \begin{equation}G_D(-s) = \tfrac12\Pr\Big[\mathrm{Beta}\big(\tfrac12, \tfrac{D-2}{2}\big) > \frac{s^2}{D-1}\Big]\label{eq:gd}\end{equation} This is the coordinate distribution given by RaBitQ~\cite{r62}, written as a Beta distribution. For $s < 0$, by symmetry, $G_D(s) = 1 - G_D(-s)$.

ADSampling~\cite{r64} performs the distance comparison that decides whether a candidate vector lies within the $k$-th distance found so far by reading the randomly orthogonally transformed coordinates little by little from the front. If the distance estimated from some of the coordinates exceeds that distance even after allowing a slack, it discards the vector without reading the remaining coordinates, and it bounds the probability of wrongly discarding a vector within the distance by adding up, with a union bound, the failure probabilities at all points where reading can stop. Moreover, since the result is the same as the one computed with exact distances as long as the comparisons for the true $k$-nearest neighbors are correct, it restricts the union bound to the $k$ true neighbors instead of all candidates~\cite{r64}. This paper inherits these two ideas, namely the progressive test that reads the transformed coordinates from the front and the union over the $k$ true neighbors. Section~\ref{sec:theory} determines the slack with the exact distribution of Eq.~\eqref{eq:gd} and extends the argument to vectors of unvisited regions tested with residual codes and to a radius that depends on the rotation, stating it as a recall guarantee for the query.

\bigskip

\section{Per-Query Guarantees and Cost}\label{sec:theory}

In this section, we present the theory of the proposed per-query guarantees and the law that determines their cost. The error identity and the union argument of Section~\ref{sec:prelim} alone cannot provide per-query guarantees at low cost. For codes of which only a prefix is read, there is no formula that sets the failure probability of a single test to a target value, and there is no guarantee for $\tau < 1$ or for a radius shrunk while unvisited vectors are being tested. Section~\ref{sec:rule} presents a pruning rule that reads codes from the front together with its failure probability, and Section~\ref{sec:guarantee} presents theorems that transfer it to the per-query $(\tau, \delta)$-recall guarantee and the expected-recall guarantee. These theorems also determine which vectors may be excluded without testing (the triangle-inequality range) and the condition under which the radius may be shrunk during testing (Theorem 5). Section~\ref{sec:costlaw} derives a formula that predicts the fraction of vectors passing the test without data-fitted constants and without reading the rotation, and identifies the factors that reduce this fraction. Section~\ref{sec:search} designs the index and the search algorithm according to these results.

\subsection{Prefix-Truncated Codes and the Pruning Rule}\label{sec:rule}

We extend the code of Section~\ref{sec:prelim} to a staged test that reads the code from the front. Let $\bar{\boldsymbol{x}}^{(\ell)}$ be the vector that keeps only the first $m_\ell$ coordinates of the code $\bar{\boldsymbol{x}}$, and let the code vector be $\bar{\boldsymbol{o}}^{(\ell)} = P\bar{\boldsymbol{x}}^{(\ell)}$ ($m_1 < \dots < m_L \le D$). The test passes through stages $\ell = 1, \dots, L$ in turn, and stage $\ell$ reads only the first $m_\ell$ coordinates and estimates the inner product $\rho$ as follows. \begin{equation}\hat\rho^{(\ell)} = \frac{\langle\bar{\boldsymbol{o}}^{(\ell)}, \boldsymbol{q}\rangle}{\langle\bar{\boldsymbol{o}}^{(\ell)}, \boldsymbol{o}\rangle} = \frac{\langle\bar{\boldsymbol{x}}^{(\ell)}, P^{-1}\boldsymbol{q}\rangle}{\langle\bar{\boldsymbol{x}}^{(\ell)}, P^{-1}\boldsymbol{o}\rangle}\label{eq:est}\end{equation} When $m_\ell = D$, this estimator coincides with that of RaBitQ~\cite{r62}, and its denominator and numerator are computed in the same way. Since a prefix of the rotated coordinates acts as a random projection, a short prefix can discard most vectors, and longer codes need to be read only for the remaining vectors. This is the same idea as the progressive dimension sampling of ADSampling~\cite{r64}. Hereafter, we call such codes read from the front \textit{prefix-truncated codes}. The magnitude of the estimation error is determined by \begin{equation}\kappa^{(\ell)} = \frac{\sqrt{m_\ell/D - \langle\bar{\boldsymbol{o}}^{(\ell)}, \boldsymbol{o}\rangle^2}}{\langle\bar{\boldsymbol{o}}^{(\ell)}, \boldsymbol{o}\rangle\sqrt{D-1}}\label{eq:kappa}\end{equation} which is precomputed for each vector at indexing time. In the denominator of $\kappa^{(\ell)}$, $\langle\bar{\boldsymbol{o}}^{(\ell)}, \boldsymbol{o}\rangle$ is the component of the code vector in the direction of $\boldsymbol{o}$, and the numerator is the length of the remainder of the code vector after this component is subtracted. For 1-bit prefix codes, the squared norm of the code vector is $m_\ell/D$, so this length is $\sqrt{m_\ell/D - \langle\bar{\boldsymbol{o}}^{(\ell)}, \boldsymbol{o}\rangle^2}$. The estimation error arises from this remaining component. When $m_\ell = D$, the code vector is a unit vector, so $\kappa^{(\ell)} = \sqrt{1-\langle\bar{\boldsymbol{o}}^{(\ell)}, \boldsymbol{o}\rangle^2}\,/\,(\langle\bar{\boldsymbol{o}}^{(\ell)}, \boldsymbol{o}\rangle\sqrt{D-1})$, which equals the scale of the error bound of RaBitQ~\cite{r62}.

\begin{thmblock}{\textbf{Lemma 1} (Error Identity):}Let $D \ge 3$ and $\lvert\rho\rvert < 1$, let the code be a function only of the vector's own $P^{-1}\boldsymbol{o}$ (and of random numbers independent of $P$), and let $\langle\bar{\boldsymbol{o}}^{(\ell)}, \boldsymbol{o}\rangle > 0$. If $\kappa^{(\ell)}$ is taken as Eq.~\eqref{eq:kappa} with $m_\ell/D$ replaced by the squared norm of the code vector (Eq.~\eqref{eq:kappa} itself for 1-bit prefix codes), then the estimator of Eq.~\eqref{eq:est} satisfies \begin{equation}\hat\rho^{(\ell)} - \rho = \sqrt{1-\rho^2}\,\kappa^{(\ell)}\,\zeta^{(\ell)}\label{eq:err}\end{equation} where $\zeta^{(\ell)}$ is independent of $P^{-1}\boldsymbol{o}$ (and of those random numbers) and follows the distribution of Eq.~\eqref{eq:gd}.\end{thmblock}

The proof of Lemma 1 is given in Appendix~\ref{app:proofstatic}. For 1-bit codes with $m_\ell = D$, Eq.~\eqref{eq:err} is equivalent to the error identity and the error distribution of RaBitQ~\cite{r62} combined. Lemma 1 extends this to arbitrary codes, including 1-bit codes of which only a prefix is read and 2-bit codes. Furthermore, RaBitQ~\cite{r62} states that the error is independent of $\langle\bar{\boldsymbol{o}}, \boldsymbol{o}\rangle$, and the proof in its technical report (arXiv:2405.12497, Appendix B.3) conditions on $P^{-1}\boldsymbol{o}$; Lemma 1 states this independence from all of $P^{-1}\boldsymbol{o}$ for any code that is a function of $P^{-1}\boldsymbol{o}$, including prefix codes and 2-bit codes. Consequently, even for prefix-truncated codes and 2-bit codes, in which $\kappa^{(\ell)}$ is a function of $P^{-1}\boldsymbol{o}$, the failure probability of the test (Theorem 1) and the pass probability (Section~\ref{sec:costlaw}) can be computed independently of the data.

\begin{figure}[t]
\centering
\includegraphics{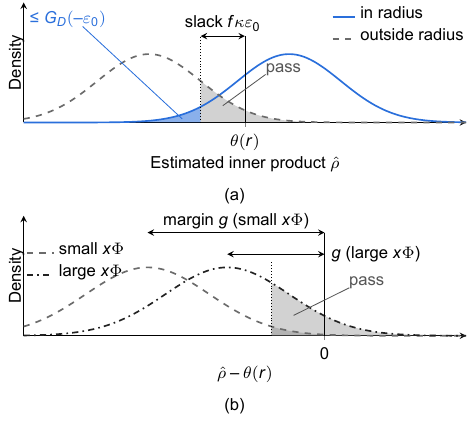}
\refstepcounter{figure}\label{fig:judgment}
\par\vspace{3pt}{\small\setstretch{1.15}\IEEEfigcap{Figure \thefigure. Pruning failure and pass (schematic). (a) Estimate distributions of a vector within and outside the radius. (b) Pass of an outside vector versus the residual scale $x\Phi$.}\par}
\end{figure}

The test prunes vectors that cannot lie within radius $r$ of the query. As shown in Section~\ref{sec:prelim}, a vector lies within radius $r$ if and only if $\rho \ge \theta(r)$. However, $\rho$ is unknown for a vector whose exact distance has not been computed, so the estimate $\hat\rho^{(\ell)}$ obtained from the code is used instead. To prevent a vector within the radius from being discarded because its estimate comes out smaller than the true value, we prune every vector that falls short of the threshold even after an error slack is added to its estimate, that is, every vector satisfying \begin{equation}\hat\rho^{(\ell)} + f\,\kappa^{(\ell)}\varepsilon_{0,\ell} < \theta(r)\label{eq:rule}\end{equation} Vectors with zero residual and vectors with $\langle\bar{\boldsymbol{o}}^{(\ell)}, \boldsymbol{o}\rangle \le 0$ pass without testing, and if $\Phi = 0$, the form of Eq.~\eqref{eq:rule} multiplied by $2x\Phi$ is used, so that a vector is pruned only when $x > r$. Here, $f\,\kappa^{(\ell)}\varepsilon_{0,\ell}$ is the slack added to the estimate to guard against the estimation error. By Eq.~\eqref{eq:err}, $\rho = \hat\rho^{(\ell)} - \sqrt{1-\rho^2}\,\kappa^{(\ell)}\zeta^{(\ell)}$; hence, when the error satisfies $\zeta^{(\ell)} \ge -\varepsilon_{0,\ell}$ and $f \ge \sqrt{1-\rho^2}$, and $\varepsilon_{0,\ell} \ge 0$ so that $(f - \sqrt{1-\rho^2})\,\kappa^{(\ell)}\varepsilon_{0,\ell} \ge 0$, a pruned vector satisfies $\rho \le \hat\rho^{(\ell)} + f\,\kappa^{(\ell)}\varepsilon_{0,\ell} < \theta(r)$, so as long as the error stays within this range, no vector within the radius is pruned. Therefore, a vector within the radius is pruned only when $\zeta^{(\ell)} < -\varepsilon_{0,\ell}$ (Fig.~\ref{fig:judgment}(a)). The \textit{slack factor} $f$ is a value that replaces $\sqrt{1-\rho^2}$, which cannot be computed because $\rho$ is unknown. Since pruning a vector outside the radius does not affect the result, $f$ need only be an upper bound of this factor for vectors within the radius, and $f = 1$ always satisfies this. In addition, since $\zeta^{(\ell)}$ follows the data-independent distribution $G_D$ (Lemma 1), the probability of this event is $G_D(-\varepsilon_{0,\ell})$. Therefore, to assign a failure probability $\delta_\ell$ to stage $\ell$, it suffices to set the error-bound parameter $\varepsilon_{0,\ell}$ by $G_D(-\varepsilon_{0,\ell}) = \delta_\ell$. We take $\delta_\ell \le 1/2$, so that $\varepsilon_{0,\ell} \ge 0$, which the argument above uses; every setting in our experiments satisfies this.

\begin{thmblock}{\textbf{Theorem 1} (Failure Probability of Pruning):}Let $D \ge 3$, let the code satisfy the conditions of Lemma 1, let the stage configuration and the error-bound parameters $\varepsilon_{0,\ell} \ge 0$ (i.e., $G_D(-\varepsilon_{0,\ell}) \le 1/2$) be chosen independently of $P$, and let the slack factor satisfy $f \ge \sqrt{1 - \max(\theta(r), 0)^2}$ (the radius $r$ may depend on $P$). (a) If a vector within radius $r$ is pruned at stage $\ell$, then $\zeta^{(\ell)} < -\varepsilon_{0,\ell}$. (b) The event in (a) has probability at most $G_D(-\varepsilon_{0,\ell})$. (c) Hence, a vector within the radius is pruned at some stage with probability at most $\sum_{\ell=1}^{L} G_D(-\varepsilon_{0,\ell})$.\end{thmblock}

The proof of Theorem 1 is given in Appendix~\ref{app:proofstatic}. ADSampling~\cite{r64} also bounds the probability of the same event, but its bound contains a constant of unknown value, so its actual implementation uses an error-bound parameter determined empirically. In contrast, Theorem 1 sets the error-bound parameter from the exact distribution of Eq.~\eqref{eq:gd}, so the failure probability is bounded by the target value itself, with no empirically determined constant. In addition, the error bound of RaBitQ~\cite{r62} sets the factor $\sqrt{1-\rho^2}$ of Eq.~\eqref{eq:err} to 1, but since vectors within the radius satisfy $\rho \ge \theta(r)$, when $\theta(r) > 0$ one can choose $f$ smaller than 1 and make the test tighter (the specific value is given in Section~\ref{sec:process}).

\subsection{Per-Query Guarantees}\label{sec:guarantee}

Transferring Theorem 1 to a guarantee for a query requires premises about the search. The per-query guarantees hold for any search that satisfies the following four conditions, and they do not depend on the structure of the index or the traversal order.

\begin{description}[font=\normalfont,leftmargin=3.6em,labelwidth=3.2em,labelsep=0.4em,align=left]
\item[C1.] The reference point (the centroid $\boldsymbol{c}$ that normalizes codes), the stage configuration, and the error-bound parameters are chosen independently of the random orthogonal transformation $P$.
\item[C2.] The test radius $r$ is the $k$-th smallest of the exact distances of vectors chosen independently of $P$ (hereafter, visited vectors) (infinity if there are fewer than $k$ visited vectors).
\item[C3.] Unvisited vectors outside the triangle-inequality range are excluded without testing, and the other vectors are tested at every stage by Eq.~\eqref{eq:rule} with a slack factor $f \ge \sqrt{1-\max(\theta(r), 0)^2}$.
\item[C4.] The exact distances of the vectors that pass all stages are computed, and the $k$ nearest among the vectors whose exact distances have been computed are returned as the result.
\end{description}

Each condition is needed for the guarantee for the following reasons. C1 allows the distribution of the test error to be computed as $G_D$ of Theorem 1. If the reference point or the error-bound parameters were chosen by looking at the same $P$, a selection bias that picks out favorable rotations would arise. By C2, the radius is at least the true $k$-th distance, so all true neighbors lie within the radius. C3 ensures that unvisited vectors are also checked without omission (the condition on the slack factor is for applying Theorem 1), and since the vectors excluded without testing are certainly outside the radius, true neighbors can be missed only in the test. By C4, the true neighbors whose exact distances have been computed are among the $k$ results, so the only true neighbors missing from the result are those whose exact distances were not computed. In contrast, no condition is imposed on the data distribution, the quality of the partitions, or the number of visited vectors; these change only the cost (Section~\ref{sec:costlaw}). Our IVF search takes the vectors of the partitions with the nearest centroids as the visited vectors, sets the initial radius by C2, and satisfies C1, C3, and C4; during the search, it shrinks this radius according to Theorem 5 below (Section~\ref{sec:process}).

To transfer the failure probability of a single test to a guarantee for a single query, the probabilities of Theorem 1 must be summed over all vectors that must not be missed (union bound). To this end, we call the probability allowed for a single vector to be pruned at any stage the \textit{failure probability budget} $\beta$, and divide it into per-stage shares, writing $\beta = \sum_\ell\delta_\ell$. By Theorem 1, if $G_D(-\varepsilon_{0,\ell}) = \delta_\ell$, the probability that a single vector within the radius is pruned is at most $\beta$. Summing over all $n$ vectors would shrink the budget to $\beta = \delta/n$, which enlarges the error-bound parameters and lets far more vectors pass the test (our measurements are in Section~\ref{sec:ablation}). However, the recall of a query result drops only when true neighbors are missed, so, as in ADSampling~\cite{r64}, it suffices to sum over only the $k$ true neighbors. The following lemma shows this.

\begin{thmblock}{\textbf{Lemma 2}:}If the $k$ vectors with the smallest distances in a candidate set $C$ ($\lvert C\rvert \ge k$) are returned, then $\mathrm{Recall}(q) \ge 1 - \lvert\mathcal N_k^*(q) \setminus C\rvert/k$, even with ties.\end{thmblock}

The proof of Lemma 2 is given in Appendix~\ref{app:proofstatic}. The reason is that true neighbors in the candidate set are also among the $k$ nearest within the candidate set. That is, recall falls below 1 only when some true neighbors fail to enter the candidate set, and it falls only by their number. Hereafter, writing this number, i.e., the number of missed true neighbors, as $M = \lvert\mathcal N_k^*(q) \setminus C\rvert$, we have $\mathrm{Recall}(q) \ge 1 - M/k$. From this we obtain the guarantee of exact search.

\begin{thmblock}{\textbf{Theorem 2} (Exact Search):}Let $D \ge 3$, let the code satisfy the conditions of Lemma 1, and in a search satisfying C1–C4, set $G_D(-\varepsilon_{0,\ell}) = \delta_\ell \le 1/2$ and $\sum_\ell\delta_\ell = \beta$. If $V$ denotes the set of unvisited true neighbors, then \begin{equation}\Pr[\text{result is not exact}] \le \lvert V\rvert\sum_{\ell=1}^{L} G_D(-\varepsilon_{0,\ell}) \le k\beta\label{eq:union}\end{equation} and therefore, if $\beta = \delta/k$, the $(\tau, \delta)$-recall guarantee of Eq.~\eqref{eq:goal} holds for $\tau = 1$.\end{thmblock}

The proof of Theorem 2 is given in Appendix~\ref{app:proofstatic}. Visited true neighbors enter the candidate set with their exact distances, and unvisited true neighbors also pass all stages and enter the candidate set unless they are pruned; hence, by Lemma 2, the result is not exact only if some vector of $V$ is pruned. Applying Theorem 1 to each vector of $V$ for this event and summing yields Eq.~\eqref{eq:union}. The argument of taking the union over only the $k$ true neighbors is that of ADSampling~\cite{r64}. Theorem 2 applies this argument with exact quantiles when the radius is set from the exact distances of visited vectors and the other vectors are tested with codes, and shows that the same upper bound holds in the following two cases. When some vectors are excluded without testing by the triangle-inequality range before the test, the excluded vectors are certainly outside the radius, so no true neighbor is lost at this step. Also, when several vectors are at the same distance as the true $k$-th neighbor, by Lemma 2 the recall is determined only by the number of true neighbors missing from the result, so the same upper bound is obtained.

When $\tau < 1$, up to $m$ true neighbors may be missed, so for the same $\delta$ the failure probability budget $\beta$ can be set larger. A larger $\beta$ enlarges the per-stage shares $\delta_\ell$, which makes the error-bound parameters $\varepsilon_{0,\ell}$ smaller; the slack of the test shrinks and more vectors are pruned, so the cost decreases. Theorem 3 below shows how large it can be set. That is, since $\mathrm{Recall}(q) < \tau$ requires $M \ge m + 1$, $\beta$ may be set to $\delta(m+1)/k$, which is $m + 1$ times that of exact search.

\begin{thmblock}{\textbf{Theorem 3} ($(\tau, \delta)$-Recall Guarantee):}In the same setting as Theorem 2, when $\tau < 1$, if $\beta = \delta(m+1)/k$, then $\Pr[\mathrm{Recall}(q) < \tau] \le \delta$ for every query.\end{thmblock}

The proof of Theorem 3 is given in Appendix~\ref{app:proofstatic}. By the same calculation as Eq.~\eqref{eq:union}, $\mathbb E[M] \le k\beta = \delta(m+1)$, and since $\mathrm{Recall}(q) < \tau$ is equivalent to $M \ge m + 1$, Markov's inequality~\cite{r76} gives $\Pr[M \ge m + 1] \le \mathbb E[M]/(m+1) \le \delta$. This Markov step is tight only for analyses that use the per-neighbor pruning probabilities alone: given only these probabilities, it cannot be improved, because Markov's inequality holds with equality when the number of missed true neighbors $M$ takes only the two values 0 and $m + 1$. Such simultaneous failures are realized by placements with exact duplicate vectors. With a single test stage and a positive test threshold, $m + 1$ identical vectors in one unvisited partition make the failure probability arbitrarily close to this bound when $k = m + 1$ (which means $\tau \le 1/k$); with $m + 1$ identical vectors in each of several unvisited partitions ($k$ a multiple of $m + 1$), the gap from the bound is only a small quantity of second order in $\beta$ (first order in relative terms) (Proposition 1 in Appendix~\ref{app:proofstatic}). The bound is thus attained exactly only in that special case and otherwise within a second-order term. This tightness is a statement about this proof step on duplicate-heavy data, not about the algorithm on other data or about every per-query method; the lower bounds of Appendix~\ref{app:lowerbound} (Theorems 7–9) concern the cost of a per-query guarantee, not the Markov step. Therefore, as long as the same allowed probability is set for every neighbor independently of the data, $\beta$ cannot be set larger than this, and Theorem 3 uses only as much slack as an analysis based on per-neighbor probabilities requires.

An expected-recall guarantee, which states that the expectation of recall over $P$ is at least $\tau$ for every query, can also be provided. This guarantee is obtained from the linearity of expectation alone, without any assumption on the relationship among the failures of neighbors, and a much smaller slack than that of the $(\tau, \delta)$-recall guarantee suffices.

\begin{thmblock}{\textbf{Theorem 4} (Expected-Recall Guarantee):}Let $D \ge 3$, let the code satisfy the conditions of Lemma 1, and in a search satisfying C1–C4 with every $\varepsilon_{0,\ell} \ge 0$, let $V$ be the set of unvisited true neighbors. If \begin{equation}\lvert V\rvert\sum_{\ell=1}^{L} G_D(-\varepsilon_{0,\ell}) \le (1-\tau)k\label{eq:ex}\end{equation} holds (in particular, if $G_D(-\varepsilon_{0,\ell}) = \delta_\ell \le 1/2$ and $\beta = \sum_\ell\delta_\ell \le 1-\tau$), then $\mathbb E[\mathrm{Recall}(q)] \ge \tau$ for every query.\end{thmblock}

The proof of Theorem 4 is given in Appendix~\ref{app:proofstatic}. If the probability of missing each individual true neighbor is at most $1 - \tau$, then the expectation of the missed fraction $M/k$ is also at most $1 - \tau$, and hence the expectation of recall is at least $\tau$. That is, a failure probability budget of $\beta = 1 - \tau$ suffices for the expected-recall guarantee. This value is much larger than the $\beta = \delta(m+1)/k$ of the $(\tau, \delta)$-recall guarantee, so the error-bound parameters are small. For example, in our experimental setting ($\tau = 0.9$, $\delta = 0.01$, $k \in \{10, 100\}$, two test stages), the expected-recall guarantee needs a slack of only 1.65 times the standard deviation of the error, whereas the $(\tau, \delta)$-recall guarantee with the same $\tau$ needs 3.08–3.26 times. By the same reasoning, replacing the right-hand side of Eq.~\eqref{eq:ex} with $\alpha k$ yields $\mathbb E[\max(0, \tau - \mathrm{Recall}(q))] \le \alpha$ for every query, and setting $\alpha = \alpha_{\rm CTR}(1-\tau)$ for the target $\alpha_{\rm CTR}$ of CTR~\cite{r11} bounds, for every query, the loss that CTR bounds on average over the query distribution. Unlike TPU-KNN~\cite{r168}, Theorem 4 holds for every query, including unvisited vectors, without any assumption on the placement of neighbors. The same reasoning applies to range queries, i.e., searches that return all vectors within a radius $r$ given independently of $P$. In a search satisfying C1, C3, and C4, setting $\sum_\ell G_D(-\varepsilon_{0,\ell}) \le \delta/n$ makes the probability of missing even one vector within the radius at most $\delta$, by Theorem 1 (c) and the union over the unvisited vectors within the radius (at most $n$ of them).

The $(\tau, \delta)$-recall guarantee bounds $\Pr[M \ge m + 1]$ and the expected-recall guarantee bounds $\mathbb E[M]$; in general, neither implies the other. For example, with $k = 10$ and $\tau = 0.9$, if recall is 0.9 with probability $1-\delta$ and 0 with probability $\delta$, the $(\tau, \delta)$-recall guarantee holds but the expectation is $0.9(1-\delta) < \tau$; if recall is 1.0 and 0.8 with probability 1/2 each, the expected-recall guarantee holds but $\Pr[\mathrm{Recall}(q) < \tau] = 1/2$. However, Theorem 3 bounds $\mathbb E[M]$ before applying Markov's inequality, so the following corollary shows that the same execution also satisfies the expected-recall guarantee.

\begin{thmblock}{\textbf{Corollary 1}:}In the $(\tau, \delta)$-recall guarantee mode of Theorem 3, if $\delta(m+1) \le (1-\tau)k$, then $\mathbb E[\mathrm{Recall}(q)] \ge \tau$ also holds for every query.\end{thmblock}

The proof of Corollary 1 is given in Appendix~\ref{app:proofstatic}. This condition is equivalent to the $\beta = \delta(m+1)/k$ of Theorem 3 satisfying the condition $\beta \le 1 - \tau$ of Theorem 4, and if $(1-\tau)k \ge 1$, it holds whenever $\delta \le 1/2$. All settings with $\tau < 1$ in our experiments fall into this case, so the single $(\tau, \delta)$-recall guarantee mode yields both guarantees at once. In contrast, for $\tau = 1$, $\mathbb E[\mathrm{Recall}(q)] \ge 1$ means that the failure probability is zero, so no method with a positive failure probability can satisfy the expected-recall guarantee.

If the visited vectors contain few true neighbors, the radius set from their distances can be much larger than the true $k$-th distance. In this case, it is more efficient to shrink the radius to the $k$-th distance as more vectors have their exact distances computed while unvisited vectors are being tested, and to test subsequent vectors with the shrunk radius (hereafter called the \textit{adaptive radius}). C2 fixes the radius at the value set from the visited vectors and thus does not cover this case, but all that the guarantee actually requires is that the radius be at least the true $k$-th distance. However, which vectors pass the test and have their exact distances computed depends on the test results, and the test results depend on the rotation $P$. Hence, a radius shrunk in this way also depends on $P$. Since Theorems 2–4 presuppose a radius determined independently of the test errors, it must be shown separately that the guarantee is preserved even when the radius and the test errors depend on the same $P$. The following theorem shows that the guarantee remains intact even when C2 is relaxed in this way.

\begin{thmblock}{\textbf{Theorem 5} (Adaptive Radius):}Let $D \ge 3$, let the search satisfy C1, C3, and C4, and let the test radius be set as follows instead of by C2. After an initial radius is set from visited vectors chosen independently of $P$, the radius may be changed each time an exact distance is computed while unvisited vectors are tested, provided that (i) the radius used for the triangle-inequality range and for each test is the maximum of the distances of some $k$ distinct data vectors whose exact distances have been computed (infinity until $k$ such vectors have been collected), and (ii) the threshold $\theta(r)$ and the slack factor $f$ of each test are computed with the radius $r$ used in that test, with $f \ge \sqrt{1-\max(\theta(r), 0)^2}$. The order of tests, the choice of vectors whose exact distances are computed first, and whether the same vector is tested again may depend on $P$. Then, with $V$ defined with respect to the visited vectors that set the initial radius, Theorems 2–4 and Corollary 1 hold as they are.\end{thmblock}

The proof of Theorem 5 is given in Appendix~\ref{app:proofstatic}. The guarantee is preserved even though the radius depends on $P$ for the following reasons. By (i), the radius is always the maximum of the distances to $k$ actual vectors, so it is at least the true $k$-th distance, and the true neighbors lie within the radius at every point in time. Theorem 1 (a) holds for any radius, so whenever and with whatever radius a true neighbor was pruned, its error satisfies $\zeta^{(\ell)} < -\varepsilon_{0,\ell}$. Since this condition does not involve the radius, the union bounds and the expectation calculations of Theorems 2–4 do not use the $P$-dependent radius or order at all. Testing the same vector again at the same stage yields the same error, so the failure probability does not increase. Moreover, the radius in (i) need not be the $k$-th of all distances computed so far. Therefore, the guarantee remains intact even when multiple threads share the processing of a single query and each thread sets the radius from its own collected $k$ vectors, or when a test uses the value from before another thread shrank the radius (Section~\ref{sec:process}). RaBitQ~\cite{r62} also selects re-ranking targets within IVF using the distance of the nearest neighbor found so far as the threshold, and explains that, by its error bound, the true nearest neighbor in the visited partitions is sent to re-ranking with high probability. The step that converts the pruning event into a condition on the error $\zeta$ alone, without the radius, also appears in ADSampling~\cite{r64}, whose per-comparison argument uses only that the distance is within the threshold; ADSampling regards earlier tests changing the threshold of later tests as a bias in its verification study and fixes the threshold there at the distance of the true $k$-th neighbor. Theorem 5 makes explicit that the recall guarantee for a query, including unvisited vectors, is preserved even when the threshold depends on the rotation, also with per-thread radii and stale radii.

\subsection{Residual Cost Law}\label{sec:costlaw}

The theorems of Section~\ref{sec:guarantee} only set an upper bound on the failure probability; they do not determine the cost of providing the same guarantee. In this section, we present the law that determines this cost; in this paper, the law is used in two ways. First, it shows that the cost is determined by the residual scale, the test radius, and the error scale, which grounds the residual codes and the index and search design of Section~\ref{sec:search}. Second, since it predicts the cost without reading the rotation, the parameters chosen by this prediction (Section~\ref{sec:devices}) and the decisions for maintaining an efficient index in dynamic environments (Section~\ref{sec:maintenance}) do not break the premises of Theorem 1 and Theorem 6 (Section~\ref{sec:dynguarantee}). Therefore, the cost can be adjusted without retuning while the guarantee is kept. Moreover, since this law is independent of the guarantee, a wrong prediction changes only the cost. We first show that the fraction of unvisited vectors that pass the first-stage ($\ell = 1$) test, i.e., the first-stage pass rate, determines the cost; we then derive a formula that predicts this fraction without data-fitted constants, and finally summarize the implications of this formula for designs that reduce the cost.

Unvisited vectors not pruned at the first stage are tested again with longer codes at later stages, and if they pass all stages, their exact distances are computed with the original vectors. Hence, the more vectors pass the first stage, the higher the cost of the later-stage tests and of the exact distance computations. In our measurements, the vectors whose exact distances were computed amounted to a median of only 0.17\% of all vectors. Therefore, most of the time is spent on the first-stage tests and on processing the vectors that pass the first stage, and the main variable that determines the cost is the first-stage pass rate. The first-stage test of our implementation is Eq.~\eqref{eq:rule} with an added deterministic slack $\sigma_{\det}$ that covers quantization and floating-point errors (Section~\ref{sec:process}). That is, a vector $\boldsymbol{o}_r$ passes when $\hat\rho^{(1)} \ge \theta(r) - f\kappa^{(1)}\varepsilon_{0,1} - \sigma_{\det}$. Substituting Eq.~\eqref{eq:err} into this condition and solving for $\zeta^{(1)}$, the pass condition becomes $\zeta^{(1)} \ge z(\kappa)$, where $\kappa = \kappa^{(1)}$ and $z(\kappa) = (\theta(r) - \rho - f\kappa\,\varepsilon_{0,1} - \sigma_{\det})/(\sqrt{1-\rho^2}\,\kappa)$. By Lemma 1, $\zeta^{(1)}$ is independent of $P^{-1}\boldsymbol{o}$ and $\kappa$ is a function of $P^{-1}\boldsymbol{o}$, so given $\kappa$, the probability of passing is $G_D(-z(\kappa))$ by the symmetry of $G_D$. Therefore, the probability that a vector $\boldsymbol{o}_r$ in the triangle-inequality range passes the first stage is \begin{equation}\Pr[\text{pass}] = \mathbb E_{\kappa}\big[G_D(-z(\kappa))\big]\label{eq:passprob}\end{equation} (Fig.~\ref{fig:judgment}(a)). Here, the expectation is over the distribution of $\kappa$ induced by $P$, and this distribution is determined by $D$ and $m_1$ alone. The first-stage pass rate of a query is the average of this probability over the triangle-inequality range. The only inputs are $x$, $\Phi$, $\rho$, $r$, $\sigma_{\det}$, $(D, k, \tau, \delta)$, and the stage configuration, and there is no data-fitted constant.

When the radius $r$ and the deterministic slack $\sigma_{\det}$ are fixed, Eq.~\eqref{eq:passprob} is an equality, not an upper bound (for vectors with $x, \Phi > 0$ and $\lvert\rho\rvert < 1$). Since the slack of the implementation varies slightly with the rotation, for the measured pass rate the formula is an approximation that plugs in the average slack. Although $\rho$ is unknown during search, it is known for calibration queries, whose exact distances can be computed; hence, this formula is used to predict the pass rate with the true $k$-th distance of the calibration queries as the radius, and the parameter selection rule of Section~\ref{sec:devices} uses a light approximation of it. In our measurements, this prediction agreed well with the measured pass rates without any data-fitted constant. Over 256 points with up to 10 partitioning methods at 1M scale, the log correlation was 0.9995, and over 38 points at 10M scale it was 0.999 (Fig.~\ref{fig:passrate}; validation under updates and at 100M scale is also in Section~\ref{sec:ablation}).

The effect of the centroid on the cost appears in the term $\theta(r) - \rho$ of Eq.~\eqref{eq:passprob}. Rearranging it using Eq.~\eqref{eq:theta} and Eq.~\eqref{eq:dist} gives \begin{equation}g = \theta(r) - \rho = \frac{\lVert \boldsymbol{o}_r - \boldsymbol{q}_r\rVert^2 - r^2}{2x\Phi}\label{eq:gap}\end{equation} This margin $g$ indicates how far the $\rho$ of a vector outside the radius lies below the test threshold. Since $\lVert\boldsymbol{o}_r - \boldsymbol{q}_r\rVert$ in the numerator is the distance between the original vectors, it does not depend on the centroid, and the entire effect of the centroid on the margin lies in the residual scale $x\Phi$ in the denominator. With the other inputs of Eq.~\eqref{eq:passprob} ($f$, $\sigma_{\det}$, $\rho$) fixed, the pass probability decreases in the margin; therefore, when the centroid lies farther from the vector and the query so that $x\Phi$ grows, the margin shrinks and more vectors outside the radius pass the first stage (Fig.~\ref{fig:judgment}(b)). Conversely, when the radius shrinks from $r$ to $r'$, the margin grows by $(r^2 - r'^2)/(2x\Phi)$, so the adaptive radius (Theorem 5) reduces the cost without changing the guarantee. This is why the cost of residual coding depends on the centroid, and Section~\ref{sec:maintenance} uses this relation in the decisions for maintaining an efficient index in dynamic environments.

The error scale of a stage that reads $m_\ell$ code coordinates is typically $\kappa^{(\ell)} \approx \sqrt{\pi D/(2m_\ell) - 1}/\sqrt{D-1}$, and for $m_\ell = D$ it is about $0.755/\sqrt D$, which corresponds to the $\langle\bar{\boldsymbol{o}}, \boldsymbol{o}\rangle \approx 0.8$ reported by RaBitQ~\cite{r62}. If the first-stage code length $m_1$ is fixed, $\kappa^{(1)}$ converges to $\sqrt{\pi/(2m_1)}$ as the dimension grows and is nearly independent of the dimension. For example, with $m_1 = 128$, it is 0.067 for $D = 128$ and 0.106 for $D = 1024$, and the limit is 0.111. Therefore, the cost of the first stage is determined by the number of bits, and the error scale does not exceed $\sqrt{\pi/(2m_1)}$ even as the dimension grows (Section~\ref{sec:scale}).

Eq.~\eqref{eq:passprob} shows how to reduce the cost under the same guarantee. To enlarge the margin $g$, one must reduce the residual scale $x\Phi$ or shrink the radius $r$ toward the true $k$-th distance (Theorem 5); to reduce the error scale $\kappa^{(1)}$, one must read longer codes at the first stage, which increases the first-stage cost accordingly; and the deterministic slack $\sigma_{\det}$ becomes smaller as the quantization error of the implementation is reduced. IVF partitions contribute to both components of the margin. The centroid, as the reference point that normalizes the 1-bit codes, reduces the residual scale, and visiting the partitions near the query first brings the initial radius close to the true $k$-th distance through the true neighbors in those partitions (Section~\ref{sec:ablation}). Section~\ref{sec:search} designs the index and the search algorithm according to these terms.

Comparing this result with exhaustive search gives an intuitive understanding of why the proposed method provides the same guarantee at low cost. Exhaustive search computes the distances of all vectors and thus reads all $D$ coordinates of the original of every vector. However, what most vectors need for the result to be determined is not the value of their distance but only a test showing that they cannot lie within the radius $r$, and for vectors far from the radius, even a coarse estimate makes it clear that they are outside the radius. The first stage makes this test with an $m_1$-bit code per vector. For example, with $m_1 = 128$, this takes 23 bytes including the coefficients, about 1/130 of a 768-dimensional fp32 original (3,072 bytes), and SIMD tests 64 vectors at once (Section~\ref{sec:process}). The risk of discarding a true neighbor because of a wrong coarse estimate is prevented by a slack on the error, whose distribution is fixed thanks to the random orthogonal transformation. Since this slack grows only roughly in proportion to $\sqrt{\ln(1/\delta)}$ as the failure probability $\delta$ is reduced, strengthening the guarantee increases the cost only slightly. Residual codes increase the number of vectors filtered out under the same guarantee. Since the error is proportional to the magnitude of the residual, coding with respect to a nearby centroid enlarges the margin so that more vectors are filtered out at the first stage (Eq.~\eqref{eq:gap}, Fig.~\ref{fig:judgment}(b)); vectors outside the triangle-inequality range are discarded without even reading their codes; and visiting the partitions near the query first starts the radius small. As a result, for most vectors it suffices to read a few tens of bytes at the first stage, and only the few vectors that pass all stages have their originals read for distance computation. For the same first-stage code length, the size ratio between the original and the code is proportional to the dimension, so the gain grows with the dimension. IVF without guarantees also reduces the number of vectors whose originals are read, but since it does not look at unvisited partitions, it can miss the true neighbors there, and to answer all test queries correctly it must scan many partitions with originals to accommodate the hardest query. The proposed method examines only a few partitions with originals and handles the rest with this cheap test, while still providing a guarantee for every query.

\bigskip

\section{Index and Search}\label{sec:search}

In this section, we describe the index and the search algorithm designed according to Section~\ref{sec:theory}. Section~\ref{sec:index} describes the structure and construction of the index and then explains how and why it differs from standard IVF; Section~\ref{sec:process} describes how queries are processed so as to satisfy the conditions of Theorem 5 (Section~\ref{sec:guarantee}), together with the implementation; and Section~\ref{sec:devices} describes the rule that chooses, by the cost law of Section~\ref{sec:costlaw}, the options that change only the query processing cost. Whatever the rule chooses, the guarantee remains intact, and its effect is measured in Section~\ref{sec:ablation}.

\subsection{Index Structure and Construction}\label{sec:index}

The index divides the data vectors into IVF partitions, normalizes each vector as a residual with respect to its partition centroid (Section~\ref{sec:prelim}), and stores it as a prefix-truncated code (Section~\ref{sec:rule}). The per-vector values needed for the test (Eq.~\eqref{eq:rule}), namely the residual norm $x$ and, for each stage, the inner product of the code and the unit vector $\langle\bar{\boldsymbol{o}}^{(\ell)}, \boldsymbol{o}\rangle = \langle\bar{\boldsymbol{x}}^{(\ell)}, P^{-1}\boldsymbol{o}\rangle$, the inner product of the code and the centroid $\langle\bar{\boldsymbol{x}}^{(\ell)}, P^{-1}\boldsymbol{c}\rangle$, and the error scale $\kappa^{(\ell)}$ (Eq.~\eqref{eq:kappa}), are computed at indexing time. Since the test radius must be an exact distance to actual vectors (C2 and C4 in Section~\ref{sec:guarantee}), the original vectors are also stored. Vectors with zero residual (vectors equal to the centroid) have no defined unit vector or code, so they are marked separately.

\begin{figure}[t]
\centering
\includegraphics{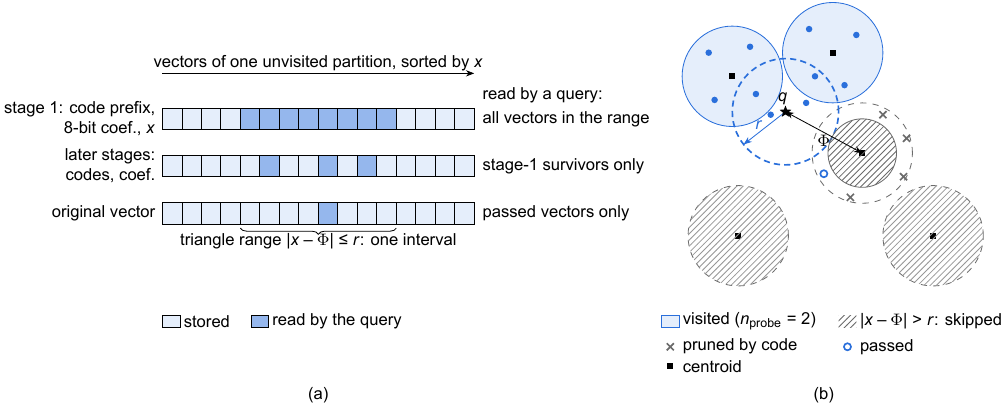}
\refstepcounter{figure}\label{fig:structure}
\par\vspace{3pt}{\small\setstretch{1.15}\IEEEfigcap{Figure \thefigure. Index structure and query processing (schematic). (a) Storage layout of one partition. (b) Geometry of query processing.}\par}
\end{figure}

Algorithm~\ref{alg:build} is the construction procedure of the index. The number and sizes of the partitions and the assignment rule are determined without reading $P$, so they are irrelevant to the guarantee and change only the cost. A large number of partitions $N$ increases the routing cost of computing the distances to all centroids for each query, and a small $N$ increases the number of vectors in the visited partitions and hence the exact distance computations. In contrast, the first-stage test of unvisited vectors is performed for all vectors, so its cost is nearly independent of $N$, and a larger $N$ reduces the residual scale and thereby the first-stage pass rate (Section~\ref{sec:costlaw}). The centroids are obtained by the k-means~\cite{r159} of FAISS~\cite{r23} (line 1). The assignment places an upper bound on the partition size, following the same idea as the balanced clustering of SPANN~\cite{r24} (line 2). This is because every partition is tested for each query and batch queries are processed in parallel per partition, so the largest partition determines the load imbalance among threads. However, vectors assigned to a centroid other than the nearest one because of the cap have larger residuals. Accordingly, vectors are assigned in decreasing order of the difference between the squared distances to their two nearest centroids, i.e., the increase in the squared residual when assigned to the second-nearest centroid, and the destinations are limited to a few nearest centroids (the nearest one if all are full). Lines 1–3 do not read $P$, and $P$ is drawn afterward in line 4, once and independently of the data, uniformly from the set of all $D$-dimensional orthogonal matrices. Lines 5–6 apply the Encode function (lines 7–11) to every vector and the Layout function (lines 12–15) to every partition. The Encode function normalizes the residual of one vector to make its code (line 8), computes for each stage the inner products of the code and the error scale $\kappa^{(\ell)}$ (Eq.~\eqref{eq:kappa}) (line 9), and, if the 2-bit final stage (Section~\ref{sec:devices}) is used, also computes that code (line 10). The Layout function sorts the vectors of one partition in order of the residual norm $x$ and stores them divided into the first-stage data, the later-stage codes and coefficients, and the original vectors (Fig.~\ref{fig:structure}(a); see the storage layout paragraph below). The main costs of construction are k-means, applying the rotation (matrix multiplication, $O(nD^2)$), and computing the codes ($O(nD)$); Section~\ref{sec:cost} compares index construction costs. The segment structure for updates is described in Section~\ref{sec:dynguarantee}, and insertion, deletion, and merging in Section~\ref{sec:dynops}; these operations reuse the Encode and Layout functions as they are (Algorithm~\ref{alg:update}).

\begin{algorithm}[t]
\caption{Index Construction}\label{alg:build}
\KwIn{$n$ data vectors $\boldsymbol{o}_r$, number of partitions $N$ and size cap, stage configuration $m_1 < \dots < m_L \le D$, and whether to use the 2-bit final stage}
\KwOut{partition centroids, rotation $P$, and per-partition sorted codes, coefficients, and original vectors}
compute $N$ centroids by k-means\;
in decreasing order of the difference between the squared distances to the two nearest centroids, assign each vector to the first of a few nearest centroids that has not reached the size cap (to the nearest one if all are full)\;
centroid $\boldsymbol{c}$ of each partition $\leftarrow$ mean of its assigned vectors\;
$P \leftarrow$ matrix drawn uniformly from the set of all $D$-dimensional orthogonal matrices\;
\lFor{each vector $\boldsymbol{o}_r$}{\textit{Encode}($\boldsymbol{o}_r$, centroid $\boldsymbol{c}$ of its assigned partition, $P$)}
\lFor{each partition}{\textit{Layout}(partition)}
\Fn{Encode($\boldsymbol{o}_r$, $\boldsymbol{c}$, $P$)}{
$x \leftarrow \lVert\boldsymbol{o}_r - \boldsymbol{c}\rVert$, $\boldsymbol{o} \leftarrow (\boldsymbol{o}_r - \boldsymbol{c})/x$, $\bar{\boldsymbol{x}} \leftarrow \mathrm{sign}(P^{-1}\boldsymbol{o})/\sqrt D$\;
\lFor{$\ell = 1, \dots, L$}{compute $\langle\bar{\boldsymbol{x}}^{(\ell)}, P^{-1}\boldsymbol{o}\rangle$, $\langle\bar{\boldsymbol{x}}^{(\ell)}, P^{-1}\boldsymbol{c}\rangle$, $\kappa^{(\ell)}$}
if the 2-bit final stage is used, compute the 2-bit code and its inner products and $\kappa$\;
return $x$ and the per-stage codes and coefficients\;
}
\Fn{Layout(partition)}{
sort the vectors of the partition by $x$, and quantize the first-stage coefficients to 8 bits with the minimum and step\;
lay out the first-stage codes in blocks of 64 vectors to match the SIMD register width\;
store the original vectors and the later-stage codes and coefficients in sorted order\;
}
\end{algorithm}

The proposed index is used differently from standard IVF~\cite{r19,r23}. Standard IVF visits only $n_{\rm probe}$ partitions and tolerates missing the true neighbors left in the rest, whereas the proposed index tests all vectors of the unvisited partitions as well and discards only the vectors that are certainly outside the radius. Accordingly, residuals and codes serve different purposes. IVF-PQ encodes residuals because smaller residuals can be quantized more accurately with the same code length~\cite{r19}, whereas in the proposed index a smaller residual scale enlarges the margin of the test so that more vectors are discarded (Section~\ref{sec:costlaw}). Codes are read from the front, and the failure probability of the test at each stage is set by Theorem 1. Making the distribution of the test error data-independent by a random orthogonal transformation and keeping the original vectors for re-ranking by exact distances are the same as in the IVF index of RaBitQ~\cite{r62}; in the proposed index, they are the premises of, respectively, the computation of the failure probability (Lemma 1, Theorem 1) and the test radius determined by exact distances (Section~\ref{sec:guarantee}). In addition, in the proposed index, an insertion is a partition assignment, a deletion is a bitmap mark, and maintenance takes place per centroid, so the conditions of the guarantee remain simple under updates (Section~\ref{sec:dynamic}).

Since most of the search time is spent on the first-stage tests and on processing the vectors that pass the first stage (Section~\ref{sec:costlaw}), the storage layout is designed to reduce the amount of data read in these two kinds of processing. Fig.~\ref{fig:structure}(a) shows the storage layout of one partition, which is built by the Layout function of Algorithm~\ref{alg:build} (lines 12–15). The vectors in a partition are sorted in order of the residual norm $x$. Then, as in CLIP~\cite{r14}, which skips parts of inverted lists, the triangle-inequality range becomes a single contiguous interval in each partition and can be found by binary search. The storage is divided into three parts according to which vectors they are read for. The first-stage data at the top are read for all unvisited vectors in the triangle-inequality range for every query, so they are kept much smaller than the originals. The first-stage data consist of the code prefix, three coefficients for computing Eq.~\eqref{eq:rule} with a single comparison per vector, and $x$. The coefficients are determined by $x$, $\langle\bar{\boldsymbol{o}}^{(1)}, \boldsymbol{o}\rangle$, $\kappa^{(1)}$, and the inner product of the code and the centroid $\langle\bar{\boldsymbol{x}}^{(1)}, P^{-1}\boldsymbol{c}\rangle$ (the use of this inner product is described in Section~\ref{sec:process}), and they are stored quantized to 8 bits with the minimum and step of each partition. For example, with $m_1 = 128$, the first-stage data take 23 bytes per vector. Below them, the later-stage codes and coefficients are read only for the vectors that pass the preceding stage, and the original vectors (integer data in their original 8-bit integers) are read only for the visited vectors and the vectors that pass all stages.

\subsection{Query Processing}\label{sec:process}

Algorithm~\ref{alg:query} and Fig.~\ref{fig:structure}(b) show the processing of a single query. Lines 1–3 visit $n_{\rm probe}$ partitions in order of the exact distance $\Phi$ to the centroid, compute the exact distances between the query and all vectors in those partitions, and set the $k$-th of these distances as the initial radius $r$. In the unvisited partitions, vectors outside the triangle-inequality range are excluded without testing (line 6), and vectors within the range are tested progressively with binary codes (line 8). The 1-bit test stages start at the first-stage code length $m_1$ and end at $D$ bits, and if the 2-bit final stage is used, one more stage is added after them (Section~\ref{sec:devices}). However, vectors with zero residual have no code, so if they fall within the triangle-inequality range, their exact distances are computed without a code test (lines 8–9). The vectors that pass all stages are re-ranked by computing their exact distances; with these distances, the radius is shrunk to the $k$-th distance among the candidates, and subsequent tests use the shrunk radius (line 9, adaptive radius). Line 10 returns the $k$ nearest candidates (Lemma 2), and vectors near the $k$-th distance are ranked after their distances are recomputed in double precision (fp64).

Fig.~\ref{fig:structure}(b) illustrates this process in two dimensions. The two partitions nearest to the query $q$ are visited, and the radius $r$ (dashed circle) is set from the vectors in them (lines 1–3). If the distance between the query and the centroid of an unvisited partition is $\Phi$, vectors closer to the centroid than $\Phi - r$ have $x < \Phi - r$ and hence lie farther than $r$ from the query, so they are excluded without testing (hatched inner circle, line 6). The remaining vectors are tested with their codes; most of them are pruned ($\times$), and only a few pass and have their exact distances computed ($\bigcirc$, lines 8–9). Partitions all of whose vectors lie outside the triangle-inequality range, like the two partitions at the bottom of the figure, are excluded entirely.

\begin{algorithm}[t]
\caption{Query Processing}\label{alg:query}
\KwIn{query vector $\boldsymbol{q}_r$, $k$, the index and rotation $P$ of Algorithm~\ref{alg:build}, error-bound parameters $\varepsilon_{0,\ell}$ ($\ell = 1, \dots, L$)}
\KwOut{$k$ vectors close to $\boldsymbol{q}_r$}
compute the distances $\Phi$ to all centroids, and take the $n_{\rm probe}$ partitions with the smallest $\Phi$ as the visited partitions\;
$C \leftarrow$ vectors of the visited partitions with their exact distances\;
$r \leftarrow$ $k$-th distance in $C$ ($\infty$ if $\lvert C\rvert < k$)\;
build LUTs from $P^{-1}\boldsymbol{q}_r$\;
\For{each unvisited partition (a front part with small $\Phi$ first)}{
find the interval with $\lvert x - \Phi\rvert \le r$ by binary search\;
\For{each vector $\boldsymbol{o}_r$ in the interval}{
\If{$x = 0$, or $\boldsymbol{o}_r$ passes all tests (Eq.~\eqref{eq:rule}) of stages $\ell = 1, \dots, L$ with the current $r$}{
compute the exact distance of $\boldsymbol{o}_r$; if it is at most $r$, add $\boldsymbol{o}_r$ to $C$ and set $r \leftarrow$ $k$-th distance in $C$\;
}
}
}
return the $k$ vectors with the smallest distances in $C$\;
\end{algorithm}

Algorithm~\ref{alg:query} satisfies the conditions of Theorem 5. The centroids are determined before $P$ (Algorithm~\ref{alg:build}), the stage configuration is determined without reading the rotation (Section~\ref{sec:devices}), and the error-bound parameters $\varepsilon_{0,\ell}$ are fixed in advance, independently of the rotation and the query, by $G_D(-\varepsilon_{0,\ell}) = \delta_\ell$ from the per-stage shares $\delta_\ell$ of the failure probability budget $\beta$ (Theorems 2–4); hence C1 holds. The visited vectors are the vectors of the $n_{\rm probe}$ partitions chosen in order of $\Phi$, which is independent of $P$, so the initial radius satisfies C2 (lines 1–3). Lines 6 and 8 satisfy C3 and lines 9–10 satisfy C4, and in addition, line 9 shrinks the radius only to distances of $k$ distinct vectors whose exact distances have been computed. Only the test in line 8 reads the rotation $P$ (the LUTs in line 4 are a preparation for this test); the radius shrunk in line 9 depends on $P$ through the test results of line 8, and by Theorem 5 the guarantees of Theorems 2–4 hold in this case as well. Why the remaining lines do not discard true neighbors was shown in Section~\ref{sec:guarantee}.

In standard IVF, $n_{\rm probe}$ determines recall. In contrast, the proposed index tests all unvisited partitions as well, so $n_{\rm probe}$ is merely the number of candidates from which the initial radius is set, and the guarantee that the failure probability is at most $\delta$ (Theorems 2–5) holds regardless of the value of $n_{\rm probe}$. Increasing $n_{\rm probe}$ brings the initial radius closer to the true $k$-th distance and reduces the number of vectors that pass the test, but increases the number of vectors whose exact distances are computed in the visited partitions.

The implementation described below serves to speed up query processing, and the theorems of Section~\ref{sec:theory} permit it while preserving the guarantee. Since most of the time of Algorithm~\ref{alg:query} is spent in lines 8–9 (Section~\ref{sec:costlaw}), we speed up the first-stage test by computing it for many vectors at once with SIMD instructions. Since Theorem 1 sets the failure probability of the test per stage, the guarantee remains intact even if the first stage tests with short codes and the later stages process only the few vectors that pass the first stage. Query processing is implemented in C++ with AVX-512. The LUTs are built once per query from the rotated coordinates $P^{-1}\boldsymbol{q}_r$ of the original query vector, and $\langle\bar{\boldsymbol{x}}, P^{-1}(\boldsymbol{q}_r - \boldsymbol{c})\rangle$, which is needed for the numerator $\langle\bar{\boldsymbol{x}}, P^{-1}\boldsymbol{q}\rangle$ of Eq.~\eqref{eq:est}, is obtained from them by subtracting the inner product of the code and the centroid stored in the per-vector coefficients. As in fast-scan~\cite{r177}, the first stage quantizes a 16-entry LUT for every 4 code coordinates (4 bits) to 8 bits, keeps it in SIMD registers, and looks it up with shuffle instructions; it accumulates 64 vectors at a time, the number that fits in one 512-bit register, and tests them in the integer domain. Moreover, since Theorem 5 does not restrict the order of tests, a single query scans a front part of the unvisited partitions first and then the rest in stored order so that memory is read sequentially, and batch queries are sorted in order of their nearest centroids and processed in groups per partition so that the first-stage data are reused from the cache.

Since Theorem 1 requires only $f \ge \sqrt{1-\max(\theta(r), 0)^2}$ of the slack factor, the first stage uses $f = \min(1, r/\Phi)$, a single value per partition. This value is an upper bound of $\sqrt{1-\max(\theta(r), 0)^2}$ over $x$, so it satisfies the condition of Theorem 1 for all vectors in the partition. Since it is a single value per partition, the test can be computed from per-vector coefficients and per-query, per-partition values alone, which suits the SIMD test above; and in unvisited partitions with $r < \Phi$, the slack is smaller than when 1 is used as in RaBitQ~\cite{r62}, so more vectors are pruned. The later stages use the per-vector value $\sqrt{1-\max(\theta(r), 0)^2}$. Testing with 8-bit quantized LUTs and coefficients keeps the guarantee intact. If the rounding errors of quantization and accumulation are replaced by deterministic bounds so that the left-hand side of the test is computed to be no smaller than its exact value and the right-hand side no larger than its exact value (with a bound that holds for every rotation, even if the slack depends on the rotation), then any vector pruned by the implementation has an exact left-hand side smaller than the exact right-hand side and is therefore also pruned by Eq.~\eqref{eq:rule}. Hence, the test of the implementation passes every vector that Eq.~\eqref{eq:rule} passes, and Theorem 1 applies as is. The sum of these bounds is the deterministic slack $\sigma_{\det}$ of Section~\ref{sec:costlaw} (the list of errors to which this argument applies is in Appendix~\ref{app:proofpass}).

Since the radius of Theorem 5 need only be the maximum of the distances of some $k$ distinct vectors whose exact distances have been computed, synchronization among threads can be reduced when a single query is processed by multiple threads. Each thread keeps a local heap, the radius is shared through atomic minimum updates, and the vectors in a local heap are handed over to the shared heap whenever the lock can be acquired without waiting and when the thread finishes its processing. On the other hand, counting the same vector twice could make the radius smaller than the true $k$-th distance; hence, if a vector that has already entered the candidates through early exact computation (Section~\ref{sec:devices}) passes the test again during the scan, it is filtered out by its id so that it is not inserted into the candidates twice.

\subsection{Parameter Selection by the Cost Law}\label{sec:devices}

The proposed index has options that change only the query processing cost while preserving the guarantee: the configuration of the test stages (the first-stage code length $m_1$ and the use of a final 2-bit test stage) and the use of \textit{optimization techniques} that shrink the radius early or reduce the deterministic slack. Which choice is faster depends on which term of Eq.~\eqref{eq:passprob} is large, and thus differs by dataset and metric. In this section, we describe the parameter selection rule based on the cost law of Section~\ref{sec:costlaw}. The rule chooses the stage configuration and global centering by the prediction of Eq.~\eqref{eq:passprob}, which does not read the rotation, and the use of the other two techniques by times measured on calibration queries; since all choices lie within the conditions of Theorem 1 and Theorem 5, the per-query guarantee remains intact whatever is chosen. Therefore, calibration queries affect only speed, and the guarantee is maintained even if the distributions of calibration queries and actual queries differ. This distinguishes our approach from standard IVF, which tunes $n_{\rm probe}$ on calibration queries to meet an average recall, and from CTR~\cite{r11} and ConANN~\cite{r1}, whose guarantees hold under the assumption that calibration queries and actual queries follow the same distribution.

The targets of the selection are the three terms of Eq.~\eqref{eq:passprob} that determine the first-stage pass rate: the margin $g$, the error scale $\kappa^{(1)}$, and the deterministic slack $\sigma_{\det}$. Since the margin grows as the radius shrinks, the adaptive radius of Section~\ref{sec:process} is always used, and two techniques that shrink the radius early are added for cases in which the initial radius is larger than the true $k$-th distance, such as out-of-distribution queries. Early exact computation computes, before line 5 of Algorithm~\ref{alg:query}, the exact distances of the vectors with small first-stage estimated distances in a few partitions at the front in $\Phi$ order; estimate-ordered processing, when a single query is processed by multiple threads, first processes to the end a few vectors with small estimated distances among those that pass the first stage in a partition, thereby shrinking the radius, and then processes the rest (lines 7–9). The error scale $\kappa^{(1)}$ is determined by the first-stage code length $m_1$; longer codes reduce the pass rate, but the first-stage cost increases proportionally. Adding, after the last 1-bit stage, a stage that tests with a 2-bit code that quantizes each coordinate of $P^{-1}\boldsymbol{o}$ to a 4-level uniform grid symmetric about 0 makes the error scale of that stage about half that of the 1-bit code, at the expense of more memory. The LUT quantization slack, which accounts for most of the deterministic slack, is proportional to the norm of the query vector, so for data whose mean is far from the origin, it is reduced by \textit{global centering}, which builds the LUTs in coordinates relative to the mean $\boldsymbol{c}_0$ of the entire data. The benefits of these choices differ by dataset. For example, the configuration that tests the first stage with short codes is fast on in-distribution data, but on datasets with out-of-distribution queries, where many vectors pass the first stage, the configuration that reads the entire code at once is faster (Section~\ref{sec:ablation}). The operation and implementation of each technique are summarized in the appendix (Appendix~\ref{app:techniques}).

The per-query guarantee is preserved no matter how the first-stage code length $m_1$, the 2-bit final stage, global centering, early exact computation, and estimate-ordered processing are chosen or turned on and off. The stage configuration is determined without reading the rotation, so it lies within the assumptions of Theorem 1. The 2-bit code, like Extended RaBitQ~\cite{r63}, determines the code, including the scale of the grid, from each vector's $P^{-1}\boldsymbol{o}$ alone, so it satisfies the conditions of Lemma 1, and its stage is also included when the failure probability budget $\beta$ is divided into per-stage shares. Global centering does not change the test value and only reduces the slack, since $\langle\bar{\boldsymbol{x}}, P^{-1}(\boldsymbol{q}_r - \boldsymbol{c}_0)\rangle - \langle\bar{\boldsymbol{x}}, P^{-1}(\boldsymbol{c} - \boldsymbol{c}_0)\rangle = \langle\bar{\boldsymbol{x}}, P^{-1}(\boldsymbol{q}_r - \boldsymbol{c})\rangle$. The adaptive radius and the two techniques that assist it change only the radius, the order of tests, and the choice of vectors whose exact distances are computed first, so they lie within the conditions of Theorem 5.

The parameter selection rule is run once before queries are processed. After the per-stage values and the storage layout are built for each candidate configuration with the Encode and Layout functions of Algorithm~\ref{alg:build}, the first-stage code length $m_1$ and the use of the other techniques are determined for each dataset, query setting ($k$ and the guarantee level), and performance metric. This is because, even on the same data, the error-bound parameters differ with the guarantee level, and the fastest configuration can differ between processing queries one at a time and processing them together. The metrics are the latency when a single query is processed by a single thread, the latency when a single query is divided among multiple threads, and the throughput when many queries are processed together by multiple threads. The stage configuration and global centering are chosen as the configuration with the shortest predicted time on calibration queries (the simpler one if the difference is small). The predicted time is obtained by converting into time, with machine constants, a pass-rate prediction that uses the true $k$-th distance of the calibration queries as the radius and lightly approximates Eq.~\eqref{eq:passprob} (setting each vector's $\rho$ to 0, $\kappa$ to a per-stage representative value, and the deterministic slack to an expression proportional to the query norm); the candidates are the configurations that combine the first-stage code length (the default or $D$), the use of the 2-bit final stage, and the use of global centering. The default is set so that a longer code is chosen when the first-stage pass rate is higher, measured with some data vectors as queries on an index built with a separate rotation independent of the index. The inputs of these rules are the true $k$-th distances of the calibration queries, the query–centroid distances, the residual norms, the error-bound parameters, per-stage representative values obtained from the universal distribution of $\kappa$, machine constants, and the pass rate measured with the separate rotation; they do not read the rotation or the codes of the index.

Whether to turn on early exact computation (when a single query is processed by a single thread, and for batches) and estimate-ordered processing (when a single query is processed by multiple threads) is decided by directly measuring the times of the enabled and disabled paths on calibration queries, and a technique is turned on only when the enabled path is sufficiently faster (hereafter, the enabling rule). These times depend on the rotation of the index, so this amounts to choosing a path by looking at the rotation, yet the guarantee remains intact. The two paths share the visited partitions, the stage configuration, and the error-bound parameters, so on either path a true neighbor is missed only when the estimation error of an unvisited true neighbor exceeds the error-bound parameter ($\zeta^{(\ell)} < -\varepsilon_{0,\ell}$) (Theorem 5). This condition is determined independently of the path, so whichever path is chosen, the failure probability does not exceed the bounds of Theorems 2–4. In our measurements, the configurations chosen by the rule were 1.18$\times$ faster in geometric mean than the configuration with the default first-stage code length and all techniques other than the adaptive radius turned off, over 96 settings on eight 1M datasets, with no setting slowed down; the gains were concentrated on laion, yandex, and SSNPP, where one term of Eq.~\eqref{eq:passprob} is large, reaching up to 3.8$\times$ (Section~\ref{sec:ablation}). The values and procedure of the rule are summarized in the appendix (Appendix~\ref{app:negative}).

\bigskip

\section{Guarantees under Updates}\label{sec:dynamic}

In this section, we extend the index and search algorithm of Section~\ref{sec:search} to a dynamic environment with ongoing insertions and deletions. The extension adds three components: a delta segment that holds new vectors, a bitmap that marks deletions, and merges that combine the two segments. Update operations reuse the encoding function and the layout function of Algorithm~\ref{alg:build}, search runs Algorithm~\ref{alg:query} unchanged, and neither the implementation of Section~\ref{sec:process} nor the parameter selection rule of Section~\ref{sec:devices} changes. Without updates, the dynamic index is identical to the index of Section~\ref{sec:search}. Section~\ref{sec:dynguarantee} defines the structure of the dynamic index, gives the conditions the dynamic index must satisfy, shows that the static guarantees are maintained unchanged under these conditions, and shows the risk when updates are chosen after observing search results. Section~\ref{sec:dynops} explains what the update operations and search reuse from Section~\ref{sec:search} and shows that they satisfy these conditions. Section~\ref{sec:maintenance} addresses how to maintain an efficient index without affecting the guarantee and explains why the overhead added by the dynamic extension is small.

\subsection{Guarantee under Updates}\label{sec:dynguarantee}

Like the long-term index, temporary index, and delete list of FreshDiskANN~\cite{r93}, the dynamic index consists of a \textit{main segment}, a \textit{delta segment}, and a \textit{deletion bitmap} (Fig.~\ref{fig:dynamic}(a)). The main segment is the index built by Algorithm~\ref{alg:build}; new vectors are inserted into the delta segment, deletions are marked in the bitmap, and accumulated updates are folded into the main segment by merges. As updates continue, new vectors lie farther from their assigned centroids, so residuals grow, and partition sizes become uneven, which raises search cost (Section~\ref{sec:costlaw}). To reverse this, at merges we maintain an efficient index by applying \textit{recentering}, which moves a centroid to the mean of its live members; \textit{split}, which divides a partition in two by 2-means on the original vectors; and \textit{dissolution}, which removes a small partition and moves its members to the nearest other partitions. Hereafter we call these three operations \textit{maintenance operations}. The procedure for applying the operations at a merge is described in Section~\ref{sec:dynops}, and the method for choosing them in Section~\ref{sec:maintenance}. A vector not marked as deleted is called a \textit{live vector}; let $L_t$ be the set of live vectors at time $t$, $\mathcal N_{k,t}^*(q)$ the set of true $k$ nearest neighbors in it, and $\mathrm{Recall}_t(q)$ the recall with it as the ground truth.

Probabilities are defined as in Section~\ref{sec:problem}. We call the insertions and deletions that have occurred from the initial index build up to time $t$, listed in the order in which they occurred, the \textit{update sequence}. The times of merges and of rotation changes are determined by values that do not use the rotation, such as vector counts and a predetermined schedule, and are regarded as fixed together with the update sequence. The probability at time $t$ is then taken only over $P$, with the data, the update sequence, and the query fixed. This extends the “given query” of the static guarantee to updates, and means that the update sequence and the query are given independently of the rotation. In this frame, for the dynamic index to keep the guarantee, it must satisfy the following three conditions in addition to the conditions of Section~\ref{sec:guarantee}.

\begin{description}[font=\normalfont,leftmargin=3.6em,labelwidth=3.2em,labelsep=0.4em,align=left]
\item[C5.] When centroids and assignments are changed to maintain an efficient index, which partitions are changed and how, and the resulting new centroids and assignments, are determined only from values computed without using the rotation, such as residual norms of original vectors, partition sizes, 2-means on original vectors, and predictions by Eq.~\eqref{eq:passprob}. The number of vectors passing stage 1 as actually counted by testing under the current rotation (hereafter the \textit{observed pass count}) and search results are not used.
\item[C6.] The two segments use the same rotation and the same centroids, and each stored code is a function only of its vector, the current centroid, and the current rotation.
\item[C7.] Vectors marked as deleted are used neither as candidates nor for the radius (also when the radius is shrunk while unvisited vectors are tested), and at each time C1, C3, and C4 hold and the radius used for testing is set so as to satisfy C2 or the conditions of Theorem 5.
\end{description}

Each condition is needed for the following reasons. If C5 is violated, centroids become functions of the rotation, and the distribution of the test error cannot be computed by Theorem 1. C6 ensures that, even after updates, the stored codes represent residuals with respect to the current centroids. If codes computed with old centroids remain after recentering, the residuals used by the pruning rule and the codes disagree, and the test error does not follow the distribution of Theorem 1. If C7 is violated, the distance of a vector marked as deleted can enter the radius, and the radius can become smaller than the true $k$-th distance among live vectors.

\begin{thmblock}{\textbf{Theorem 6} (Per-Query Guarantee under Updates):}In this frame, a dynamic index satisfying C5–C7 has the following properties. (a) For a query at any time $t$, the result of the dynamic index satisfies, unchanged, the guarantees obtained by applying Theorems 2–5 to a static index with $L_t$ as its data. For example, in the $(\tau, \delta)$-recall guarantee mode, \begin{equation}\Pr\big[\mathrm{Recall}_t(q) < \tau\big] \le \delta\label{eq:dyn}\end{equation} and the failure probability $\delta$ need not be split between the main segment and the delta segment. (b) When the same rotation continues to be used, let $F_s$ and $F_t$ be the failure events of the same query asked at two times $s < t$; then $\Pr[F_s \cup F_t] \le 2\delta$, and the two events are in general not independent. In particular, if a true neighbor pruned at time $s$ remains a true neighbor at time $t$, its centroid and deterministic slack are unchanged, it is still unvisited and not a target of early exact computation, and the radius at the moment it is tested satisfies $r_t \le r_s$, then that neighbor is pruned at time $t$ as well. (c) If the rotation is changed on a predetermined schedule at times when all live vectors are re-encoded with a new rotation (e.g., merges), failures at two times that use different rotations are mutually independent (given the live vectors, the centroids, the assignments, the query, and the decisions of the parameter selection rule).\end{thmblock}

The proof of Theorem 6 is given in Appendix~\ref{app:proofdynamic}. Eq.~\eqref{eq:dyn} has the same form as Eq.~\eqref{eq:goal}, except that recall is measured against $\mathcal N_{k,t}^*(q)$, the set of true $k$ nearest neighbors among the vectors live at time $t$. The proof starts from the following: since the update sequence is given and, by C5, the state at time $t$ (live vectors, centroids, assignments) is independent of the rotation, the visited set is determined independently of $P$; and by C7, the radius is set from exact distances of live vectors. Since an unvisited true neighbor is live, by C7 the triangle-inequality range and the test path of its own segment process that vector, and by C6 the premise of Lemma 1 holds. The terms of the union bound and Markov's inequality are only the $k$ vectors of $\mathcal N_{k,t}^*(q)$, so each neighbor is counted only once, whichever segment it resides in. In a simple design that sets the radius of the delta segment to the $k$-th distance of the main segment's result, two union bounds are needed, and the failure probability $\delta$ must be divided between the two segments. What Theorem 6 shows is that such a split is unnecessary if the radius and the error-bound parameters are set from the whole of $L_t$ from the start. In addition, Theorem 6 identifies the conditions (C5–C7) under which the guarantee is maintained unchanged as updates continue, and distinguishes how failures at different times are related when the same rotation continues to be used and when it is changed.

Adaptive updates, in which updates are chosen after observing the rotation or search results, fall outside this frame, and the guarantee can then be broken. An adaptive adversary~\cite{r175}, which chooses inputs after observing the rotation or test results, can insert a new nearest vector of query $q$ into an unvisited partition, choosing it so that Eq.~\eqref{eq:rule} surely prunes it; the failure probability of exact search then becomes 1 (for $\tau < 1$, it suffices to insert $m+1$ such vectors). In our synthetic experiment ($D = 32$, $\delta = 0.05$), insertions chosen in this way by observing the rotation caused failures on 67 of 68 queries under that same rotation, but when the same vectors were evaluated under 24 independent rotations, the average failure rate was 0.0049, within $\delta$. Selection bias also arises when decisions made to maintain an efficient index observe the rotation (that is, when C5 is violated). If the targets of recentering or split are chosen by observed pass counts, centroids become functions of the rotation, and if the configuration with the smallest observed pass count is chosen among several candidate configurations, the upper bound on the failure rate is multiplied by the number of candidates.

Updates decided from exact-search results (e.g., inserting an upload when no matching copy is found), however, see the rotation only through failures. If the decision depends only on the distances of the results (e.g., whether the nearest distance is at most a threshold) and the deciding queries are exact searches ($\tau = 1$), a query at time $t$ fails with probability at most $(1 + j)\delta$, where $j$ is the number of earlier queries whose results decided updates since the last rotation change. The update sequence that would arise if every result were exact (the ideal sequence) is a function of the data, the query sequence, and the decision rule alone, so it is independent of the rotation, and every exact result has the same distances and hence yields the same decision. The actual state can therefore depart from the ideal state only if one of the $j$ earlier queries fails on the ideal state, which by Theorem 6 (a) has probability at most $\delta$ for each query. A failure at time $t$ is either a failure on the ideal state or such a departure, so the union bound gives $(1 + j)\delta$. If the rotation is replaced at predetermined times, the updates before a replacement form a fixed sequence independent of the new rotation, so only queries after the replacement are counted. Hence a pipeline that, like copy detection, decides insertions from results and makes at most $j_{\max}$ such decisions between merges that replace the rotation can meet a target failure probability by setting $\delta$ to the target divided by $1 + j_{\max}$. The cost depends on how small $\delta$ becomes. The error-bound parameters grow only like $\sqrt{\ln(1/\delta_\ell)}$, and a 10$\times$ smaller $\delta$ took only 1.08–1.14$\times$ the time (geometric means), but the first-stage pass rate grows faster than these parameters, so on six 1M datasets $\delta = 10^{-4}$ and $10^{-8}$ took 1.37$\times$ and 2.25$\times$ the 1-thread time of $\delta = 0.01$ in geometric mean (up to 5.7$\times$), still less than that of FAISS IndexFlatL2 in every setting, and even at $\delta = 10^{-8}$ the time of copy-detection range queries changed by at most 6.2\% (Section~\ref{sec:cost}).

A \textit{filtered} $k$-NN query returns the $k$ nearest vectors among those that satisfy a predicate $\varphi_q$ on metadata. Let $S_q$ be the live vectors at time $t$ that satisfy $\varphi_q$, and measure recall, $\mathrm{Recall}_\varphi(q)$, against the true $\min(k, \lvert S_q\rvert)$ nearest neighbors in $S_q$. No new machinery is needed.

\begin{thmblock}{\textbf{Corollary 2} (Filtered Search):}Suppose that $\varphi_q$ is determined independently of the rotation (by the query, the metadata, and the update sequence, not by the results of searches that use $P$), and that the search marks the vectors in $L_t \setminus S_q$ in the deletion bitmap, in addition to the deleted vectors, and then runs the dynamic search unchanged. Then, for a dynamic index satisfying C5–C7, (a) every returned vector satisfies $\varphi_q$; (b) if $\lvert S_q\rvert \le k$, the result is $S_q$ itself, so it is exact; and (c) if $\lvert S_q\rvert > k$, the guarantees of Theorems 2–5 hold with $S_q$ as the data, for example $\Pr[\mathrm{Recall}_\varphi(q) < \tau] \le \delta$, for any selectivity and without splitting $\delta$ among predicates or segments.\end{thmblock}

To prove it, append to the update sequence the deletion of $L_t \setminus S_q$ at time $t$ without a merge. Since $\varphi_q$ is independent of $P$, this sequence is still given independently of $P$; deletions only set bits, so C5 and C6 continue to hold; and the filtered search is the search just after time $t$, whose live set is $S_q$. Theorem 6 (a) then gives (c), and C7 gives (a). For (b), C2 sets the radius to infinity while fewer than $k$ vectors have been visited, and the search visits partitions until $k$ live candidates are collected or the partitions are exhausted, so every vector of $S_q$ is computed exactly. Predicates built from the results of other searches with the same rotation (e.g., a semi-join of two $k$-NN operators) fall outside the corollary, as do the result-decided updates above. The guarantee does not depend on selectivity, but the cost does: stage 1 still reads the codes of masked vectors, and the radius, the $k$-th distance within $S_q$, grows as selectivity falls. Section~\ref{sec:scale} measures this cost and a plan that switches to exact search over $S_q$ when $S_q$ is small.

\subsection{Update Operations and Search}\label{sec:dynops}

This section describes the update operations and search designed to satisfy C5–C7 of Section~\ref{sec:dynguarantee}. Algorithm~\ref{alg:update} gives the update operations and search of the dynamic index. The main segment is exactly the index of Section~\ref{sec:search}, and the delta segment is an index of the same format using the same partitions. Insertion (lines 1–4) assigns a new vector to the partition of the nearest current centroid, computes the code and coefficients of that single vector with the encoding function of Algorithm~\ref{alg:build} (Section~\ref{sec:index}), inserts them into the delta segment, and then re-lays out only that partition of the delta segment with the layout function. Neither k-means (line 1 of Algorithm~\ref{alg:build}) nor rotation generation (line 4) is redone, and the codes of other vectors are not changed. When several vectors are inserted together, each vector is assigned and encoded, and then the layout function needs to run only once for each partition of the delta segment that received vectors. Deletion (lines 5–6) ends with marking the bitmap; neither the codes nor the 8-bit quantization ranges of the pruning-rule coefficients kept per partition (line 13 of Algorithm~\ref{alg:build}) are recomputed. These ranges are the reference for restoring the stored coefficients, and the coefficients of live vectors are quantized within these ranges; so if the ranges are kept, the restored values do not change, and the deterministic slack only becomes slightly larger, to the extent that the interval is wider than necessary. Hence, even without physically removing vectors marked as deleted, the bounds used by the test do not become wrong.

\begin{figure}[t]
\centering
\includegraphics{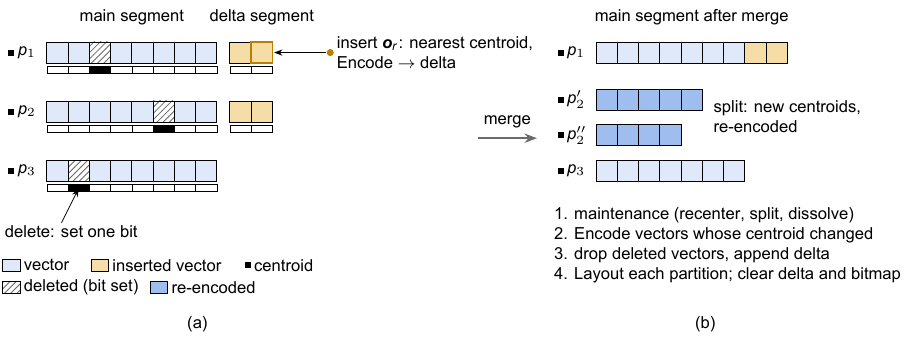}
\refstepcounter{figure}\label{fig:dynamic}
\par\vspace{3pt}{\small\setstretch{1.15}\IEEEfigcap{Figure \thefigure. Structure and updates of the dynamic index (schematic). (a) Before a merge. (b) After a merge.}\par}
\end{figure}

The delta segment must be scanned separately in each unvisited partition, and vectors marked as deleted are still read as stage-1 data until they are removed; therefore, when accumulated updates grow, we perform a merge that folds the delta segment into the main segment (lines 7–13, Fig.~\ref{fig:dynamic}(b)). A merge is performed when the number of vectors in the delta segment plus the number of vectors marked as deleted in the main segment reaches a fixed fraction (the \textit{merge threshold}) of the number of live vectors in the main segment. A merge removes vectors marked as deleted and empty partitions, folds the vectors of the delta segment into the partitions of the main segment, and rebuilds the layouts with the layout function (lines 11–12). Codes are not recomputed, since they were already computed with the current centroids at insertion. Since a merge is also the time when layouts are rebuilt, maintenance operations that reduce residuals grown by updates are applied at the same time (line 8). Which operations to apply and when is addressed in Section~\ref{sec:maintenance}. After maintenance operations, only vectors whose assigned centroid changed, that is, vectors of partitions that underwent recentering or split and vectors moved to another partition by dissolution, need to be re-encoded (line 10), so this work is handled together with rebuilding the layouts. Hence a merge skips k-means, rotation generation, and most code computation in Algorithm~\ref{alg:build}, and its cost consists of the linear time of rebuilding the layouts and the encoding of the vectors changed by maintenance operations. Graph indexes lose search quality unless the connections around deleted nodes are repaired~\cite{r93}, whereas for our index, postponing merges only increases cost.

A merge may also change the rotation (lines 9–10). If the same rotation continues to be used, by Theorem 6 (b) the failure of a query may persist after a merge. By Theorem 6 (c), if a new rotation is drawn at each merge on a predetermined schedule and all live vectors are re-encoded, failures in different merge intervals become mutually independent. Since the failure probability at each time is at most $\delta$ either way, this choice matters for applications in which the same query is repeated over time, such as continual copy checks, to keep failures from persisting. The price is the cost of re-encoding all live vectors at each merge (Section~\ref{sec:expdynamic}). Since the new rotation is drawn according to a predetermined schedule and random sequence, it does not depend on the update contents or search results, and since the choice of maintenance operations does not read the rotation, this choice does not violate C5. The seed of this random sequence must be kept secret from whoever issues queries and updates; anyone who knows it can compute the current and future rotations and choose queries or updates against them, which is the adaptive setting below.

\begin{algorithm}[t]
\caption{Updates and Search of the Dynamic Index}\label{alg:update}
\KwIn{main segment (the index of Algorithm~\ref{alg:build}), delta segment, deletion bitmap, rotation $P$}
\Op{Insert($\boldsymbol{o}_r$)}{
let $\boldsymbol{c}$ be the current centroid nearest to $\boldsymbol{o}_r$, and assign $\boldsymbol{o}_r$ to its partition\;
put the result of \textit{Encode}($\boldsymbol{o}_r$, $\boldsymbol{c}$, $P$) into that partition of the delta segment\;
\textit{Layout}(that partition of the delta segment)\;
}
\Op{Delete($\boldsymbol{o}_r$)}{
mark $\boldsymbol{o}_r$ in the deletion bitmap\;
}
\Op{Merge(if the number of vectors in the delta segment plus the number of deletion marks in the main segment is at least the number of live vectors in the main segment $\times$ the merge threshold)}{
the maintenance policy applies to each partition whichever of recentering, split, dissolution, and doing nothing has the largest gain in Eq.~\eqref{eq:maint}\;
if a new rotation is used, $P \leftarrow$ an orthogonal matrix drawn from a predetermined random sequence\;
\lFor{each vector $\boldsymbol{o}_r$ whose assigned centroid changed (all live vectors if a new rotation is used)}{\textit{Encode}($\boldsymbol{o}_r$, the current centroid $\boldsymbol{c}$ of $\boldsymbol{o}_r$, $P$)}
remove vectors marked as deleted and empty partitions, and fold the vectors of the delta segment into the partitions of the main segment\;
\lFor{each partition of the main segment}{\textit{Layout}(partition)}
clear the delta segment and the deletion bitmap\;
}
\Op{Search($\boldsymbol{q}_r$, $k$, $\varepsilon_{0,\ell}$)}{
run Algorithm~\ref{alg:query}, but exclude vectors marked as deleted from the visited vectors and the test targets, and test the intervals of the main segment and the delta segment separately in each unvisited partition\;
}
\end{algorithm}

\begin{figure}[t]
\centering
\includegraphics{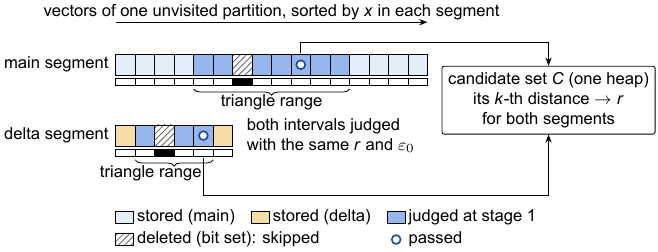}
\refstepcounter{figure}\label{fig:dynsearch}
\par\vspace{3pt}{\small\setstretch{1.15}\IEEEfigcap{Figure \thefigure. Search in the dynamic index (schematic): tests of the main segment and the delta segment in one unvisited partition.}\par}
\end{figure}

Search in the dynamic index (lines 14–15) runs Algorithm~\ref{alg:query} unchanged and differs only in the following points (Fig.~\ref{fig:dynsearch}). The visited vectors in line 2 and the test targets in line 7 of Algorithm~\ref{alg:query} are restricted to live vectors, and lines 6–9 are performed separately on the main segment and the delta segment in each unvisited partition. If the visited partitions contain fewer than $k$ live vectors, more partitions are visited in the order of $\Phi$; these visits are also independent of the rotation and thus within C2. Since the two segments share the set of partitions, routing (line 1), the initial radius (line 3), and the LUTs (line 4) are computed once per query; the intervals of the two segments are tested with the same radius and the same error-bound parameters, and vectors whose exact distances are computed are put into the same candidate set $C$ and shrink the radius together. Accordingly, the implementation of Section~\ref{sec:process} (SIMD tests, the slack factor, multithreading, id deduplication) and the early exact computation, estimate-ordered processing, 2-bit final stage, and global centering of Section~\ref{sec:devices} are all used unchanged, and the deletion bitmap is applied once, with a bitwise operation, to the stage-1 SIMD test results. Since vectors marked as deleted are excluded at stage 1, the vectors chosen by early exact computation and the distances used to shrink the radius are all restricted to live vectors (C7). The parameter selection rule (Section~\ref{sec:devices}) is applied once on the initial index.

Algorithm~\ref{alg:update} satisfies C5–C7. The choice of maintenance operations is made by the policy of Section~\ref{sec:maintenance} only from values computed without using the rotation, so C5 holds. Insertion computes codes with the current centroids and the current rotation (line 3), a merge re-encodes vectors whose assigned centroid changed (line 10), and when a new rotation is used, all live vectors are re-encoded (lines 9–10), so C6 holds. Search excludes vectors marked as deleted from the visited vectors and the test targets (line 15) and shrinks the radius only with live vectors whose exact distances are computed, and is otherwise identical to Algorithm~\ref{alg:query}; hence C7 holds, together with C1, C3, and C4 and the radius condition of Theorem 5 (Section~\ref{sec:process}).

\subsection{Residual-Based Maintenance}\label{sec:maintenance}

Since Theorem 6 holds however the index is maintained, as long as C5 is respected, decisions made to maintain an efficient index do not affect the guarantee and change only search cost. This section addresses the principle and policy for choosing the maintenance operations that reduce search cost at the merges of Section~\ref{sec:dynops}.

As Eq.~\eqref{eq:gap} shows, the residual growth mentioned in Section~\ref{sec:dynguarantee} shrinks the margin and lets more vectors outside the radius pass stage 1, so the timing of maintenance operations can be set by balancing this cost increase against the cost of the operations. At a merge, for each partition $p$, one of the candidate maintenance operations $\alpha$ (recentering, split, dissolution, or doing nothing) is chosen. Let $Q_H$ be the number of queries to be processed until the next merge, $C_{\rm m}(\alpha, p)$ the cost of applying operation $\alpha$ to partition $p$, and $T(q)$ and $T^{\alpha,p}(q)$ the search cost of query $q$ before and after the operation, respectively; then the gain obtained by applying operation $\alpha$ to partition $p$ is \begin{equation}B_p(\alpha) = Q_H\,\mathbb E_q\big[T(q) - T^{\alpha,p}(q)\big] - C_{\rm m}(\alpha, p)\label{eq:maint}\end{equation} where $\mathbb E_q$ is the average over calibration queries and $T(q)$ is the cost predicted by Eq.~\eqref{eq:passprob}. That is, the gain is the total search cost saved until the next merge minus the cost of the operation. At a merge, the gains of the four candidate operations are computed for each partition, the operation $\alpha_p^*$ with the largest gain is applied (line 8 of Algorithm~\ref{alg:update}), and only vectors whose assigned centroid changed are re-encoded (line 10). Doing nothing has no cost and does not change search cost, so its gain is 0; hence, if the gains of the other three operations are all 0 or less, partition $p$ is left as is. Since the inputs of Eq.~\eqref{eq:passprob} are independent of the rotation, $T^{\alpha,p}$ can be predicted before the operation is applied, without reading the rotation.

Computing the gain in Eq.~\eqref{eq:maint} requires predicting the search cost $T^{\alpha,p}$ after the operation before the operation is applied, and maintenance policies are distinguished by this prediction method. We call policy R the policy that regards the number of vectors passing stage 1 in a partition as proportional to a power of the partition's mean residual, with the exponent measured per dataset, and that predicts $T^{\alpha,p}$ by approximating the residual after a split with Zador's law of quantization error~\cite{r165}; we compare it in Section~\ref{sec:expdynamic}. Our default policy (policy K) predicts the cost of unvisited partitions with a light approximation of Eq.~\eqref{eq:passprob}, so its pass-rate prediction has no data-fitted constants. LIRE~\cite{r48}, which splits partitions according to a size threshold and merges small partitions into their neighbors, uses partition size as the maintenance signal. Using the residual (reconstruction error) as the maintenance signal also appears in Ada-IVF~\cite{r49}. Our policy takes as its signal the margin (Eq.~\eqref{eq:gap}), which directly determines cost; because it predicts the pass rate by a formula with no data-fitted constants, it need not be refitted when the data change, and because it separates maintenance from the guarantee, a wrong prediction does not increase the failure probability. It did not, however, search faster than a size-based policy (Section~\ref{sec:expdynamic}).

Policy K predicts $T^{\alpha,p}$ by approximating Eq.~\eqref{eq:passprob} on sample vectors drawn from each partition. As the radius it uses the initial radii of calibration queries (C2); it approximates the distribution of $\kappa$, which is determined only by $D$ and $m_1$ (Section~\ref{sec:costlaw}), by a few representative values and $G_D$ by a normal distribution, and it omits the deterministic slack. This prediction uses neither the rotation, the codes, nor observed pass counts. Candidate operations are limited to partitions changed since the last merge and small partitions, and the cost of visited partitions is computed from partition sizes. For the constants of the maintenance cost $C_{\rm m}(\alpha, p)$ (the time to encode one vector and the time of 2-means), we use values measured once on the initial index with a separate rotation independent of the index, and the 1-bit and 2-bit encoding of vectors whose centroids were changed by maintenance operations is done all at once at the merge.

A single insertion consists of finding the nearest centroid, encoding, and re-laying out the delta interval of that partition, and modifies neither the codes of other vectors nor the connections of the index. A deletion changes one bit. In search, the only additions are scanning one more interval, that of the delta segment, in each unvisited partition and applying the deletion bitmap to the stage-1 test results with a bitwise operation, and the delta segment stays below the merge threshold. A merge skips most of Algorithm~\ref{alg:build} (Section~\ref{sec:dynops}). Moreover, since the guarantee is maintained unchanged as updates continue (Theorem 6), there is no need to retune parameters or re-check recall with calibration queries after updates. An index tuned for average recall sees per-query failures increase as updates continue and must be retuned (Fig.~\ref{fig:motivation}(b)). On the official streaming runbook of Section~\ref{sec:expdynamic}, the time of the last merge was 0.12 times the time to rebuild from scratch, and the search speed of the maintained index was 0.81–1.16 times that of an index rebuilt from scratch at the same point.

\bigskip

\section{Experiments}\label{sec:experiments}

In this section, we answer the following questions. Is the guarantee of our method, PQR-Search, actually kept (Section~\ref{sec:verification})? How much faster is it than existing methods that give the same guarantee (Section~\ref{sec:cost})? What happens to the guarantee and cost under updates (Section~\ref{sec:expdynamic})? Is the pass-rate prediction of the cost law accurate, how much do the centroids and each component contribute to cost, and does the parameter selection rule work on unseen data as well (Section~\ref{sec:ablation})? Finally, we present results by application (copy detection and database operators), scale, and dimension (Section~\ref{sec:scale}).

\subsection{Setup}\label{sec:setup}

All experiments were run alone on a server with an AMD Threadripper PRO 9975WX (32 cores, 64 threads, 128 MB L3) and 256 GB of memory. Table~\ref{tab:datasets} lists the datasets used and the application each represents. Reflecting the applications mentioned in Section~\ref{sec:intro}, we centered the datasets on SSNPP~\cite{r39}, a copy detection benchmark; laion, yandex, and imagenet-align~\cite{r38}, for cross-modal search that finds images from text; and hotpotqa~\cite{r38}, which finds documents from real user questions. Cross-modal queries and real questions differ in distribution from the data vectors~\cite{r38}, so they are real cases in which an index tuned for average recall meets a distribution gap between queries and data. In addition, for comparison with prior work, we also used SIFT and GIST~\cite{r19}, the image retrieval dataset YFCC~\cite{r40}, and the text datasets arxiv, agnews, and msmarco~\cite{r38}, which use held-out documents as queries. imagenet-align was never used in designing or calibrating the parameter selection rule and served only to verify generalization; LAION1M was used for the comparison under the same conditions as the CTR paper~\cite{r11}, and msturing-10M, a web search embedding dataset, for the official NeurIPS'23 streaming runbook. We also added the widely used benchmarks DEEP1M~\cite{r39} and GloVe~\cite{r180}, and for the verification at scale we used DEEP-100M, MSTuring-100M, and SIFT100M, which take the first 100M vectors each of DEEP, MSTuring, and BIGANN (SIFT) of the Big-ANN'21 benchmark~\cite{r39}. All ground-truth distances were computed in fp64. For SIFT, arxiv, and msmarco, which have published ground truth, we recomputed fp64 distances for the ids of the published ground truth and re-sorted them (confirming that the $k$-th distance matches that of fp64 exhaustive search); for the others we used fp64 exhaustive search. For GIST and agnews, the published ground truth had queries that missed a closer point at nearly the same distance at the boundary (6 and 2 queries, respectively, at $k = 100$), so we used exhaustive search.

\Needspace{16\baselineskip}
{\centering \textbf{Table \refstepcounter{table}\label{tab:datasets}\thetable{}. Datasets and applications. $R_{\mathrm{vis}}$ is the query median ($k = 10$, Appendix~\ref{app:relradius}) of the relative radius, that is, the initial radius (C2) divided by the residual scale $\sqrt{x\Phi}$; for msturing-10M it is measured on an index built from all 10M base vectors.}\par}\nopagebreak

{\small\setstretch{1.1}
\begin{xltabular}{\linewidth}{YYYYY}
\toprule
\textbf{Dataset} & \textbf{Size $\times$ dimension} & \textbf{Queries} & \textbf{$R_{\mathrm{vis}}$} & \textbf{Application} \\
\midrule
\endfirsthead
\multicolumn{5}{l}{\emph{Table~\thetable\ (continued)}} \\
\toprule
\textbf{Dataset} & \textbf{Size $\times$ dimension} & \textbf{Queries} & \textbf{$R_{\mathrm{vis}}$} & \textbf{Application} \\
\midrule
\endhead
SSNPP-1M, SSNPP-10M~\cite{r39} & 1M, 10M $\times$ 256 (uint8) & public queries, range ground truth & 1.24, 1.22 & image copy detection \\
laion~\cite{r38} & 1M $\times$ 512 & CLIP~\cite{r170} text $\rightarrow$ image & 1.50 & cross-modal image retrieval \\
yandex~\cite{r38} & 1M $\times$ 200 & text $\rightarrow$ image & 1.16 & cross-modal image retrieval \\
imagenet-align~\cite{r38} & 1.28M $\times$ 640 & caption $\rightarrow$ image & 1.54 & cross-modal retrieval (unseen data, rule verification) \\
hotpotqa~\cite{r38} & 5.23M $\times$ 640 & real user questions & 1.19 & question-answering document retrieval \\
msmarco~\cite{r38} & 8.84M $\times$ 1024 & held-out passages & 0.94 & passage retrieval, 10M scale \\
YFCC-1M, YFCC-10M~\cite{r40} & 1M, 10M $\times$ 192 (uint8) & public queries & 0.84, 0.81 & image retrieval, scale \\
arxiv~\cite{r38} & 1.34M $\times$ 768 & held-out documents & 0.91 & arXiv paper abstract retrieval (nomic-embed-text~\cite{r171} embeddings) \\
agnews~\cite{r38} & 0.77M $\times$ 1024 & held-out documents & 0.94 & news article retrieval (mxbai-embed-large~\cite{r172} embeddings) \\
SIFT1M, GIST1M~\cite{r19} & 1M $\times$ 128 (uint8 values), 960 & public queries & 0.81, 0.97 & classic benchmarks \\
DEEP1M~\cite{r39}, GloVe~\cite{r180} & 1M $\times$ 96, 1.18M $\times$ 100 & public queries & 0.79, 1.02 & image descriptors, word embeddings \\
DEEP-100M, MSTuring-100M, SIFT100M~\cite{r39} & 100M $\times$ 96, 100, 128 (uint8) & public queries & 0.69, 0.89, 0.73 & image descriptors, web search, and a classic benchmark at 100M scale \\
LAION1M~\cite{r11} & 1,000,448 $\times$ 512 & 4,000 data vectors & 0.88 & same conditions as the CTR paper \\
msturing-10M~\cite{r40} & 10M insertions, 5.4M deletions $\times$ 100 & official queries & 0.97 & web search, streaming runbook \\
\bottomrule
\end{xltabular}}

The settings of our method are the same on all datasets. A query visits the two partitions with the smallest $\Phi$ ($n_{\rm probe} = 2$); the candidates for the stage-1 code length $m_1$ are 128 bits, 256 bits, and $D$ bits, and the default is 256 bits if the stage-1 pass rate predicted by the formula of Section~\ref{sec:devices} exceeds 5\%, and 128 bits otherwise. For example, the $m_1$ candidates for laion ($D = 512$) are 128, 256, and 512 bits. One partition is created per 1,000 vectors, with a size cap of 1.1 times the mean and the 8 nearest centroids as assignment candidates. At 100M scale, k-means training is reduced to 64 samples per centroid and 10 iterations, the 8 nearest centroids of each vector are found with HNSW~\cite{r55} built over the centroids, and FAISS IndexIVFFlat also uses the same centroids and assignment. The merge threshold of the dynamic index is 10\% of the number of live vectors in the main segment; when choosing maintenance operations, the number of queries until the next merge, $Q_H$, is assumed to be 10 queries per updated vector, and policy K uses 64 calibration queries and 32 samples per partition. These values are unrelated to the guarantee and change only cost (the remaining values are in Appendix~\ref{app:negative}).

Baselines with the same kind of guarantee are FAISS~\cite{r23} IndexFlatL2, Tribase~\cite{r15}, and PDX-BOND~\cite{r194} for exact search and CTR~\cite{r11} and ConANN~\cite{r1} for distribution-level guarantees, and we compare with them only at the same guarantee level. IndexFlatL2 was used with its default settings (pip wheel faiss-cpu 1.15.1), and a batch of 1,000 queries is processed through the matrix-multiplication path of the OpenBLAS bundled with the wheel. In addition, FAISS IndexIVFFlat serves as a reference for measuring the cost of the guarantee. The number of partitions is max(1000, n/1000), and $n_{\rm probe}$ was set in two ways. One is the minimum value that answers all test queries exactly (for the expected-recall guarantee, the minimum for which every test query has recall at least $\tau$), hereafter called the \textit{oracle}. The oracle is chosen by looking at the test ground truth and thus cannot be obtained in actual operation; it represents the post-hoc optimal cost for IVF to get only the test queries right without a per-query guarantee. Using such a post-hoc optimal configuration as a reference rather than as a competitor is also done in related work~\cite{r13,r50}. The other is a value set with separate calibration queries, which corresponds to actual operation but has no per-query guarantee. The latency of processing a single query with 32 threads was also measured with settings provided by FAISS. IndexFlatL2 automatically uses a path that splits the data across threads when the number of queries is smaller than the number of threads; bound by memory bandwidth, it was 2.0–2.5 times faster with 32 threads than with 1 thread. IndexIVFFlat by default parallelizes over queries and processes a single query on one thread, so we used the setting that parallelizes over the partitions to be scanned (parallel\_mode = 1). At 100M scale, FAISS 1.15.1 IndexIVFFlat returned wrong results with its default setting at 32 threads, so batch was measured with the setting that parallelizes over queries only when there are multiple queries (parallel\_mode = 3). For CTR~\cite{r11}, ConANN~\cite{r1}, and Tribase~\cite{r15}, we used the official code (ConANN built with AVX-512, as is our method), and the numbers of calibration queries and tuning queries of ConANN followed the official settings (5,000 and 1,000). PDX-BOND~\cite{r194}, the exact search of the PDX vertical layout, prunes a vector once its partial distance reaches the current $k$-th distance. It has been removed from the current official PDX repository, so we built the last official version that contains it (Python package pdxearch 0.3, commit 4a2e65e) with the official build procedure (Clang 18, -O3 -march=native) and used its official Python API, which processes one query at a time on a single thread. ANN methods without guarantees are not baselines with the same guarantee; they serve to expose per-query failures and the resulting wrong operator answers (Sections~\ref{sec:motivation}, \ref{sec:expdynamic}, and \ref{sec:scale}) and, timed in the same run, to price the guarantee (end of Section~\ref{sec:cost}). For static indexes, we used the standard FAISS~\cite{r23} implementations of three representative families: partition-based IVF-Flat, graph-based HNSW32 ($M = 32$, efConstruction 200), and quantization-based IVF-RaBitQ (including a stage that re-ranks with exact distances). For the update experiments, we used only implementations that support insertion and deletion through a standard API. Since FAISS HNSW and IVF-RaBitQ with the re-ranking stage do not support deletion, for updates involving deletions we used hnswlib~\cite{r55}, which implements the same algorithm, instead of HNSW, excluded IVF-RaBitQ, and added FreshDiskANN~\cite{r93}, a graph index designed for updates. We also ran Quake~\cite{r50}, which is designed to meet a recall target under updates, with its public implementation. For Quake, we gave a recall target of 0.99 to its feature that determines the number of partitions to scan per query (APS) and invoked its maintenance at each step. However, with the default value (10\%) of the fraction of partitions that APS initially takes as candidates, the average recall on out-of-distribution queries stayed at 0.92, so we set this fraction by the same rule as for the other methods (the smallest value for which the average recall of the calibration queries is at least 0.99), which gave 50\%.

Since average recall is dominated by easy queries~\cite{r47}, we do not use it as the main metric. Moreover, since the probability of the guarantee is over the rotation of the index, not over the query distribution, measuring the fraction of queries that meet the target in a single run~\cite{r1,r11,r13} cannot verify a per-query guarantee. Therefore, to verify the guarantee, we use new rotations and test queries not used in the design, and report the failure fraction measured for each query over several independent rotations and the number of queries whose Clopper–Pearson lower bound~\cite{r167} on the per-query failure rate exceeds $\delta$ (significant violations). Hereafter this procedure is called an \textit{audit}. Whether a result is exact was judged strictly by whether the $j$-th of the sorted result distances is at most the true $j$-th distance (distances re-measured in fp64, with a relative tolerance of $10^{-9}$ on squared distances). For contrast with methods without guarantees, we use the fraction of queries that fall short of the target and the recall of the worst query. The fraction of queries that fall short of the target is 1 minus Robustness-$\delta$@K~\cite{r79}, which measures robustness as the fraction of queries whose recall is at least a threshold, and that threshold corresponds to our $\tau$ (the $\delta$ in this name is a recall threshold, different from our failure probability $\delta$).

For timing we used 1,000 distinct test queries (300 of them only for the 1-thread latency at 10M scale and above and on hotpotqa), because a batch that repeats the same queries shares partition visits and inflates throughput. Batch throughput is the median over 3 rounds of the median of measurements taken, after one warm-up run, at least 5 times and until 1 s accumulates, and the order of the compared methods was changed in each round. The batch time of our method includes the time to sort the queries in the order of their nearest centroids. Measured in the original unsorted order, the geometric mean of the main results was 26.7$\times$, almost the same as the 26.4$\times$ that includes the sorting time. Latency is the median time of processing the queries one at a time. Since FAISS's per-query time is nearly constant, its latency was measured with 50 (1-thread) and 100 (32-thread) queries sampled evenly from the 1,000, and with 10 and 20 at 5M scale and above. At 10M scale and on hotpotqa, the IVF times were measured in only 1 round. Of the three metrics of Section~\ref{sec:devices}, the two that use multiple threads are measured with 32 threads, and the three are hereafter called 1-thread, 32-thread, and batch, respectively. Most related work reports only 1-thread latency~\cite{r11,r62,r64}, but in interactive applications it is natural to share cores to reduce the response time of a single query, so we also measured 32-thread latency. The thread count for batch was set to 32 after measuring the throughput of our method and FAISS at 24–64 threads, as the count closest to peak throughput for both methods, and the FAISS batch was also measured in the same run with the same thread count (the selection procedure is in Appendix~\ref{app:negative}). The Python binding of FAISS installs an interrupt callback at import, which makes the HNSW batch search process queries in chunks of about $10^8/(\text{levels} \cdot d \cdot \text{efSearch})$, so at high dimension and large efSearch only some of the threads work. We therefore re-timed the HNSW32 batches of Tables~\ref{tab:price}, \ref{tab:abs}, and \ref{tab:tail} in a separate run with the callback disabled, as in the C++ library; this raised their throughput by up to 4.9$\times$ (msmarco, efSearch 4,096) and left the returned results bit-identical. The callback did not affect IndexIVFFlat, IVF-RaBitQ, or single-query latency.

\subsection{Verification of the Guarantees}\label{sec:verification}

Table~\ref{tab:verification} summarizes the results of verifying the guarantees (in tables and figures, the exact-search guarantee is abbreviated as T1 and the expected-recall guarantee with $\tau = 0.9$ as E). Audits were performed by running the rule-selected configuration used for measurement on the processing path of that metric (batch, 32-thread, 1-thread). The dynamic index was audited with all techniques of Section~\ref{sec:devices} enabled and timed with the configuration chosen by the enabling rule. The two configurations are paths of Theorem 5 that use the same error-bound parameters, so their guarantees are the same. The failure rates of the exact-search guarantee ($\delta = 0.01, 0.1$) were at most $\delta$ in all settings, and on static indexes the setting with the highest failure rate was at 16\% of $\delta$. Over the 168 settings in which the $(\tau, \delta)$-recall guarantee was verified on static indexes (120 from ten 1M datasets and 48 from the 10M-scale datasets and hotpotqa; 504 processing paths, among which 198 distinct configurations were chosen by the rule; 10 or 4 rotations $\times$ 1,000 queries), the total number of failures was 45 of 4,176,000, at most 0.5\% of $\delta$ per configuration. For the expected-recall guarantee, not a single query had a rotation-averaged recall below $\tau$. To check Corollary 1, we also tested the per-query rotation-averaged recall in the same runs of the $(\tau, \delta)$-recall guarantee mode. In all of the 168 static settings and the 36 LAION1M settings, no query had an average below $\tau$, and the per-query average was at least 0.025 above $\tau$ even in the lowest case (at least 0.06 on 1M data). On LAION1M, under the same conditions as the CTR paper, we tested up to significant violations with 30 rotations and 4,000 queries (0 violations). In the settings with 100M scale, with $\delta = 0.001$, and with varied $n_{\rm probe}$ and partition size, the failure rates were also all at most $\delta$. For the dynamic index as well, the failure rate was at most $\delta$ at every search step and every time point (for the exact-search guarantee, the maximum at a single time point was 7 of 3,000, 23\% of $\delta$).

\Needspace{16\baselineskip}
{\centering \textbf{Table \refstepcounter{table}\label{tab:verification}\thetable{}. Verification of the guarantees.}\par}\nopagebreak

{\small\setstretch{1.1}
\begin{xltabular}{\linewidth}{YYY}
\toprule
\textbf{Target} & \textbf{Scale} & \textbf{Result} \\
\midrule
\endfirsthead
\multicolumn{3}{l}{\emph{Table~\thetable\ (continued)}} \\
\toprule
\textbf{Target} & \textbf{Scale} & \textbf{Result} \\
\midrule
\endhead
Static $(\tau, \delta)$, T1, E (ten 1M datasets, SSNPP-10M, hotpotqa, two 10M datasets) & $(\tau, \delta)$ 168 settings, 504 paths & $(\tau, \delta)$ failures 45/4,176,000, T1 at most 16\% of $\delta$, E 0 queries short, 0 significant violations (1M) \\
LAION1M $(\tau, \delta)$ & 36 settings (108 paths), 30 rotations $\times$ 4,000 queries (excluding tied queries) & failures 20/12,846,600, 0 significant violations \\
Expected-recall guarantee in the $(\tau, \delta)$-recall guarantee mode (Corollary 1) & 168 static settings, 36 LAION1M settings & 0 queries with per-query average recall < $\tau$ (0 significantly short) \\
Ectr (same loss as CTR, Section~\ref{sec:cost}) & 27 settings, 20 rotations & 0 queries significantly exceeding the target (point-estimate excess 1/27,000) \\
imagenet-align (unseen data) & 48 settings, 3 rotations & $(\tau, \delta)$ 1/108,000, T1 1/18,000 \\
10M scale, $k = 1$ & 2 datasets $\times$ 3 paths, 4 rotations & T1 0/24,000 \\
100M scale, 3 datasets & T1, E $\times$ 2 values of $k$ $\times$ 3 paths, 10 rotations & T1 failures 23/180,000 (at most 4\% of $\delta$), E 0 queries short, 0 significant violations \\
$\delta = 0.001$ (T1 on eight 1M datasets) & 16 settings $\times$ 3 paths, 10 rotations & failures 4/480,000 (at most 10\% of $\delta$ per path) \\
$n_{\rm probe}$, partition size (T1 on four 1M datasets) & 48 settings $\times$ 3 paths, 10 rotations & failures 342/1,440,000 (at most 11\% of $\delta$ per path), 0 significant violations \\
SSNPP range queries & 50,000 trials (1,850 on queries with copies) & 0 missed queries, 0 returns outside the radius \\
Dynamic: 11 update workloads (Section~\ref{sec:expdynamic}) & 11 workloads $\times$ 6 time points $\times$ 2 values of $k$ $\times$ 3,000 (1–2 configurations, 180 time point–configuration pairs) & T1 failures 219/540,000 (at most 7/3,000 per time point), $(\tau, \delta)$ 1/270,000 \\
Dynamic: official streaming runbook & 33 steps $\times$ 3,000 & T1 failures 30/99,000 and 40/99,000 (2 configurations, at most 5/3,000 per step), $(\tau, \delta)$ 0/99,000 \\
Dynamic: arxiv streaming & 160 steps $\times$ 3,000 & T1 failures 148/480,000 (at most 4/3,000 per step), $(\tau, \delta)$ 0/480,000 \\
Dynamic: 100M scale (DEEP-100M, 10M vectors replaced) & 11 time points $\times$ 1,000 & T1 failures 0/11,000 \\
Dynamic: $\delta$ = 0.1, 0.5 & 11 workloads and 2 streaming scenarios & at most $\delta$ at all time points (at most 1/6.0 of $\delta$) \\
Dynamic: expected-recall guarantee (Corollary 1 and the expected-recall guarantee mode) & 11 workloads (468,000 trials), 2 streaming scenarios (1,356,000 trials) & 0 queries with average recall < $\tau$ (($\tau$, $\delta$) mode), 0 significantly short (expected-recall mode) \\
\bottomrule
\end{xltabular}}

The failure rates observed in the verification were at most $\delta$ in all settings, and the gap from $\delta$ varied by dataset. On static indexes, the settings closest to $\delta$ were on SSNPP and laion, where the adaptive radius sits close to the true $k$-th distance (at most 16\% of $\delta$ for the exact-search guarantee). When the radius is large, a true neighbor clears the pruning threshold by a wide margin and survives even if the error satisfies $\zeta^{(\ell)} < -\varepsilon_{0,\ell}$, but when the radius is close to the true $k$-th distance, true neighbors lie just above the pruning threshold, so such an error leads directly to pruning. Hence the observation that measured failure rates are smaller than $\delta$ stems from radii larger than the true $k$-th distance, for which the bound is not attained, not from a miscalibrated error-bound parameter, and it must not be used as grounds for relaxing the nominal $\delta$. On the other hand, since 10 rotations cannot test an upper bound on the failure rate of each individual query, on 1M data we compared the number of significant violations expected if every query's true failure rate were exactly $\delta$ (256 for the exact-search guarantee, 3,071 for the $(\tau, \delta)$-recall guarantee) with the actual number (0). For the 10M-scale datasets and hotpotqa, measured with 4 rotations, and for $\delta = 0.001$, even a query that failed only once has a lower bound on its failure rate exceeding $\delta$ (0.0127 > 0.01 with 4 rotations, 0.0051 > 0.001 with 10), so instead of this test we report only the failure rates.

Per-query upper bounds were checked separately on laion and SSNPP, where failure rates are closest to $\delta$ (exact-search guarantee, $k \in \{10, 100\}$, the rule's 1-thread path). Auditing every one of the 1,000 test queries with 1,000 fresh rotations, 1,567 of 4,000,000 trials failed, and no query failed more than 5 times (rate 0.005). The one-sided 95\% Clopper–Pearson upper bound was below $\delta$ for 3,998 of the 4,000 query–$k$ pairs; the other two were queries that failed 5 times (upper bound 0.0105). There was no significant violation, where 47.9 per setting would be expected if every query failed with probability exactly $\delta$, and no single rotation made more than 0.4\% of the queries fail. The 100 hard queries selected in advance from the 10 rotations of the earlier audit failed only 1.07–1.40$\times$ as often as all queries, and contained only 0–3 of the 10 queries that failed most often over the 1,000 rotations: a few rotations cannot identify hard queries in advance.

Whether the audit can detect violations was checked with a positive control, a synthetic placement where the bound is attained. Running the placement of Proposition 1 ($D = 64$ with the exact-search guarantee for $k = 1, 10$ and the $(0.9, 0.01)$-recall guarantee for $k = 10$; $D = 256$ with the exact-search guarantee for $k = 100$) through the same kernel and audit code over 100,000 rotations, the failure rate for $\delta = 0.01$ was 0.95–0.97\%, just below the bound. When the error-bound parameters were deliberately reduced to the values for $\delta' = 0.02$ and $0.05$, the failure rates rose to 1.87–1.98\% and 4.69–4.86\%, and the audit clearly exposed the violation of $\delta = 0.01$. Thus the failure rates far below $\delta$ on real data come from radii larger than the true $k$-th distance, not from an insensitive audit.

\begin{figure}[t]
\centering
\includegraphics{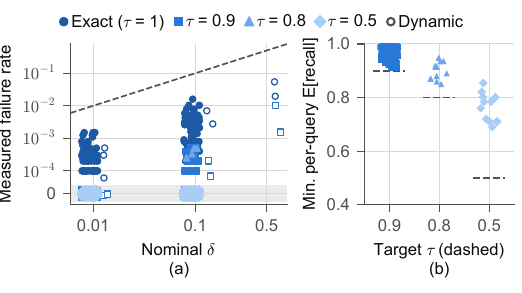}
\refstepcounter{figure}\label{fig:audit}
\par\vspace{3pt}{\small\setstretch{1.15}\IEEEfigcap{Figure \thefigure. Verification of the guarantees. (a) Nominal $\delta$ versus measured failure rate (the 0 band at the bottom marks settings with no failures). (b) Target $\tau$ of the expected-recall guarantee versus the minimum per-query average recall.}\par}
\end{figure}

Fig.~\ref{fig:audit} shows the same audit results against the nominal $\delta$. The failure rates measured in the audits of static indexes (one point per setting and processing path) and in the two runbooks of the dynamic index were all below the diagonal, that is, below the nominal $\delta$, and for $\tau < 1$ most settings had no failures (Fig.~\ref{fig:audit}(a)). For the expected-recall guarantee as well, the minimum per-query average recall (the average over rotations) was at least the target $\tau$ in all settings (Fig.~\ref{fig:audit}(b), settings with 10 or more rotations). Even when the queries were divided by difficulty, failures were not concentrated on hard queries. Difficulty was measured by the relative contrast $\mathrm{RC}_k$~\cite{r118}, the mean distance divided by the distance to the $k$-th neighbor; the smaller it is, the harder the query~\cite{r38,r47}. When the queries of eight 1M datasets were divided into deciles of $\mathrm{RC}_k$, the per-decile failure rates of the exact-search guarantee ($\delta = 0.01$) were all at most 0.088\%, including the hardest decile (Clopper–Pearson upper bound at most 0.17\%), and for the $(0.9, 0.01)$-recall guarantee there were no failures in any decile.

\subsection{Cost under the Same Guarantee}\label{sec:cost}

We first compared our method with FAISS for exact search (T1) and the expected-recall guarantee (E, $\tau = 0.9$). In 52 settings combining ten 1M datasets and three 10M datasets with $k \in \{10, 100\}$, our method was faster than FAISS IndexFlatL2 in every setting. The speedup was 3.3–88.8$\times$ (1-thread), 13.8–303$\times$ (32-thread), and 5.8–73.4$\times$ (batch), with geometric means of 22.9$\times$, 53.3$\times$, and 26.4$\times$, respectively. Among them, the 26 settings of the expected-recall guarantee were 12.7–88.8$\times$ (1-thread), 22.7–303$\times$ (32-thread), and 10.3–73.4$\times$ (batch). Our method was also faster in all 52 settings of the main comparison than IndexIVFFlat with the oracle $n_{\rm probe}$, the post-hoc optimal reference (1.2–64.4$\times$ 1-thread, 2.6–98.1$\times$ 32-thread, 13.5–486$\times$ batch); $k = 1$ and the 100M scale, where this relation partly reverses, are discussed in Section~\ref{sec:scale}. This is because a fixed $n_{\rm probe}$ must serve the hardest test query while the work of our method adapts to each query, and IVF batches scan lists per query without BLAS, so they were even slower than IndexFlatL2 batches. Table~\ref{tab:abs} gives the absolute latency and throughput behind these ratios. The batches of FAISS IndexFlatL2 use the OpenBLAS bundled with its pip wheel; an exhaustive search based on matrix multiplication that we optimized ourselves had 1.5–8.4$\times$ its batch throughput, and our method was 1.03–37$\times$ faster in batch than this baseline in every setting except laion ($k = 100$, T1), where the baseline was 1.2$\times$ faster (Appendix~\ref{app:baselines}). Exhaustive search on a GPU was faster in batch at 1M scale, where the data fit in GPU memory (Appendix~\ref{app:baselines}). Fig.~\ref{fig:speedup} shows the 26 T1 settings (numbers in Appendix Table~\ref{tab:speedup}). Bars show the speedup over FAISS IndexFlatL2 and markers the speedup over IndexIVFFlat with the oracle $n_{\rm probe}$; every bar and marker exceeds 1. The smallest speedup was 3.3$\times$ on SSNPP-1M ($k = 100$, 1-thread), followed by 3.7$\times$ on GloVe ($k = 100$). SSNPP has a high intrinsic dimension, that is, the dimension of the space over which the data actually spread, so the distance gap between true neighbors and other vectors is small and it is hard to filter vectors with the 1-bit test. The median local intrinsic dimension~\cite{r176}, estimated from the distances to the 20 nearest neighbors of a query, was 96 on SSNPP versus 16–52 on the other seven 1M datasets measured the same way, and the fraction of vectors whose exact distances were computed was at most 21\% on SSNPP, at most 7.6\% on GloVe, and at most 1.3\% on the other datasets (100\% for FAISS IndexFlatL2). In our T1 results, 2–3 of 26,000 queries, depending on the processing path, were not exact (1 of the 21,800 queries in 1-thread latency), and FAISS IndexFlatL2 (fp32) also failed to answer 22 queries exactly when ties are judged strictly.

\begin{figure}[t]
\centering
\includegraphics{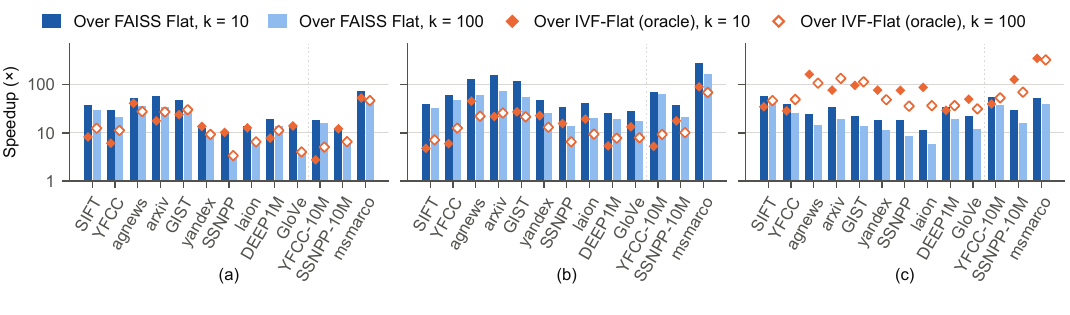}
\refstepcounter{figure}\label{fig:speedup}
\par\vspace{3pt}{\small\setstretch{1.15}\IEEEfigcap{Figure \thefigure. Speedup for exact search ($\delta = 0.01$). (a) 1-thread. (b) 32-thread. (c) Batch.}\par}
\end{figure}

An $n_{\rm probe}$ chosen with calibration queries gives no per-query guarantee. In 29 of the 52 settings (14 T1, 15 E), the calibrated $n_{\rm probe}$ differed from the oracle, and in 9 T1 and 15 E settings among them it answered test queries incorrectly (T1 numbers in Appendix Table~\ref{tab:speedup}; two T1 settings caused by ties in fp32 distances are excluded). For example, on arxiv ($k = 10$), it failed to answer 17 of the 1,000 test queries exactly. Since $n_{\rm probe}$ is fixed, these queries fail every time, and nothing bounds the fraction of queries it fails in another workload. Our method can also fail the same query repeatedly while one transformation is in use, but under the exact-search guarantee (T1), in any workload chosen without looking at $P$ the expected fraction of failing queries is at most $\delta$, and under the expected-recall guarantee (E) every query has expected recall at least $\tau$. Our method uses the same calibration queries only for parameter selection, and its guarantee holds separately for each query. IVF with the calibrated $n_{\rm probe}$ had shorter latency than our method in three settings on YFCC-10M (exact search with $k = 10$, expected-recall guarantee with $k = 10$ and $100$; ours at 0.44–0.83$\times$ at 1 thread, and 0.83$\times$ at 32 threads only for the expected-recall guarantee with $k = 10$), but in all three the calibrated $n_{\rm probe}$ failed test queries (12, 7, and 3). Since the oracle may be unfavorable to IVF, we also measured the smallest $n_{\rm probe}$ for which at most $\delta$ (10) of the 1,000 test queries miss the target (the $\delta$-oracle; for the expected-recall settings, a post-hoc ($\tau = 0.9$, $\delta = 0.01$) criterion). The $\delta$-oracle scanned up to 4$\times$ (T1) and 12$\times$ (E) fewer partitions than the oracle; our method was faster in batch in all 52 settings (5.7–248$\times$, geometric means 43.6$\times$ for T1 and 27.3$\times$ for E) and in every metric, including 1- and 32-thread latency, in 49 settings. In the remaining three, the YFCC-10M settings above, the $\delta$-oracle IVF had shorter latency (ours 0.44–0.84$\times$).

\begin{figure}[t]
\centering
\includegraphics{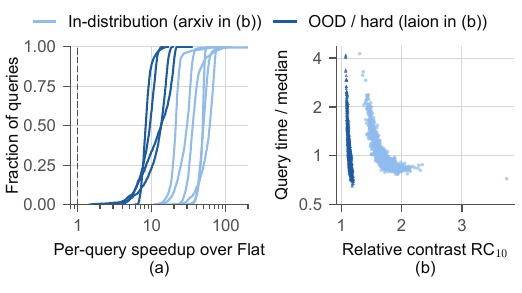}
\refstepcounter{figure}\label{fig:perquery}
\par\vspace{3pt}{\small\setstretch{1.15}\IEEEfigcap{Figure \thefigure. Per-query cost (exact search, $k = 10$, 1-thread). (a) Cumulative distribution of the per-query speedup over FAISS IndexFlatL2. (b) Relative contrast $\mathrm{RC}_{10}$ and query time.}\par}
\end{figure}

\begin{figure}[t]
\centering
\includegraphics{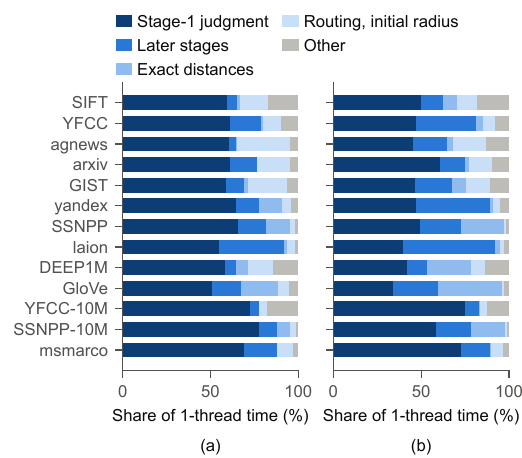}
\refstepcounter{figure}\label{fig:breakdown}
\par\vspace{3pt}{\small\setstretch{1.15}\IEEEfigcap{Figure \thefigure. 1-thread time breakdown for exact search ($\delta = 0.01$). (a) $k = 10$. (b) $k = 100$.}\par}
\end{figure}

Because our method scans a different amount for each query depending on the test outcomes, its time varies across queries, unlike FAISS IndexFlatL2, whose query cost is nearly constant (coefficient of variation of per-query time at most 1.9\%) (Fig.~\ref{fig:perquery}). Over the 52 settings, the p99/p50 of 1-thread latency was 1.09–4.86 for the exact-search guarantee and 1.04–3.73 for the expected-recall guarantee, with long tails for out-of-distribution queries (yandex, laion) and msmarco. Still, all 43,600 queries were faster than with FAISS IndexFlatL2, and the minimum per-query speedup was 1.06$\times$ (GloVe, $k = 100$). The ranking of query times nearly matched the ranking of the number of vectors that passed the first stage (median Spearman rank correlation 0.97), and among difficulty measures, query time correlated most strongly with $\mathrm{RC}_k$ (median $-0.85$). Among queries within the same dataset, the correlation with the local intrinsic dimension was weak (median 0.43), consistent with the observation of~\cite{r47} that the local intrinsic dimension correlates only weakly with empirical difficulty because of estimation noise. Breaking down the 1-thread query time by stage, the first-stage test took a median of 61\% (34–88\%), the rotation of the query a median of 0.5\% (up to 7\% on agnews with 1,024 dimensions), and exact distance computation a median of 2.6\%. Fig.~\ref{fig:breakdown} shows the 26 exact-search settings. Since the first-stage test filters out most unvisited vectors, exact distance computation takes only a few percent of the time, except in settings where the test is hard, such as SSNPP and GloVe ($k = 100$), and exact search on DEEP1M ($k = 100$) (25\%).

\Needspace{14\baselineskip}
{\centering \textbf{Table \refstepcounter{table}\label{tab:abs}\thetable{}. Absolute cost of exact search (T1, $\delta = 0.01$; ours / baseline, ranges over datasets and $k \in \{10, 100\}$). Latency is that of a single query in ms; batch throughput is for 1,000 queries with 32 threads in queries/s. FAISS IndexFlatL2 is from the runs of Fig.~\ref{fig:speedup} and Table~\ref{tab:hundredm}; HNSW32 is tuned to an average recall of 0.99 on calibration queries in the runs of Table~\ref{tab:price} (eight 1M datasets and msmarco, the 15 settings where it reached the target), with our times from the same runs and its batch re-timed as in Section~\ref{sec:setup}.}\par}\nopagebreak

{\small\setstretch{1.1}
\begin{xltabular}{\linewidth}{YlllY}
\toprule
\textbf{Data (settings)} & \textbf{Baseline} & \textbf{1-thread ms} & \textbf{32-thread ms} & \textbf{Batch queries/s} \\
\midrule
\endfirsthead
\multicolumn{5}{l}{\emph{Table~\thetable\ (continued)}} \\
\toprule
\textbf{Data (settings)} & \textbf{Baseline} & \textbf{1-thread ms} & \textbf{32-thread ms} & \textbf{Batch queries/s} \\
\midrule
\endhead
1M, 10 datasets (20) & IndexFlatL2 & 0.28–6.7 / 8.6–88 & 0.11–1.0 / 3.1–41 & 4.0K–78K / 409–1.6K \\
10M, 3 datasets (6) & IndexFlatL2 & 9.1–36 / 165–771 & 1.1–4.6 / 76–368 & 1.5K–7.6K / 51–138 \\
100M, 3 datasets (6) & IndexFlatL2 & 57–77 / 873–1,177 & 7.2–12.5 / 397–521 & 583–1,272 / 8.3–11.7 \\
1M, msmarco (15) & HNSW32 & 0.29–13.7 / 0.11–9.6 & 0.11–2.2 / 0.12–9.6 & 2.0K–78K / 1.8K–110K \\
\bottomrule
\end{xltabular}}

\begin{figure}[t]
\centering
\includegraphics{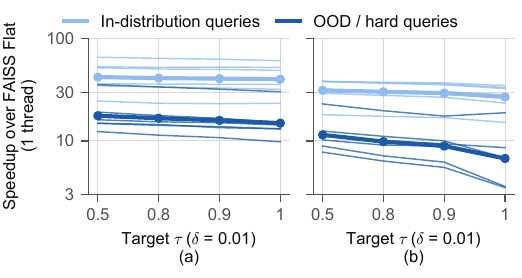}
\refstepcounter{figure}\label{fig:tauspeed}
\par\vspace{3pt}{\small\setstretch{1.15}\IEEEfigcap{Figure \thefigure. 1-thread speedup over FAISS IndexFlatL2 versus the target $\tau$ ($\delta = 0.01$). Thin lines are datasets and thick lines geometric means. (a) $k = 10$. (b) $k = 100$.}\par}
\end{figure}

The results were the same for the $(\tau, \delta)$-recall guarantee with $\tau < 1$. In 144 settings combining $\tau \in \{0.9, 0.8, 0.5\}$, $\delta \in \{0.01, 0.1\}$, and $k \in \{10, 100\}$ on the ten 1M datasets, SSNPP-10M, and hotpotqa, our method was 5.5–72.5$\times$ (1-thread), 18.4–192$\times$ (32-thread), and 6.7–61.7$\times$ (batch) faster than FAISS IndexFlatL2 (medians 24.6$\times$, 46.0$\times$, and 23.4$\times$). The gain over exact search with the same $\delta$ depended on the query distribution. For in-distribution queries, it was small at 1-thread, 0.99–1.08$\times$ for $k = 10$ and 1.05–1.25$\times$ for $k = 100$, but for real out-of-distribution queries and hard data with $k = 100$, it was large: 1.26–1.52$\times$ on laion, 1.58–2.23$\times$ on SSNPP, 1.73–2.49$\times$ on GloVe, and 1.09–1.56$\times$ on yandex. Fig.~\ref{fig:tauspeed} shows this relation as the speedup versus the target $\tau$, which corresponds to the recall–throughput curve of methods without guarantees. Lowering $\tau$ from 1 to 0.5 increased the geometric-mean speedup by 1.05–1.16$\times$ for in-distribution queries and by 1.18–1.70$\times$ for out-of-distribution queries and hard data. The gain is small for in-distribution queries because the error-bound parameter grows only like $\sqrt{\ln(1/\delta_\ell)}$ as the per-stage share $\delta_\ell$ set by $\tau$ and $\delta$ shrinks, and on such data the first-stage pass rate is already low, so most of the time is spent in the first-stage scan, which does not depend on the error-bound parameter (Fig.~\ref{fig:breakdown}). For the same reason, reducing the failure probability to $\delta = 0.001$ increased the cost only slightly. Our method was faster than FAISS IndexFlatL2 in all 16 exact-search settings on eight 1M datasets (the smallest value being 2.4$\times$ at 1-thread on SSNPP, $k = 100$), and its time was 1.08–1.14$\times$ that for $\delta = 0.01$ in geometric mean. Smaller $\delta$, which a pipeline that decides updates from search results may need (Section~\ref{sec:dynguarantee}), costs more on some data. On six 1M datasets (SIFT, GIST, laion, SSNPP, arxiv, and yandex; 12 exact-search settings, with the configuration and techniques reselected by the rule for each $\delta$), $\delta = 10^{-4}$, $10^{-6}$, and $10^{-8}$ took 1.37$\times$, 1.81$\times$, and 2.25$\times$ the 1-thread time of $\delta = 0.01$ in geometric mean (batch 1.32$\times$, 1.66$\times$, and 2.02$\times$). From $\delta = 0.01$ to $10^{-8}$ the error-bound parameter grew only 1.6–1.9$\times$, but the first-stage pass rate grew 2.9–8.3$\times$, so the time grew most where many vectors lie near the pruning threshold: at 1 thread, 1.2–2.3$\times$ on SIFT, GIST, and arxiv and 2.1–5.7$\times$ on laion, SSNPP, and yandex (the largest on SSNPP, $k = 10$). Even at $\delta = 10^{-8}$, every setting was faster than FAISS IndexFlatL2 (at least 1.46$\times$, at 1 thread on SSNPP with $k = 100$, against the IndexFlatL2 times of the main comparison). For the copy-detection range queries on SSNPP-1M and SSNPP-10M (Section~\ref{sec:scale}), the time at $\delta = 10^{-8}$ was within 6.2\% of that at $\delta = 0.01$ for all three configurations, because fewer than 0.07\% of the vectors passed the first stage and the first-stage scan, which does not depend on $\delta$, dominated. Audits cannot check such small $\delta$; the guarantee rests on the computation of the error-bound parameter, whose tail probability matched its target to a relative error of $4 \times 10^{-13}$ in a numerical check.

We compared our method with CTR, another method with guarantees, in two ways. We used SIFT and GIST, which the CTR paper~\cite{r11} used in its evaluation, and arxiv, a text-embedding dataset. First, at the same $\tau$ with our $\delta$ set equal to CTR's $\alpha$, CTR's per-query shortfall fraction exceeded $\alpha$ in 10 of the 27 settings on SIFT, arxiv, and GIST (up to 16.4\% on GIST), whereas the failure rate of our $(\tau, \delta)$-recall guarantee was at most 0.2\% of $\delta$ in every setting (20 rotations $\times$ 1,000 queries, 20 failures in 1,620,000). Next, matching the loss target $\mathbb E[\max(0, \tau - \mathrm{Recall}(q))] \le \alpha(1-\tau)$ that the CTR paper guarantees, we set our method to the Ectr guarantee, that is, the expected-recall guarantee with target $1 - \alpha(1-\tau)$ (since $\max(0, \tau - \mathrm{Recall}(q)) \le 1 - \mathrm{Recall}(q)$, this expected-recall guarantee ensures CTR's loss target for each query). In the 27 settings, no query of ours had an expected per-query shortfall significantly exceeding the target (one of 27,000 queries exceeded it in point estimate, with a shortfall of 0.01, but it was within the target under the one-sided interval over 20 rotations), whereas 0–16.4\% of CTR's queries exceeded the target, and in five settings even the query average exceeded the target (Fig.~\ref{fig:motivation}(c)). The 1-thread time depended on the data. In median latency, CTR, for which a query that passes certification finishes with a single HNSW search, was 2.6–3.2$\times$ faster on arxiv; the two were similar on SIFT (CTR at 0.89–1.31$\times$ ours); and our method was 23.9–86.0$\times$ faster on GIST (CTR was 1.4$\times$ faster in only one setting). In mean latency, which includes the rectification time of queries that fail certification, CTR was at 0.33–4.79$\times$ ours (arxiv), 0.90–20.2$\times$ (SIFT), and 23.5–85.0$\times$ (GIST). On LAION1M from the CTR paper~\cite{r11} (settings with filters were excluded), our failure rate was at most $\delta$ over the entire range of $\tau$ from 0.65 to 0.9 (30 rotations $\times$ 4,000 queries, tied queries excluded, 20 failures in 12,846,600), whereas 4.8–11.4\% of CTR's queries fell below $\tau$ and the worst query's recall was 0. The main cause was queries unreachable in the HNSW graph (about 4\%), and even excluding them, CTR's shortfall was 1.1–8.0\%. CTR's median latency was 0.38–0.44 ms in most settings, shorter than our 1-thread latency (0.64–1.44 ms), but its mean latency (0.38–95.9 ms) was 1.9–48.2$\times$ ours at $\tau$ of 0.8 and 0.85, and 0.29–66.6$\times$ ours over all settings.

For exact search, we also compared our method on eight 1M datasets with Tribase~\cite{r15}, public software that gives a deterministic guarantee. For Tribase, we built an index with $\sqrt n$ partitions that supports all three prunings, searched all partitions, and used the fastest of the seven pruning combinations on the test queries (latency is the median over 50 and 100 queries, as for the FAISS baselines). This condition favors Tribase. With ties judged strictly, Tribase failed to answer 4 of the 16,000 queries processed in batch exactly, all caused by ties in fp32 distances, the same level as FAISS IndexFlatL2. Its speed, however, was only 0.74–1.41$\times$ that of FAISS IndexFlatL2 at 1-thread and 0.11–0.77$\times$ in batch. The 1-thread gain of triangle-inequality pruning by centroid distances was 1.9–2.2$\times$ on SIFT but 0.99–1.09$\times$ on the other datasets, where it rarely took effect, and Tribase processes each query in a single thread and does not group distances into matrix multiplications even in batch. As a result, our method was faster than Tribase in all 16 settings: 3.8–63.6$\times$ (1-thread, geometric mean 22.6$\times$), 34.7–371$\times$ (32-thread), and 35.9–268$\times$ (batch). This result, that distances concentrate in high dimensions so that centroid distances alone can hardly filter vectors, is the same reason this paper adds the 1-bit test to the triangle-inequality bounds of residual norms.

We also compared with PDX-BOND~\cite{r194} on the eight 1M datasets with at least 128 dimensions (its Python API rejects fewer), in the same runs as our method and FAISS IndexFlatL2. We report BOND over an IVF index whose $\lceil\sqrt n\rceil$ buckets are all visited in centroid-distance order, as in the official example, which was its fastest exact configuration in our runs, and BOND without an index. Under the strict tie rule, 19 of the 16,000 queries were not exact, almost all from fp32 rounding at the $k$-th distance as with IndexFlatL2 (22 queries). On our server PDX-BOND was only 1.16$\times$ faster than FAISS IndexFlatL2 at 1 thread in geometric mean (0.84–1.62$\times$), and our method was faster than it in all 16 settings: 18.3$\times$ at 1 thread (geometric mean, 4.3–35.7$\times$; 22.4$\times$ against BOND without an index), with the smallest ratios on SSNPP and laion with $k = 100$. Since PDX processes one query on one thread, we compared our 32-thread latency with its 1-thread latency (ours 98.6$\times$ shorter) and our batch throughput with 32 times its sequential throughput, an ideal scaling that it does not provide (ours 14.1$\times$ higher, 6.1–22.8$\times$). PDX's ADSampling search, which gives no guarantee, was faster than PDX-BOND but still 5.7$\times$ slower than our method at 1 thread, and it answered 14–373 of the 1,000 queries inexactly in every setting.

We also compared with the simplest configuration that gives the same guarantee. By Theorem 5, if RaBitQ's~\cite{r62} IVF searches all partitions and only its error-bound parameter is replaced by the value of the exact-search guarantee, it satisfies the exact-search guarantee even when the initial radius is infinite and the radius is reduced by the exact distances of re-ranked vectors. We built this configuration inside the implementation of our method — no visited partitions (infinite initial radius), no triangle-inequality range, slack factor $f = 1$, a single test stage with the full code, and all optimization techniques off (hereafter the RaBitQ-equivalent configuration) — and measured it on the 26 exact-search settings of the main comparison in the same runs as our method. For batch, the baseline was the faster of scanning all partitions in centroid-distance order per query and the partition-wise processing of our method. The RaBitQ-equivalent configuration also kept the exact-search guarantee in every setting (0 queries not exact) and was faster than FAISS IndexFlatL2 by 12.1$\times$ (1-thread), 20.5$\times$ (32-thread), and 18.3$\times$ (batch) in geometric mean. That is, the per-query guarantee framework alone already turns RaBitQ codes into a guaranteed method much faster than exhaustive search. Our method was again faster than this configuration by 1.64$\times$ (1-thread, 1.01–4.42$\times$), 2.33$\times$ (32-thread, 1.38–5.76$\times$), and 1.24$\times$ (batch, 0.86–2.12$\times$) in geometric mean. On SIFT, SSNPP, and SSNPP-10M, where both configurations use the same single stage with the full code, the 1-thread gain of 1.44–1.72$\times$ is the effect of visited partitions, the triangle-inequality range, the slack factor, and the optimization techniques (including global centering on SSNPP); on high-dimensional embeddings (msmarco 2.86–4.42$\times$, arxiv 1.91–2.91$\times$, agnews, GIST) the multi-stage codes read from the front added a larger gain. In contrast, on the lower-dimensional YFCC, yandex, DEEP1M, and GloVe the 1-thread gain was small (1.01–1.31$\times$), and on YFCC ($k = 100$) batch our method was slower (0.86$\times$). The 32-thread comparison is unfavorable to the RaBitQ-equivalent configuration, because with an infinite initial radius several threads test the first partitions together before the radius shrinks. The official implementation of RaBitQ~\cite{r62} (C++, all partitions searched, with the same partitions and error-bound parameter) was 1.88$\times$ (1.00–3.81$\times$) slower than this configuration at 1 thread in geometric mean (our method was 3.07$\times$, 1.74–5.61$\times$, faster than it), so the comparison with the RaBitQ-equivalent configuration is the one less favorable to our method. The official implementation also had at most $\delta$ queries not exact in each setting (0–4, at most 0.4\%), but its bound does not include the error of its 4-bit query quantization or of its fp32 arithmetic.

To separate what our method adds to the RaBitQ-equivalent configuration, we started from it (L0) and enabled the components one at a time: the initial radius from visited partitions (L1), the triangle-inequality window (L2), the slack factor $f < 1$ (L3), prefix-truncated multi-stage codes (L4), and the remaining techniques chosen by the rule (2-bit last stage, global centering, early exact computation, and estimate-ordered processing; L5). Over all 26 T1 settings, L5 was faster than L0 by a geometric mean of 1.64$\times$ at 1 thread (1.01–4.53$\times$), 2.33$\times$ at 32 threads (1.33–5.88$\times$), and 1.24$\times$ in batch (0.93–2.13$\times$), the same as in the RaBitQ-equivalent comparison above. The largest step depended on the data: on high-dimensional data (agnews, arxiv, GIST, msmarco), multi-stage codes contributed most (1.65–3.19$\times$ at 1 thread); on SIFT, the triangle-inequality window gave 1.20–1.43$\times$ (0.90–1.04$\times$ on the other datasets); and on laion, yandex, SSNPP, and SSNPP-10M, the final group of techniques gave 1.46–3.64$\times$. On laion and yandex, enabling prefix stages without the 2-bit last stage (L4) made search 0.33–0.75$\times$ as fast (on laion the first-stage pass rate jumped from 0.03 to 0.73), and the configuration chosen by the rule (L5) does not use prefix stages on these two datasets. The initial radius gave a geometric mean of 1.39$\times$ (up to 2.37$\times$) for 32-thread latency but was neutral at 1 thread (0.95–1.05$\times$), and its 32-thread share includes the structural handicap of L0 starting from an infinite radius. The slack-factor step was small (0.92–1.21$\times$). L0–L3 had no misses in any setting, and L4 and L5 left at most 2 of 1,000 queries inexact in a few settings, within $\delta$.

We also compared our method on the same three datasets ($k = 10$) with ConANN~\cite{r1}, another method that gives a distribution-level guarantee. ConANN scans IVF partitions one by one, stops at a calibrated threshold, and guarantees that the average FNR (1 $-$ recall) over the query distribution is at most $\alpha$. Requiring this loss for each query yields the expected-recall guarantee with $\tau = 1-\alpha$, so we set our method to this expected-recall guarantee and to the $(1-\alpha, \alpha)$-recall guarantee with $\delta$ equal to $\alpha$ ($\alpha \in \{0.1, 0.05, 0.01\}$). Repeating with the calibration query set redrawn 20 times, ConANN's average FNR was at the target level on SIFT and GIST, where calibration and test queries come from the same distribution (0.101 and 0.097 at $\alpha = 0.1$). Per query, however, 27–31\% ($\alpha = 0.1, 0.05$) and 6–8\% ($\alpha = 0.01$) of the queries had a mean recall over the 20 repetitions significantly below $1-\alpha$, and the same queries fell short almost every time even when the calibration changed. Our method had no such query over 20 independent rotations, and the failure rate of its $(1-\alpha, \alpha)$-recall guarantee was at most 0.11\%. On arxiv, which has no separate query log, we used data vectors as calibration queries, as is common in practice; even a slight difference between the distributions of calibration queries (documents) and test queries (held-out documents) made ConANN's average FNR exceed the target for all three $\alpha$ (0.118, 0.061, and 0.0103, exceeding it in 20, 20, and 17 of the 20 calibrations, respectively). Because a distribution-level guarantee relies on the premise that calibration queries represent test queries, even the query average is not maintained when this premise is shaken. Our guarantee does not use such a premise. At 1-thread, our method was 1.1–8.1$\times$ faster except for two settings on arxiv; on arxiv, the two were similar at $\alpha = 0.05$, and ConANN was 1.7$\times$ faster at $\alpha = 0.1$ (ConANN's average FNR in that setting exceeded the target). Since ConANN processes each query in a single thread, our method was 3.0–46$\times$ faster in 32-thread latency and 2.6–30$\times$ faster in batch. We compared our method in the same way with the public implementation of PUFFINN~\cite{r179}, which bounds each true neighbor's miss probability (per-query expected recall): its default setting fell short of the target even in average recall because of a candidate filter that its proof does not cover, and with the filter disabled it was 27–76$\times$ slower than our method at 1-thread (Appendix~\ref{app:puffinn}).

\begin{figure}[t]
\centering
\includegraphics{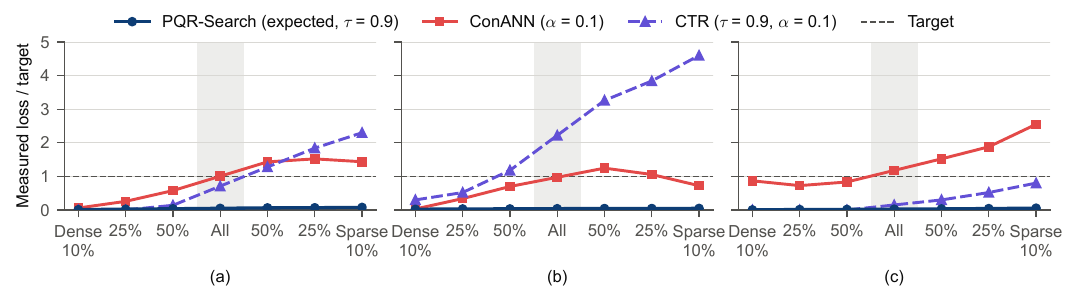}
\refstepcounter{figure}\label{fig:shift}
\par\vspace{3pt}{\small\setstretch{1.15}\IEEEfigcap{Figure \thefigure. Loss $\div$ target when test queries are selected by quantiles of the 10th-neighbor distance (calibration fixed, $k = 10$). (a) SIFT. (b) GIST. (c) arxiv.}\par}
\end{figure}

Methods that rely on calibration can lose their guarantees when queries deviate from the distribution of calibration queries. To examine this in a controlled way, we kept the calibration queries unchanged and evaluated only the test queries whose distance to the 10th neighbor is large (the top 25\% and 10\%) (SIFT, GIST, arxiv, $k = 10$). On these subsets, ConANN's average FNR at $\alpha = 0.1$ was 0.11–0.19 (up to 0.26 for the top 10\%), exceeding the target; the mean shortfall of CTR ($\tau = 0.9$, $\alpha = 0.1$) became 1.8–4.6$\times$ the target on SIFT and GIST (2.2$\times$ on GIST even over all queries); and the fraction of queries that IVF with the calibrated $n_{\rm probe}$ failed to answer exactly on arxiv rose from 1.7\% to 8.0\%. Our failure rate stayed within $\delta = 0.01$, at most 0.04\%, on all subsets. Our method also set its parameter selection rule with the same calibration queries, but since calibration queries affect only the cost, the guarantee is unchanged even when test queries deviate from the distribution of calibration queries. Fig.~\ref{fig:shift} shows the same results as each method's loss divided by its target; the value of our method (expected-recall guarantee at the same level) stayed at most 0.08 regardless of the query subset.

Finally, to measure the price of the guarantee, we compared with indexes without guarantees, tuned so that the average recall over calibration queries reached a target, in the same run (Table~\ref{tab:price}; HNSW32 batch throughput re-timed as described in Section~\ref{sec:setup}). These were FAISS HNSW32 (efSearch), IndexIVFFlat ($n_{\rm probe}$), and IVF-RaBitQ with exact re-ranking ($n_{\rm probe}$ and re-ranking factor), each with the cheapest setting whose calibration average met the target. On arxiv, agnews, and msmarco, which have no query log, data vectors served as calibration queries. With an average-recall target of 0.99 on the eight 1M datasets and msmarco (8.84M vectors), 18 settings, HNSW32 on laion ($k = 100$) and SSNPP (3 settings) and IVF-RaBitQ on laion, yandex ($k = 10$), and SSNPP (5 settings) did not reach a calibration average of 0.99 even at their most expensive settings († in Table~\ref{tab:price}). Over the settings where the baseline reached the target on calibration queries, the 1-thread latency of the exact-search guarantee (T1) was 2.5$\times$ that of HNSW32 (geometric mean, 15 settings; 1.8$\times$ over all 18 settings, since at its most expensive setting HNSW32 was slower where it missed the target): 1.7–2.7$\times$ shorter on GIST and laion ($k = 10$), 1.2–1.9$\times$ longer on SIFT, YFCC, and yandex, 3.0–11.0$\times$ longer on arxiv and agnews, and 15.5–40.4$\times$ longer on msmarco (calibration with data vectors was easier than the test queries there, so the test average recall of HNSW32 was 0.974–0.988, below the target). It was shorter than IndexIVFFlat in all 18 settings and than IVF-RaBitQ in 7 of the 13 settings where it reached the target (all but arxiv, agnews, and msmarco), and HNSW32 had 1.3$\times$ higher batch throughput (geometric mean). In exchange, even where they reached the target, these indexes left 3.3–29.2\% (HNSW32), 0.0–21.7\% (IndexIVFFlat), and 3.4–25.4\% (IVF-RaBitQ) of test queries below recall 0.99, with worst-query recall down to 0 for HNSW32, versus at most 0.1\% for our method; where they did not reach it, 32.6–96.6\% (HNSW32) and 15.4–58.7\% (IVF-RaBitQ). On arxiv and agnews ($k = 10$), calibrated with data vectors, the average recall of HNSW32 over test queries was only 0.97 (tuned afterwards on the test queries, HNSW32 was 5.2–5.8$\times$ faster, with 5.9–6.4\% of queries below 0.99). For a target of 0.9, HNSW32 was 1.57–94.6$\times$ faster at 1 thread than our expected-recall guarantee (E, $\tau = 0.9$), most on msmarco, except on SSNPP ($k = 10$) where it was 8.9$\times$ slower, but even in the 17 settings where it reached the target, 2.9–32.1\% of its queries fell below 0.9 (64.1\% on SSNPP, $k = 100$, where it did not). The speed of indexes without guarantees is bought by meeting only the average target, and an equal average does not tell in advance which queries will fall short.

\Needspace{16\baselineskip}
{\centering \textbf{Table \refstepcounter{table}\label{tab:price}\thetable{}. Price of the guarantee (1-thread latency in ms · batch throughput in thousand QPS · fraction of queries below the target recall). Ours is T1 for target 0.99 and E for 0.9; baselines are tuned so that the average recall over calibration queries meets the target; † marks cells where even the most expensive setting missed the target on calibration queries (values at that setting).}\par}\nopagebreak

{\footnotesize\setstretch{1.1}
\begin{xltabular}{\linewidth}{llllYll}
\toprule
\textbf{Target} & \textbf{Dataset} & \textbf{$k$} & \textbf{Ours} & \textbf{HNSW32} & \textbf{IndexIVFFlat} & \textbf{IVF-RaBitQ} \\
\midrule
\endfirsthead
\multicolumn{7}{l}{\emph{Table~\thetable\ (continued)}} \\
\toprule
\textbf{Target} & \textbf{Dataset} & \textbf{$k$} & \textbf{Ours} & \textbf{HNSW32} & \textbf{IndexIVFFlat} & \textbf{IVF-RaBitQ} \\
\midrule
\endhead
0.99 & SIFT & 10 & 0.29 · 78.3 · 0.0\% & 0.16 · 110.1 · 7.6\% & 0.86 · 6.1 · 4.1\% & 0.58 · 42.3 · 8.2\% \\
0.99 & SIFT & 100 & 0.40 · 50.9 · 0.0\% & 0.31 · 58.5 · 29.2\% & 0.89 · 6.1 · 21.7\% & 1.41 · 14.0 · 25.4\% \\
0.99 & GIST & 10 & 1.75 · 11.4 · 0.0\% & 2.90 · 3.4 · 8.3\% & 16.69 · 0.3 · 4.7\% & 2.40 · 7.0 · 8.4\% \\
0.99 & GIST & 100 & 2.76 · 7.4 · 0.0\% & 5.27 · 2.0 · 19.7\% & 22.99 · 0.2 · 7.2\% & 6.16 · 3.0 · 4.9\% \\
0.99 & YFCC & 10 & 0.53 · 42.3 · 0.0\% & 0.28 · 52.8 · 3.3\% & 1.08 · 4.4 · 3.2\% & 0.64 · 27.8 · 3.4\% \\
0.99 & YFCC & 100 & 0.68 · 29.3 · 0.0\% & 0.54 · 29.5 · 9.8\% & 1.12 · 4.3 · 16.8\% & 0.85 · 23.1 · 17.3\% \\
0.99 & laion & 10 & 3.57 · 10.6 · 0.1\% & 9.60 · 1.8 · 9.8\% & 31.88 · 0.1 · 1.1\% & 7.84 · 2.7 · 58.7\%† \\
0.99 & laion & 100 & 6.79 · 4.0 · 0.0\% & 9.61 · 1.8 · 32.6\%† & 31.84 · 0.1 · 4.4\% & 17.29 · 1.2 · 57.8\%† \\
0.99 & yandex & 10 & 1.36 · 21.3 · 0.0\% & 1.14 · 15.6 · 4.0\% & 4.55 · 1.1 · 5.2\% & 9.42 · 2.5 · 15.4\%† \\
0.99 & yandex & 100 & 2.10 · 13.3 · 0.0\% & 1.53 · 12.2 · 13.2\% & 6.89 · 0.7 · 8.5\% & 7.08 · 2.9 · 15.6\% \\
0.99 & SSNPP & 10 & 2.00 · 16.8 · 0.0\% & 13.31 · 1.1 · 49.9\%† & 22.23 · 0.2 · 0.0\% & 6.51 · 3.5 · 46.6\%† \\
0.99 & SSNPP & 100 & 6.19 · 7.9 · 0.0\% & 13.20 · 1.1 · 96.6\%† & 22.00 · 0.2 · 0.0\% & 13.81 · 1.6 · 43.3\%† \\
0.99 & arxiv & 10 & 1.49 · 13.9 · 0.1\% & 0.15 · 65.0 · 18.2\% & 4.50 · 1.1 · 6.7\% & 0.63 · 26.2 · 6.9\% \\
0.99 & arxiv & 100 & 2.75 · 9.4 · 0.0\% & 0.66 · 15.4 · 20.0\% & 8.91 · 0.5 · 9.5\% & 1.27 · 12.7 · 15.8\% \\
0.99 & agnews & 10 & 1.27 · 14.6 · 0.0\% & 0.11 · 68.6 · 17.7\% & 4.76 · 1.0 · 7.4\% & 0.51 · 29.3 · 8.7\% \\
0.99 & agnews & 100 & 1.92 · 9.7 · 0.0\% & 0.63 · 13.0 · 17.1\% & 9.25 · 0.5 · 18.7\% & 1.14 · 15.1 · 19.5\% \\
0.99 & msmarco & 10 & 8.86 · 2.9 · 0.0\% & 0.22 · 39.4 · 9.9\% & 25.36 · 0.2 · 7.8\% & 3.05 · 4.7 · 8.0\% \\
0.99 & msmarco & 100 & 13.68 · 2.0 · 0.0\% & 0.88 · 10.2 · 19.3\% & 71.52 · 0.1 · 15.6\% & 9.96 · 1.3 · 16.5\% \\
0.9 & SIFT & 10 & 0.28 · 81.5 · 0.4\% & 0.05 · 353.0 · 26.7\% & 0.22 · 25.2 · 19.7\% & 0.21 · 92.5 · 22.3\% \\
0.9 & SIFT & 100 & 0.33 · 61.3 · 0.0\% & 0.21 · 83.7 · 4.1\% & 0.46 · 11.7 · 14.8\% & 0.41 · 59.7 · 37.5\% \\
0.9 & GIST & 10 & 1.49 · 13.0 · 0.1\% & 0.64 · 14.8 · 24.7\% & 4.18 · 1.2 · 26.8\% & 0.76 · 24.1 · 28.3\% \\
0.9 & GIST & 100 & 2.00 · 9.6 · 0.0\% & 1.12 · 7.7 · 32.1\% & 7.87 · 0.6 · 19.4\% & 1.48 · 12.5 · 28.1\% \\
0.9 & YFCC & 10 & 0.47 · 50.5 · 0.0\% & 0.05 · 281.2 · 22.1\% & 0.15 · 34.8 · 25.6\% & 0.11 · 160.9 · 25.9\% \\
0.9 & YFCC & 100 & 0.51 · 39.7 · 0.0\% & 0.21 · 73.1 · 6.4\% & 0.30 · 17.8 · 25.8\% & 0.21 · 117.6 · 31.9\% \\
0.9 & laion & 10 & 2.30 · 16.2 · 0.7\% & 0.44 · 23.7 · 19.9\% & 15.49 · 0.3 · 11.3\% & 7.84 · 2.7 · 40.0\%† \\
0.9 & laion & 100 & 3.11 · 7.1 · 0.2\% & 0.84 · 15.5 · 27.4\% & 14.72 · 0.3 · 24.2\% & 12.41 · 1.7 · 31.6\% \\
0.9 & yandex & 10 & 0.76 · 25.1 · 0.6\% & 0.17 · 93.4 · 14.5\% & 0.56 · 8.8 · 23.9\% & 0.82 · 31.1 · 19.5\% \\
0.9 & yandex & 100 & 1.16 · 16.9 · 0.0\% & 0.32 · 50.8 · 20.5\% & 1.15 · 4.4 · 23.6\% & 1.62 · 14.4 · 26.2\% \\
0.9 & SSNPP & 10 & 1.49 · 21.5 · 0.0\% & 13.31 · 1.1 · 19.2\% & 11.10 · 0.4 · 21.6\% & 6.51 · 3.5 · 14.5\% \\
0.9 & SSNPP & 100 & 1.51 · 18.4 · 0.0\% & 13.20 · 1.1 · 64.1\%† & 16.67 · 0.3 · 0.6\% & 9.92 · 2.4 · 1.2\% \\
0.9 & arxiv & 10 & 1.21 · 16.8 · 0.1\% & 0.07 · 137.2 · 30.1\% & 1.20 · 4.0 · 12.8\% & 0.18 · 103.1 · 20.1\% \\
0.9 & arxiv & 100 & 2.35 · 12.0 · 0.0\% & 0.47 · 21.5 · 4.1\% & 1.19 · 3.9 · 32.2\% & 0.26 · 61.1 · 36.2\% \\
0.9 & agnews & 10 & 1.22 · 15.4 · 0.1\% & 0.07 · 116.9 · 20.6\% & 0.63 · 6.9 · 19.8\% & 0.12 · 127.3 · 22.9\% \\
0.9 & agnews & 100 & 1.54 · 12.7 · 0.0\% & 0.44 · 18.4 · 2.9\% & 1.23 · 3.9 · 28.2\% & 0.25 · 57.4 · 30.0\% \\
0.9 & msmarco & 10 & 8.01 · 3.8 · 0.3\% & 0.08 · 98.6 · 20.5\% & 2.35 · 2.7 · 20.3\% & 0.80 · 24.8 · 23.8\% \\
0.9 & msmarco & 100 & 12.14 · 2.8 · 0.0\% & 0.49 · 18.2 · 8.7\% & 3.71 · 1.4 · 31.9\% & 1.08 · 16.4 · 31.9\% \\
\bottomrule
\end{xltabular}}

An index without guarantees can also be tuned to the tail target itself rather than to an average. We therefore repeated the comparison with each baseline set to the cheapest setting on the same grid whose fraction of the 1,000 calibration queries below recall $\tau$ was at most $\delta$, which is the target our $(\tau, \delta)$-recall guarantee meets for every query, and also report a post-hoc tail oracle that applies the same criterion to the test queries (Table~\ref{tab:tail}; the 18 settings of Table~\ref{tab:price}, $\delta = 0.01$, timed in the same runs as our method except the re-timed HNSW32 batches). For $(0.99, 0.01)$ we used T1 for $k = 10$, where $\tau = 0.99$ allows no miss, and the $(0.99, 0.01)$-recall guarantee for $k = 100$, configured for this comparison by the same parameter selection rule; for $(0.9, 0.01)$ we used the $(0.9, 0.01)$-recall guarantee. Our method stayed within $\delta$ in all 36 pairs of setting and target (at most 0.1\% of test queries below $\tau$). At $\tau = 0.99$, HNSW32 reached the tail target on calibration queries within efSearch 4,096 in only 10 of the 18 settings (4 of the 12 on the six datasets with query logs), and the setting chosen on calibration queries still had more than $\delta$ of its test queries below 0.99 in 5 of these 10 (13 of 18 including the settings it did not reach). Where HNSW32 reached the target, our method was 1.1$\times$ faster at 1 thread and 1.9$\times$ faster in batch in geometric mean, and against the fastest baseline that reached the target in each setting, 2.1$\times$ and 4.9$\times$ faster; IndexIVFFlat reached it everywhere but was 9.0$\times$ (1-thread) and 46$\times$ (batch) slower than our method. The result depended on the calibration queries. On the six datasets with query logs our method was faster than HNSW32 in every setting where HNSW32 reached the target (2.8$\times$ at 1 thread, 3.2$\times$ in batch), whereas on arxiv, agnews, and msmarco, calibrated with data vectors, HNSW32 was 1.7$\times$ faster at 1 thread (1.5$\times$ for the post-hoc oracle) and our method remained 1.4$\times$ faster in batch. At $\tau = 0.9$, HNSW32 was faster than our method: 2.7$\times$ at 1 thread when tuned on calibration queries, whose chosen setting exceeded $\delta$ on test queries in 13 of the 18 settings, and 2.0$\times$ for the post-hoc oracle over the 14 settings it reached. This advantage came from the three datasets calibrated with data vectors (oracle 5.4$\times$ at 1 thread and 2.1$\times$ in batch); on the six datasets with query logs the two were on par at 1 thread (ours 1.03$\times$) and our method was 1.6$\times$ faster in batch. The grid ended at efSearch 4,096 (all lists for IndexIVFFlat and a re-ranking factor of 200 for IVF-RaBitQ), so some settings the baselines did not reach might be reached with larger values (we widen the grid and the baselines below), and the geometric means cover only reached settings, a filter that favors the baselines. With 1,000 queries, $\delta = 0.01$ corresponds to 10 queries, so whether a setting exceeds $\delta$ (for example, at 1.1\% or 1.2\%) is within binomial variation.

\Needspace{16\baselineskip}
{\centering \textbf{Table \refstepcounter{table}\label{tab:tail}\thetable{}. Against baselines tuned to the tail target (18 settings: eight 1M datasets and msmarco, $k \in \{10, 100\}$, 1,000 test queries). Calibrated: the cheapest setting on the grid with at most a $\delta$ fraction of calibration queries below $\tau$; oracle: the same criterion on the test queries (post hoc). Reached: settings where such a setting exists. $>\delta$: settings whose fraction of test queries below $\tau$ exceeds $\delta$, including unreached settings at the grid setting with the fewest misses. Speedup of ours: baseline time $\div$ our time, geometric mean (range) over reached settings; below 1, our method is slower.}\par}\nopagebreak

{\footnotesize\setstretch{1.1}
\begin{xltabular}{\linewidth}{YYllYY}
\toprule
\textbf{$(\tau, \delta)$} & \textbf{Method (tuning)} & \textbf{Reached} & \textbf{$>\delta$} & \textbf{1-thread speedup} & \textbf{Batch speedup} \\
\midrule
\endfirsthead
\multicolumn{6}{l}{\emph{Table~\thetable\ (continued)}} \\
\toprule
\textbf{$(\tau, \delta)$} & \textbf{Method (tuning)} & \textbf{Reached} & \textbf{$>\delta$} & \textbf{1-thread speedup} & \textbf{Batch speedup} \\
\midrule
\endhead
(0.99, 0.01) & Ours & — & 0 & — & — \\
 & HNSW32 (calibrated) & 10/18 & 13 & 1.09 (0.27–6.10) & 1.93 (0.62–5.97) \\
 & IndexIVFFlat (calibrated) & 18/18 & 3 & 9.00 (4.41–29.2) & 46.0 (18.3–175) \\
 & IVF-RaBitQ (calibrated) & 12/18 & 9 & 2.73 (1.43–8.65) & 3.28 (1.47–8.98) \\
 & Fastest reached (calibrated) & 18/18 & — & 2.11 (0.27–10.2) & 4.87 (0.62–72.8) \\
 & HNSW32 (oracle) & 10/18 & 8 & 1.27 (0.37–14.9) & 2.26 (0.90–12.5) \\
(0.9, 0.01) & Ours & — & 0 & — & — \\
 & HNSW32 (calibrated) & 14/18 & 13 & 0.37 (0.03–2.02) & 0.72 (0.10–4.12) \\
 & IndexIVFFlat (calibrated) & 18/18 & 0 & 5.15 (2.23–11.3) & 26.4 (8.93–78.3) \\
 & IVF-RaBitQ (calibrated) & 14/18 & 6 & 1.43 (0.62–4.90) & 1.76 (0.68–7.42) \\
 & Fastest reached (calibrated) & 18/18 & — & 0.69 (0.03–11.3) & 1.57 (0.10–78.3) \\
 & HNSW32 (oracle) & 14/18 & 4 & 0.49 (0.08–2.10) & 1.11 (0.25–10.1) \\
\bottomrule
\end{xltabular}}

We then widened the baselines and the scale with the same tail tuning and timing (Table~\ref{tab:tailx}; $k \in \{10, 100\}$, $\delta = 0.01$). On the four datasets whose queries come from a different distribution than the data (laion, yandex, SSNPP, and imagenet-align), we added HNSW with $M \in \{32, 48, 64\}$ (efConstruction 500 for $M > 32$) and efSearch up to 16,384, DiskANN~\cite{r56} (Vamana, $R = 64$, $\alpha = 1.2$, $L$ up to 16,384), and IVF-RaBitQ with 4- and 8-bit codes and exact re-ranking, and on laion, yandex, and SSNPP also RoarGraph~\cite{r122}, a graph built with training queries for cross-modal search. At 10M scale (YFCC-10M, SSNPP-10M) we compared with HNSW32 and DiskANN, and at 100M scale (DEEP-100M, SIFT100M) with HNSW32, both with efSearch up to 16,384. Our method stayed within $\delta$ in every setting and target (at most 0.1\% of test queries below $\tau$). At $\tau = 0.99$ on the out-of-distribution datasets, larger graphs reached the target in more settings (HNSW64 in 7 of 8), but their settings chosen on calibration queries still left more than $\delta$ of test queries below 0.99 in 5 of 8, and against the fastest baseline that reached the target in each setting, RoarGraph included, our method was 2.9$\times$ faster at 1 thread and 5.3$\times$ faster in batch. RoarGraph did not reach $\tau = 0.99$ on SSNPP even at $L = 16{,}384$ and was 3.2$\times$ (1-thread) and 6.0$\times$ (batch) slower than our method elsewhere. At $\tau = 0.9$ the fastest baseline was 1.1$\times$ faster than our method at 1 thread in geometric mean (up to 2.9$\times$ in single settings; RoarGraph 1.7$\times$) and 2.0$\times$ slower in batch. At 10M scale, neither HNSW32 nor DiskANN reached $\tau = 0.99$ or $\tau = 0.9$ on SSNPP-10M (47–96\% of test queries below $\tau$ at the largest setting), while on YFCC-10M both graphs reached $\tau = 0.9$ and were 4.8–6.0$\times$ faster than our method at 1 thread (DiskANN also up to 2.3$\times$ faster at $\tau = 0.99$). At 100M scale, where queries follow the data distribution, HNSW32 reached every target and was faster: 4.8$\times$ ($\tau = 0.99$) and 19$\times$ ($\tau = 0.9$) at 1 thread and 1.6$\times$ and 7.2$\times$ in batch, because our 1-thread latency grows to 56–71 ms; with 32 threads per query, our latency (7–9 ms) was shorter than that of HNSW32 at $\tau = 0.99$ in three of the four settings. The price of the guarantee is thus largest for in-distribution queries at a low target on large data, and small or negative for out-of-distribution queries and high targets, where tail-tuned indexes either do not reach the target or exceed $\delta$ on test queries. The HNSW batches of these runs were also timed without FAISS's Python interrupt callback.

\Needspace{16\baselineskip}
{\centering \textbf{Table \refstepcounter{table}\label{tab:tailx}\thetable{}. Wider and larger tail-tuned baselines (calibrated tuning, $\delta = 0.01$, $k \in \{10, 100\}$, 1,000 test queries). OOD: laion, yandex, SSNPP, and imagenet-align (RoarGraph: the first three). Reached: settings where the grid (efSearch or $L$ up to 16,384) has a setting with at most a $\delta$ fraction of calibration queries below $\tau$. $>\delta$: settings whose fraction of test queries below $\tau$ exceeds $\delta$, including unreached settings. Speedup of ours: baseline time $\div$ our time, geometric mean over reached settings; below 1, our method is slower.}\par}\nopagebreak

{\footnotesize\setstretch{1.1}
\begin{xltabular}{\linewidth}{llYYYYYY}
\toprule
\textbf{Data} & \textbf{Baseline} & \textbf{Reached (0.99)} & \textbf{$>\delta$ (0.99)} & \textbf{Speedup (0.99): 1-thread / batch} & \textbf{Reached (0.9)} & \textbf{$>\delta$ (0.9)} & \textbf{Speedup (0.9): 1-thread / batch} \\
\midrule
\endfirsthead
\multicolumn{8}{l}{\emph{Table~\thetable\ (continued)}} \\
\toprule
\textbf{Data} & \textbf{Baseline} & \textbf{Reached (0.99)} & \textbf{$>\delta$ (0.99)} & \textbf{Speedup (0.99): 1-thread / batch} & \textbf{Reached (0.9)} & \textbf{$>\delta$ (0.9)} & \textbf{Speedup (0.9): 1-thread / batch} \\
\midrule
\endhead
OOD, 1M & HNSW32 & 2/8 & 6 & 5.36 / 6.18 & 8/8 & 5 & 3.21 / 5.65 \\
 & HNSW48 & 5/8 & 6 & 5.52 / 8.95 & 8/8 & 4 & 1.42 / 3.25 \\
 & HNSW64 & 7/8 & 5 & 6.66 / 11.8 & 8/8 & 3 & 1.34 / 3.15 \\
 & DiskANN & 6/8 & 5 & 2.83 / 4.48 & 8/8 & 3 & 1.79 / 3.07 \\
 & IVF-RaBitQ 4-bit & 8/8 & 0 & 7.22 / 12.8 & 8/8 & 0 & 4.69 / 9.23 \\
 & IVF-RaBitQ 8-bit & 8/8 & 0 & 7.77 / 14.6 & 8/8 & 3 & 4.39 / 8.97 \\
 & RoarGraph & 4/6 & 2 & 3.15 / 6.01 & 4/6 & 4 & 0.58 / 1.23 \\
 & Fastest reached & 8/8 & — & 2.94 / 5.30 & 8/8 & — & 0.89 / 1.98 \\
10M & HNSW32 & 1/4 & 3 & 2.66 / 7.04 & 2/4 & 2 & 0.19 / 0.41 \\
 & DiskANN & 2/4 & 2 & 0.94 / 1.70 & 2/4 & 2 & 0.19 / 0.40 \\
100M & HNSW32 & 4/4 & 1 & 0.21 / 0.62 & 4/4 & 1 & 0.05 / 0.14 \\
\bottomrule
\end{xltabular}}

When isotropic Gaussian noise was added to the test queries (with noise magnitude up to twice the 10th-neighbor distance), our failure rate also stayed within $\delta$, and only the 1-thread latency increased, by 1.5–4.8$\times$. In this case, ConANN and CTR shifted toward searching more for queries whose neighbor distances grew, so their shortfall actually decreased, but CTR's 1-thread median latency increased more than 1,000-fold on SIFT and arxiv (from 0.37 ms to 444 ms on SIFT). Table~\ref{tab:guaranteed} summarizes the per-query shortfall of the compared methods that rely on calibration queries. All of these methods fell short of the target on some queries, and the fraction mostly grew on the subsets that select only queries with distant neighbors.

\Needspace{12\baselineskip}
{\centering \textbf{Table \refstepcounter{table}\label{tab:guaranteed}\thetable{}. Per-query shortfall of compared methods relying on calibration queries (fraction of queries below the target).}\par}\nopagebreak

{\small\setstretch{1.1}
\begin{xltabular}{\linewidth}{YYYYY}
\toprule
\textbf{Method} & \textbf{Guarantee} & \textbf{Measurement} & \textbf{Queries below target} & \textbf{Ours (same settings)} \\
\midrule
\endfirsthead
\multicolumn{5}{l}{\emph{Table~\thetable\ (continued)}} \\
\toprule
\textbf{Method} & \textbf{Guarantee} & \textbf{Measurement} & \textbf{Queries below target} & \textbf{Ours (same settings)} \\
\midrule
\endhead
IndexIVFFlat, calibrated $n_{\rm probe}$ & None (smallest $n_{\rm probe}$ that answers all calibration queries correctly) & 52 exact-search and expected-recall settings, single run & Shortfall in 24 settings (ties excluded), up to 1.7\% (8.0\% when only queries with distant neighbors are selected) & Failure rate at most $\delta$ (T1); 0 queries with average recall short of $\tau$ (E) \\
CTR~\cite{r11} & Risk over the query distribution $\le \alpha$ & 27 settings on SIFT, GIST, and arxiv; 36 settings on LAION1M & Up to 16.4\% (LAION1M 4.8–11.4\%) & Failures 20/1,620,000 and 20/12,846,600 \\
ConANN~\cite{r1} & Average FNR over the query distribution $\le \alpha$ & SIFT, GIST, and arxiv, $\alpha \in \{0.1, 0.05, 0.01\}$, 20 calibrations & Significantly below: 27–31\% ($\alpha = 0.1, 0.05$), 6–8\% ($\alpha = 0.01$); when only queries with distant neighbors are selected, even the average FNR exceeds the target & Significantly below: 0 \\
\bottomrule
\end{xltabular}}

We also measured index build time and size (Table~\ref{tab:build}). Since building is a batch job separate from query processing, all three methods were run with 64 threads; we report the median of three runs for the 1M datasets and a single run for the 10M scale and hotpotqa, and data loading is excluded from the time. Size counts all arrays read during query processing (including the fp64 centroids used for handling floating-point error) and the original IDs. Our build (Algorithm~\ref{alg:build}) took 1.6–2.6$\times$ longer than training and adding for FAISS IndexIVFFlat on all 13 datasets, and 62–96\% of the time went to k-means and size-capped assignment. It was 1.4–1.9$\times$ faster than the build of CTR, the other method with guarantees (HNSW construction and edge cache). Calibrating the parameter selection rule, including generating the candidate configurations, took an additional 3–41 s on the 1M datasets and 54–141 s at the 10M scale. In the default configuration without the 2-bit final stage, the index size for fp32 data was 1.04–1.11$\times$ that of IVF-Flat (up to 1.21$\times$ for configurations that use the 2-bit stage). This is because we keep the original vectors as they are for exact distances and add the first-stage data (23–39 bytes per vector) and the codes and coefficients of the later stages. On SIFT, YFCC, and SSNPP, whose original values are 8-bit integers, we keep the originals in 8 bits, so the size was 0.36–0.43$\times$ that of FAISS, which stores them as float32.

\Needspace{16\baselineskip}
{\centering \textbf{Table \refstepcounter{table}\label{tab:build}\thetable{}. Index build time (64 threads) and index size (bytes per vector, default configuration without the 2-bit stage). Eight of the 13 datasets.}\par}\nopagebreak

{\footnotesize\setstretch{1.1}
\begin{xltabular}{\linewidth}{lYYllYll}
\toprule
\textbf{Dataset} & \textbf{Vectors $\times$ dim.} & \textbf{Build: ours (s)} & \textbf{IVF-Flat (s)} & \textbf{CTR (s)} & \textbf{Size: ours (B)} & \textbf{IVF-Flat (B)} & \textbf{CTR (B)} \\
\midrule
\endfirsthead
\multicolumn{8}{l}{\emph{Table~\thetable\ (continued)}} \\
\toprule
\textbf{Dataset} & \textbf{Vectors $\times$ dim.} & \textbf{Build: ours (s)} & \textbf{IVF-Flat (s)} & \textbf{CTR (s)} & \textbf{Size: ours (B)} & \textbf{IVF-Flat (B)} & \textbf{CTR (B)} \\
\midrule
\endhead
SIFT & 1M $\times$ 128 & 4.9 & 1.9 & 7.3 & 225 & 521 & 793 \\
GIST & 1M $\times$ 960 & 19.1 & 8.1 & 35.9 & 4,030 & 3,852 & 4,121 \\
arxiv & 1.34M $\times$ 768 & 28.5 & 12.1 & 39.8 & 3,242 & 3,083 & 3,353 \\
laion & 1M $\times$ 512 & 11.1 & 4.8 & — & 2,214 & 2,058 & — \\
SSNPP & 1M $\times$ 256 & 6.8 & 3.1 & — & 371 & 1,033 & — \\
YFCC-10M & 10M $\times$ 192 & 372 & 173 & — & 289 & 777 & — \\
SSNPP-10M & 10M $\times$ 256 & 442 & 205 & — & 370 & 1,033 & — \\
msmarco & 8.84M $\times$ 1024 & 1,005 & 510 & — & 4,332 & 4,108 & — \\
\bottomrule
\end{xltabular}}

\subsection{Guarantees and Cost under Updates}\label{sec:expdynamic}

First, to see what happens after updates to parameters tuned at the start, we built 11 workloads that build an index on 50\% of the data and then update it in five steps of 10\% each. There are three update patterns: random insertion, cluster-ordered insertion (distribution shift), and a sliding window that inserts 10\% in cluster order and deletes the oldest 10\%; all three were applied to arxiv, SIFT, and laion, and the sliding window to GIST and yandex. Public dynamic software (IVF-Flat and HNSW of FAISS~\cite{r23}, hnswlib~\cite{r55}, and FreshDiskANN~\cite{r93}; hnswlib instead of FAISS's HNSW for the sliding window, which involves deletions) was tuned at the first step and not retuned afterward, whereas our method needs no tuning to meet recall (calibration queries are used only for the cost choices of Section~\ref{sec:devices}). After tuning for an average recall of 0.99, we measured at each step the fraction of queries with recall below 0.99. As a result, the per-query failures of IVF-Flat increased substantially as insertions and deletions continued. On the SIFT sliding window, this fraction rose from 6.1\% to 14.6\% for $k = 10$ and from 15.8\% to 31.6\% for $k = 100$; on GIST, it rose from 9.6\% to 21.3\% for $k = 100$; and on arxiv, queries with recall 0 appeared. Graph indexes degraded gradually even under random insertion alone. On SIFT ($k = 100$), the fraction rose from 21.0\% to 26.9\% for HNSW and from 17.5\% to 24.0\% for FreshDiskANN. On the yandex sliding window, which receives real out-of-distribution queries, IVF-Flat, hnswlib, FreshDiskANN, and Quake all reached the average target at first, but their failure fractions increased (Section~\ref{sec:motivation}, Fig.~\ref{fig:motivation}(b)), and for Quake the worst query's recall dropped from 0.73 to 0.42. On laion, hnswlib failed to reach an average recall of 0.99 from the start (0.982 at the largest search parameter, ef = 4,096). In contrast, under cluster-ordered insertion (distribution shift), the failure fractions of the public software actually decreased. On the same 11 workloads, our T1 failures were 219 in 540,000 trials (at most 7 in 3,000 at any time point), and our time was shorter than that of FAISS IndexFlatL2 in every setting (3.4–53.8$\times$ 1-thread, 11.3–116$\times$ 32-thread, 3.2–67.1$\times$ batch; the smallest value is the batch on laion).

\begin{figure}[t]
\centering
\includegraphics{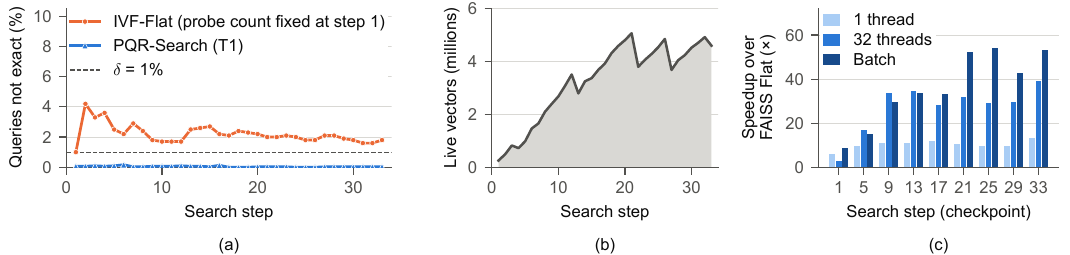}
\refstepcounter{figure}\label{fig:streaming}
\par\vspace{3pt}{\small\setstretch{1.15}\IEEEfigcap{Figure \thefigure. Official NeurIPS'23 streaming runbook (msturing-10M, $k = 10$). (a) Failure rate per search step. (b) Live vectors at search time. (c) Speedup over FAISS IndexFlatL2 at checkpoints.}\par}
\end{figure}

Next, we used, as is, the delete runbook of msturing-10M-clustered (\texttt{delete\_\allowbreak{}runbook.\allowbreak{}yaml}) from the operation sequences (runbooks) provided by the official repository of the NeurIPS'23 streaming track~\cite{r40}. This runbook performs 32 insertions, 10 deletions, and 33 searches, and the number of live vectors reaches up to 5.3M over the whole sequence. The competition's final evaluation scored a different runbook by the average recall over the entire sequence under limits of 8 GB of memory and one hour~\cite{r40}, but our experiment imposed no such limits, used 1,000 queries, and computed the ground truth directly in fp64. Searches and updates were executed serially. Without retuning, the failure rate was at most $\delta$ in all 33 search steps (Eq.~\eqref{eq:dyn}), and the per-step failure rate of the exact-search guarantee ($\delta = 0.01$) was at most 0.17\% (Fig.~\ref{fig:streaming}(a)). In contrast, IVF-Flat with $n_{\rm probe}$ fixed at 2612 (of 4096) so that 99\% of queries were exact at the first time point failed, on the same sequence, to answer 1.0\% of the queries exactly at the first search and 1.6–4.2\% at each later step, exceeding its target by up to 4.2$\times$ (Fig.~\ref{fig:streaming}(a)). Meanwhile, the number of live vectors at search time rose and fell between 260K and 5.07M with insertions and deletions (Fig.~\ref{fig:streaming}(b)). At the timed checkpoints (searches 1, 5, \ldots{}, 33), our method was 5.9–13.2$\times$ (1-thread), 3.0–39.3$\times$ (32-thread), and 8.7–53.9$\times$ (batch) faster than FAISS IndexFlatL2 (Fig.~\ref{fig:streaming}(c)), with the smallest values at the first search with 260K live vectors (our method and FAISS were measured alternately in one run on the same update sequence). As live vectors grew from 260K to 5.07M, our 1-thread latency increased from 0.38 ms to 4.52 ms and that of FAISS IndexFlatL2 from 2.22 ms to 46.9 ms. Compared with IndexIVFFlat with the oracle $n_{\rm probe}$, the post-hoc optimal reference, our method was also 10.0–14.3$\times$ (1-thread), 5.9–18.8$\times$ (32-thread), and 16.9–92.8$\times$ (batch) faster. We measured two dynamic indexes without guarantees on the same runbook. FreshDiskANN~\cite{r93} (R = 64, search list L = 384, the smallest giving average recall 0.99 on calibration queries at the first search) had up to 2.7$\times$ lower 1-thread latency and up to 1.27$\times$ higher batch throughput than ours once live vectors exceeded 2.4M, but 18.5\% of its queries fell below recall 0.99 (worst recall 0; average recall per step down to 0.926), and its update throughput was 33,127 vectors/s, 1/3.1 of ours (consolidating deletions took 45\% of its update time); with L re-chosen afterwards at each checkpoint for average recall 0.99, we were 1.31–3.09$\times$ (1-thread) and 2.73–3.12$\times$ (batch) faster. Quake~\cite{r50} (APS target 0.99, default maintenance) split partitions from 1,000 into up to 29,211, so its 1-thread search became slower than FAISS IndexFlatL2 (7.3–38.0$\times$ slower than ours), its update throughput was 1,473 vectors/s (1/69 of ours), and 2.45\% of its queries fell below recall 0.99. On a 160-step arxiv 1M streaming scenario generated with the official procedure~\cite{r40}, the failure rate also stayed at most $\delta$ to the end (148 T1 failures in 480,000 trials, at most 4 in 3,000 per step), and our method was 6.7–34.6$\times$ (1-thread), 6.6–52.0$\times$ (32-thread), and 3.6–11.0$\times$ (batch) faster than FAISS IndexFlatL2 at nine checkpoints.

\begin{figure}[t]
\centering
\includegraphics{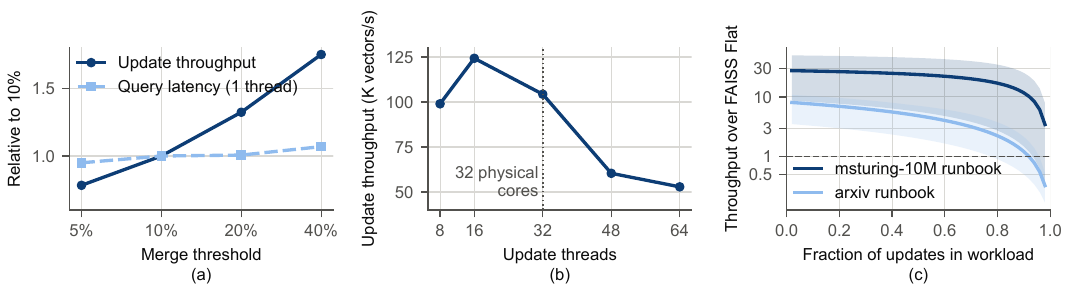}
\refstepcounter{figure}\label{fig:dynsens}
\par\vspace{3pt}{\small\setstretch{1.15}\IEEEfigcap{Figure \thefigure. Sensitivity of the dynamic index. (a) Merge threshold (official runbook). (b) Number of update threads (official runbook). (c) Total throughput versus update ratio (official runbook and arxiv runbook).}\par}
\end{figure}

We also measured the effect of maintenance. Without residual-based maintenance, the mean residual grew by 21–23\% on msturing and by 26–34\% on arxiv, and partition sizes became skewed; at the checkpoints after merges, the maintained index was faster by 1.29–2.33$\times$ (1-thread), 2.18–3.97$\times$ (32-thread), and 3.69–6.66$\times$ (batch) on msturing and by 2.42–4.84$\times$, 2.16–5.62$\times$, and 2.94–8.97$\times$, respectively, on arxiv. Compared with an index rebuilt from scratch, the search speed of the maintained index was 0.81–1.16$\times$ on msturing and 0.76–1.16$\times$ on arxiv. With the same search implementation, policy K, which decides by Eq.~\eqref{eq:passprob}, achieved 4.0$\times$ higher update throughput than policy R on laion with 1/5.8 of the maintenance time; query latency was at the same level (0.97–1.11$\times$ 1-thread, 0.96–1.01$\times$ 32-thread), and batch throughput was 0.87–1.11$\times$ depending on the configuration (step 5, a single run). We also compared, with the same search implementation and the same merge points, a LIRE-style~\cite{r48} size-based policy (splitting partitions larger than twice the mean size by 2-means, dissolving those smaller than 0.25 times, and reassigning vectors of nearby partitions, in variants with and without recomputing all centroids at every merge), and found no evidence that policy K is faster. On the official runbook, at checkpoints with at least 2.79M live vectors, policy K had 1.03–1.14$\times$ shorter 1-thread and 1.01–1.17$\times$ shorter 32-thread latency than the size-based policy (recomputing variant) and similar batch throughput (0.93–1.04$\times$), but early on it was slower (0.80–0.85$\times$, 0.98M live vectors) because it splits a large partition only in two per merge. On the arxiv runbook, its 1-thread latency was 0.93–1.01$\times$ and its batch throughput 0.78–1.01$\times$, lower at 8 of 9 checkpoints, and the variant without recomputation had 1.6$\times$ the update throughput of policy K. For all three policies the failure rate was at most $\delta$ (at most 2/1,000 exact-search failures per step), the decisions of the size-based policy were identical under two rotations at all 87 merges, and the prediction of Eq.~\eqref{eq:passprob} was 0.97–1.09$\times$ the measurement on all three indexes. The maintenance gains above thus come from maintenance itself; the contribution of policy K is not speed but predicting cost without data-fitted constants while keeping maintenance separate from the guarantee. Maintenance decisions were executed separately for each rotation, and the result (centroids, assignment) was identical at every merge. This agrees with C5, which states that maintenance decisions do not depend on the rotation. In addition, searching with the four nearest live vectors of each query only marked as deleted gave the same output as the index from which those vectors were physically removed by a merge. Varying the merge threshold from 5\% to 40\%, the failure rate was at most $\delta$ in all 33 search steps (9–14 exact-search failures in 33,000), the search time at the checkpoints was 0.94–1.07$\times$ that with a 10\% threshold, and larger thresholds reduced merges from 35 to 13, raising update throughput by 2.2$\times$ (Fig.~\ref{fig:dynsens}(a)).

A single update in our method includes encoding and maintenance, so it is 7–66$\times$ more expensive than an addition or removal in FAISS IndexFlatL2, which only appends vectors to an array or removes them (66$\times$ on the official runbook; the update throughput of the default configuration, including merges and maintenance, is 102,016 vectors/s). Since this cost is recovered by queries, it is appropriate to compare the total cost of queries and updates, $Q\,t_q + U\,t_u$. Here $Q$ is the number of queries, $U$ is the number of updated vectors, and $t_q$ and $t_u$ are their respective unit costs. The ratio of queries to updates at which our method becomes cheaper than FAISS is \begin{equation}\frac{Q}{U^*} = \frac{t_{u}^{\rm ours} - t_{u}^{\rm Flat}}{t_{q}^{\rm Flat} - t_{q}^{\rm ours}}\label{eq:breakeven}\end{equation} and the measured values on the 11 workloads and the official runbook were 0.0021–0.035 (batch) and 0.0002–0.0022 (1-thread) (0.134 for batch and 0.0037 for 1-thread on the arxiv runbook, whose index is small). That is, on these workloads, the total cost is lower than that of FAISS once the number of queries reaches only 0.2–3.5\% (batch) of the number of updated vectors. In this calculation, $t_u$ is the time when updates are grouped and processed with 32 threads. Update throughput was highest at 16 threads (124,000 vectors/s versus 104,000 at 32 threads); beyond the number of physical cores (32), the two hardware threads of a core are used together, which slows the maintenance stage of merges, and throughput dropped to 53,000 at 64 threads (Fig.~\ref{fig:dynsens}(b)). For updates alone, 16 threads gave 1.19$\times$ higher throughput, but we used 32 threads, the same as batch search. The 27 merges took 4.74 s on average (up to 8.9 s), and the last merge (6.95 s) took 0.12$\times$ the time to rebuild the index from scratch with the 4.6M live vectors at that point (58.8 s). In our replay of the official runbook (1,000 queries per search step), queries are only 0.22\% of the number of updated vectors, a workload near the break-even point. Indeed, dividing the time of the whole runbook, 98\% of our 151.0 s went to updates (91.3 s of which was maintenance), and with queries processed in batch, FAISS IndexFlatL2 (105.0 s in the same run) took 1.44$\times$ less total time. With queries processed one at a time on 1 thread, our method took 4.1$\times$ less. Varying the ratio of updates to queries, our total throughput relative to FAISS IndexFlatL2 was 7.7–48.6$\times$ at an update-to-query ratio of 2:8, 5.6–45.1$\times$ at 5:5, and 2.7–34.8$\times$ at 8:2, and 3.0–9.6$\times$, 2.0–7.1$\times$, and 0.9–3.8$\times$, respectively, on the arxiv runbook (queries in batch, at each checkpoint, Fig.~\ref{fig:dynsens}(c)). On arxiv, the throughput measured by actually running alternating updates and queries in equal numbers (with the implementation before the insertion path was fixed) agreed with the value computed from unit costs to within 0.5\%. In a deletion-heavy workload that halves the live vectors and then restores them (arxiv, from 540K through 270K back to 540K), the failure rate was at most $\delta$ at every step (8 exact-search failures in 27,000), and the mean residual stayed at 0.98–1.08$\times$ its initial value. On the arxiv sliding window, inserting updates one at a time still gave an update throughput of 5,282 vectors/s (median 0.29 ms per insertion), which grew from 11,573 to 47,345 vectors/s when updates were inserted in batches of 100 to 100,000. The guarantee did not depend on the batch size, and search time varied with batch size within 22\% in 1-thread latency and within 7\% in batch throughput.

We also measured updates at the 100M scale. After building an index with 90M vectors of DEEP-100M, we repeated 10 times a step that deletes 1M of the initial vectors and inserts 1M new vectors, replacing 10M vectors in total. At all 11 time points, there was no exact-search failure (0/11,000), and search was 10.2–13.3$\times$ (1-thread), 29.0–48.4$\times$ (32-thread), and 86–142$\times$ (batch) faster than FAISS IndexFlatL2. At this scale, the maintenance policy did not split partitions but dissolved some of them, reducing the number of partitions from 90,000 to 81,000, and search right after each of the two merges was as fast as or faster than with the initial index (1-thread 58.3–58.7 ms versus 60.4 ms; batch 1,457–1,479 versus 1,379 QPS). Update throughput, including merges and maintenance, was 23,244 vectors/s, 18.6$\times$ slower than addition and removal in IndexFlatL2, and the total time of the two merges (572 s) was 0.33$\times$ the time to rebuild from scratch with 90M live vectors (1,711 s).

\begin{figure}[t]
\centering
\includegraphics{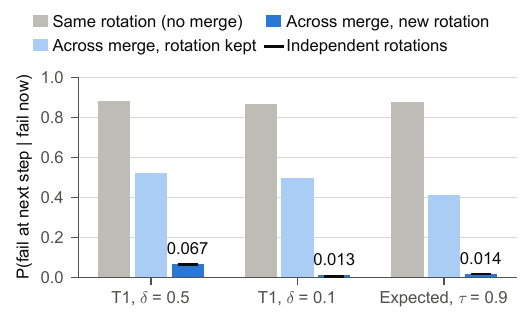}
\refstepcounter{figure}\label{fig:persistence}
\par\vspace{3pt}{\small\setstretch{1.15}\IEEEfigcap{Figure \thefigure. Conditional probability that the same query also fails in the next search step (msturing official streaming, $k = 10$).}\par}
\end{figure}

Finally, we verified Theorem 6 (b) and (c): if the same rotation is kept, a query's failures can persist, and if a new rotation is used at merges, the failures are broken. Fig.~\ref{fig:persistence} shows, for three configurations with enough failure events (the exact-search guarantee with $\delta = 0.5$ and $0.1$, and the expected-recall guarantee with $\tau = 0.9$), the conditional probability that a query failing in one search step also fails in the next, split by whether a merge occurs in between and whether the rotation is changed. Keeping the rotation couples failures strongly over time. For pairs of search steps with no merge in between, the conditional probability that the same query also fails in the next step was 0.87–0.88 with a fixed rotation, one to two orders of magnitude larger than the unconditional failure rate (0.7–5.5\%). With a new rotation at every merge, the conditional failure probability for pairs of search steps across a merge dropped from 0.41–0.52 with a fixed rotation to 0.013–0.067. For pairs without a merge, it remained high at 0.77–0.91 even with new rotations, confirming that it is the new rotation that breaks the failures. The remaining 0.067 under new rotations (the $\delta = 0.5$ configuration) is larger than the unconditional failure rate of 0.054 in the same configuration, but it is not a violation. Since hard queries tend to fail under any rotation, the conditional probability measured between mutually independent rotations (0.005–0.063) serves as the baseline for this query heterogeneity, and the value under new rotations is at the same level as this baseline. After a new rotation, the stored codes were bit-identical to codes computed from scratch; re-encoding cost 0.20 $\mu$s per vector and reduced update throughput by 9.7\%.

\subsection{Cost Law and Parameter Selection}\label{sec:ablation}

\begin{figure}[t]
\centering
\includegraphics{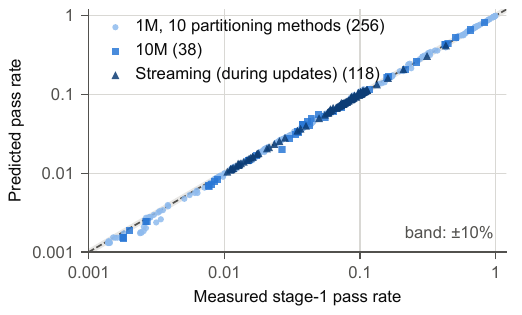}
\refstepcounter{figure}\label{fig:passrate}
\par\vspace{3pt}{\small\setstretch{1.15}\IEEEfigcap{Figure \thefigure. First-stage pass rate predicted by Eq.~\eqref{eq:passprob} versus measured.}\par}
\end{figure}

In this section, we first verify the pass-rate prediction given by the cost law of Section~\ref{sec:costlaw}, and then measure how cost is affected by the design choices of Section~\ref{sec:index} that differ from standard IVF, by the techniques of Section~\ref{sec:devices}, and by the parameter selection rule. The first-stage pass-rate prediction of Eq.~\eqref{eq:passprob} agreed well with measurements across partition methods, scales, and updates. On six 1M datasets, over 256 points varying the partition method (up to 10 kinds), $k$, and the settings of the radius and the error-bound parameter, the log correlation was 0.9995 (median of prediction $\div$ measurement 0.997, with 96\% within $\pm$10\%); over 38 settings at 10M scale it was 0.999; and at 118 time points of two streaming runbooks (msturing and arxiv) with continuing insertions and deletions, the ratio of prediction to measurement was 0.98–1.08 (Fig.~\ref{fig:passrate}). For six settings on three 100M-scale datasets, the prediction was estimated from sampled vectors for 20 queries; over 10 independent per-query samples, the mean ratio of prediction to measurement was 0.98–1.02 (0.86–1.10 for a single sample). The prediction matched the pass rate over a range of three orders of magnitude without constants fitted to the data.

Next, to confirm the role of centroids explained in Section~\ref{sec:costlaw}, we fixed the radius at the true $k$-th distance and removed two components one at a time. The residual effect is the time ratio when the centroids are replaced by a single global centroid, and the theory effect is the time ratio when the error-bound parameter is changed from that of our method (a union over the $k$ true neighbors) to the value determined by a union over all vectors. The ratio when both are removed, divided by the product of the ratios when each is removed separately, is called the interaction. If the two effects are independent of each other, the interaction is 1; if it is greater than 1, the loss from removing both together is greater than the product of the losses from removing each separately. That is, the presence of one component compensates for part of the loss from removing the other. As Table~\ref{tab:ablation} shows, the interaction was 1.13–1.40 on the text data (arxiv, agnews), GIST, and SIFT, larger at 1.54–2.18 on YFCC and msmarco at 10M scale, and near 1 at 0.95–1.03 on SSNPP and laion ($k = 100$). In terms of first-stage passes, the residual effect is up to 13.0$\times$, much larger than in time. When the pass rate is as low as 0.002–0.009, as on YFCC-10M ($k = 10$), time is bound by the first-stage scan and the time effect appears small, so the effect of residuals on cost is judged more accurately by the number of passes, i.e., by Eq.~\eqref{eq:passprob}. On SSNPP, which has a high intrinsic dimension, the residual effect was 1.02–1.07 at both 1M and 10M. The theory effect comes from reducing the terms of the union from $n$ vectors to $k$ true neighbors.

\Needspace{16\baselineskip}
{\centering \textbf{Table \refstepcounter{table}\label{tab:ablation}\thetable{}. Contributions of residuals and of the error-bound parameter (1-thread time ratios; radius fixed at the true $k$-th distance).}\par}\nopagebreak

{\footnotesize\setstretch{1.1}
\begin{xltabular}{\linewidth}{llYYlY}
\toprule
\textbf{Data} & \textbf{$k$} & \textbf{Residual effect} & \textbf{Theory effect} & \textbf{Interaction} & \textbf{First-stage passes: residual / theory} \\
\midrule
\endfirsthead
\multicolumn{6}{l}{\emph{Table~\thetable\ (continued)}} \\
\toprule
\textbf{Data} & \textbf{$k$} & \textbf{Residual effect} & \textbf{Theory effect} & \textbf{Interaction} & \textbf{First-stage passes: residual / theory} \\
\midrule
\endhead
arxiv & 10 & 1.45 & 1.96 & 1.40 & 2.75 / 6.92 \\
arxiv & 100 & 1.28 & 1.38 & 1.27 & 2.49 / 4.00 \\
agnews & 10 & 1.05 & 1.18 & 1.15 & 2.62 / 11.23 \\
agnews & 100 & 1.23 & 1.58 & 1.26 & 1.78 / 4.06 \\
GIST & 10 & 1.24 & 1.57 & 1.29 & 2.99 / 6.13 \\
GIST & 100 & 1.40 & 1.61 & 1.13 & 2.23 / 2.89 \\
SIFT & 10 & 2.07 & 1.16 & 1.20 & 13.00 / 5.63 \\
SIFT & 100 & 2.13 & 1.29 & 1.14 & 7.14 / 2.82 \\
laion & 10 & 1.23 & 2.28 & 1.24 & 1.19 / 1.35 \\
laion & 100 & 1.60 & 2.46 & 0.96 & 1.27 / 1.48 \\
SSNPP & 10 & 1.07 & 3.47 & 0.95 & 1.07 / 5.26 \\
SSNPP & 100 & 1.04 & 1.51 & 0.97 & 1.04 / 2.21 \\
YFCC-10M & 10 & 1.08 & 1.09 & 1.58 & 7.06 / 8.76 \\
YFCC-10M & 100 & 1.43 & 1.23 & 1.54 & 5.05 / 4.19 \\
msmarco & 10 & 1.17 & 2.53 & 2.18 & 2.29 / 8.23 \\
SSNPP-10M & 10 & 1.02 & 5.47 & 1.03 & 1.13 / 10.08 \\
\bottomrule
\end{xltabular}}

Measurements with 10 different partition methods lead to the same conclusion (per-method numbers in Appendix~\ref{app:partition}). At the same radius, the k-means family and anisotropic quantization~\cite{r25} yielded almost the same cost (0.95–1.03$\times$) as the partitioning of our method (k-means with size-capped assignment), and the order of pass rates was determined by the relative residual size (rank correlation across methods 0.88–0.96). In contrast, when each method used the radius it obtained by itself, methods with poor centroids (a global centroid, random projection trees~\cite{r163}) had visited partitions containing fewer neighbors, which worsened the radius as well, so they were 1.6–18.3$\times$ slower on 1M data with in-distribution queries; the adaptive radius reduced this penalty to 1.2–2.5$\times$. This agrees with the explanation in Section~\ref{sec:costlaw} that partitions play two roles: providing centroids and obtaining the radius.

We first validated the parameter selection rule on imagenet-align, which was never used in design or calibration. Across 48 settings, the configuration chosen by the rule achieved on average 0.903$\times$ (minimum 0.42) the performance of the fastest fixed configuration for each setting (hereafter the best fixed configuration). The settings that fell furthest short of the best fixed configuration were those with a small error-bound parameter, such as the expected-recall guarantee and $\delta = 0.1$, in which the rule chose the default first-stage code length rather than $D$. Even so, the guarantee held ($(\tau, \delta)$-recall guarantee failures 1/108,000), and our method was faster than FAISS IndexFlatL2 in all settings (6.8–20.6$\times$ in batch). On DEEP1M and GloVe, which were added after the rule was fixed and were likewise not used in design or calibration, the configuration chosen by the rule achieved on average 0.959$\times$ (minimum 0.81) the performance of the best fixed configuration over 24 settings. The rule chose the default configuration in 22 settings, so no setting became slower than the default, and the shortfall came mainly from not enabling centering and 2-bit records (the minimum was 32-thread latency on DEEP1M with $k = 100$). On the 96 settings of the eight 1M datasets used in design and calibration, the configuration chosen by the rule achieved on average 0.957$\times$ the performance of the best fixed configuration, and was faster, by a geometric mean of 1.18$\times$ over the three metrics (1-thread 1.22$\times$, 32-thread 1.15$\times$, batch 1.17$\times$), than the configuration with the default first-stage code length and all techniques other than the adaptive radius disabled, with no setting slower. The gains concentrated on laion (geometric mean 2.05$\times$, up to 3.80$\times$), yandex (1.32$\times$, up to 2.38$\times$), and SSNPP (1.31$\times$, up to 1.71$\times$); on the remaining data the rule mostly chose the default configuration. How the value on the design data came to be higher is explained in Appendix~\ref{app:negative}. Reducing the calibration queries used by the rule from 200 to 100 or 50 changed the chosen configuration in only one of the 48 (setting, metric) pairs on the eight 1M datasets in each case. The time differences were within measurement noise except in one setting (1.25$\times$ at 32 threads on yandex, $k = 10$), and the time to determine the rule decreased in proportion to the number of calibration queries, from 132 s to 34 s in total over 16 settings.

We also varied the number of visited partitions $n_{\rm probe}$ and the partition size (the average number of vectors per partition, $l$) (SIFT, arxiv, GIST, laion, $k = 10$, exact search). In all 48 settings of $l \in \{500, 1000, 2000\}$ and $n_{\rm probe} \in \{1, 2, 4, 8\}$, the failure rate was at most $\delta$ (Table~\ref{tab:verification}) and our method was faster than FAISS IndexFlatL2 (the smallest value was 4.7$\times$ in batch for GIST, $l = 2000$, $n_{\rm probe} = 8$). $n_{\rm probe} = 1$ was as fast as or faster than the default value 2 except at 32 threads for SIFT with $l = 500$ (0.92$\times$) (up to 1.37$\times$ in batch, on GIST), whereas at $n_{\rm probe} = 8$ the batch was up to 5$\times$ slower than at $n_{\rm probe} = 1$. This is because visiting more partitions reduces the vectors whose exact distances are computed in unvisited partitions (from 0.098\% to 0.020\% on GIST), but the cost of computing exact distances for all $n_{\rm probe} \cdot l$ vectors of the visited partitions grows more. On arxiv and GIST, larger partitions made the 1-thread case faster and the batch slower, and for $n_{\rm probe} \le 2$ the difference among the three sizes was within 1.3$\times$ in 17 of the 24 combinations of (data, metric, $n_{\rm probe}$) (at most 1.61$\times$, batch on GIST). All other experiments used the default values ($l = 1000$, $n_{\rm probe} = 2$).

Table~\ref{tab:techniques} summarizes the effect of disabling or replacing each technique of Section~\ref{sec:devices} one at a time. Reading codes prefix-first was faster on in-distribution high-dimensional data with low first-stage pass rates, whereas reading the entire code at once was faster on data where many vectors pass the first stage because the queries are out of distribution; the parameter selection rule therefore chooses the faster of the two for each dataset. The triangle-inequality range reduced the vectors to be tested on data with low intrinsic dimension, and thanks to sorting by $x$ the cost of finding the range is one binary search per partition. The effect of the adaptive radius was larger the more the radius determined from visited vectors exceeded the true $k$-th distance. Enabling techniques regardless of the data sometimes made search slower (for example, 0.89–0.94$\times$ at 1 thread on yandex). This is why the rule decides the use of each technique separately for each dataset and metric.

\Needspace{16\baselineskip}
{\centering \textbf{Table \refstepcounter{table}\label{tab:techniques}\thetable{}. Effect of each technique (time ratios between fixed configurations measured in the same run; a value above 1 means the technique is faster).}\par}\nopagebreak

{\small\setstretch{1.1}
\begin{xltabular}{\linewidth}{YYYY}
\toprule
\textbf{Technique} & \textbf{Compared with} & \textbf{Data} & \textbf{Effect} \\
\midrule
\endfirsthead
\multicolumn{4}{l}{\emph{Table~\thetable\ (continued)}} \\
\toprule
\textbf{Technique} & \textbf{Compared with} & \textbf{Data} & \textbf{Effect} \\
\midrule
\endhead
Reading codes prefix-first (128 or 256 bits first) & Whole code in one stage ($m_1 = D$) & arxiv, agnews, GIST & 1-thread 1.6–3.0, batch 1.2–2.0 \\
Whole code in one stage ($m_1 = D$) & Reading codes prefix-first & laion, yandex & Up to 3.2 \\
Triangle-inequality range & Testing all vectors without the range & SIFT, YFCC-10M & Vectors tested: 40–47\% and 70–80\% of all (fractions, not time ratios) \\
Adaptive radius (Theorem 5) & Radius determined from visited vectors & laion / SSNPP / yandex & 1-thread 5.6–10.6 / 1.45–2.09 / 1.12–1.36, batch 3.8–5.8 / 1.23–1.59 / 1.02–1.17 \\
Global centering of LUTs & Off & SSNPP (settings chosen by the rule) & 1-thread 1.36–1.83, 32-thread 1.37–1.62, batch 1.22–1.46 \\
First-stage 256 bits when the pass rate exceeds 5\% & 128 bits & Text data with $k = 100$, GIST (exact search) & 1-thread 1.19–1.80 \\
2-bit final stage & Off & laion, yandex (exact search, 1-thread) & 1.16–1.59 \\
Parameter selection rule & Default configuration (adaptive radius only) & 96 settings on eight 1M datasets & Geometric mean 1.18 (1-thread 1.22, 32-thread 1.15, batch 1.17), 0 slower settings \\
\bottomrule
\end{xltabular}}

\subsection{Application, Scale, and Dimension}\label{sec:scale}

As an application of range queries, we experimented with image copy detection. For the SSNPP range ground truth of the Big-ANN'21 benchmark~\cite{r39} (squared L2 radius 96,237), we required the probability of missing even one copy for a query to be at most 0.01 (the error-bound parameter follows the range-query setting of Section~\ref{sec:guarantee}). Over 40,000 trials at 1M scale and 10,000 trials at 10M scale (1,024 and 826 of them, 1,850 in total, on queries that have copies), our method had no query that missed a copy, and it never returned a vector outside the radius. This corresponds to 1 in mean average precision (mAP)~\cite{r39}, the official metric of the competition. In time, it was faster than the range search of FAISS IndexFlatL2 at 10M scale by 20$\times$ at 1 thread, 132–133$\times$ at 32 threads, and 54$\times$ in batch (at 1M scale, 23$\times$ at 1 thread, 222–247$\times$ at 32 threads, and 43–44$\times$ in batch). In contrast, methods without guarantees missed copies. At 10M scale, of the 826 queries with copies, HNSW32 missed copies for 16.7\% (138) of the queries even at efSearch 4,096 (185/6,956 copies missed), and missed all copies for 68 queries. Its average recall in this case was as high as 0.97. IVF tuned to an average recall of 0.99 also missed copies for 5.7\% of the queries.

We also measured whether missed neighbors actually change the answers of database operators (Section~\ref{sec:intro}). On YFCC, which has tag metadata (1M and 10M, $k \in \{10, 100\}$, 1,000 test queries), we defined operators that aggregate the tags of the $k$-NN result $S$. The tags are the metadata of the filtered track of the Big-ANN benchmarks, with a vocabulary of about 200,000 tags and 8.8 tags per vector on average. The operators are the most frequent tag (\texttt{GROUP BY tag ORDER BY COUNT(*) DESC LIMIT 1}), the most frequent tag excluding common tags attached to more than 1\% of the base vectors (informative), the top-5 tags with their counts, the number of neighbors that have all filter tags of the query (\texttt{COUNT(*) WHERE}, the AND filter of the filtered track), and whether that number is positive. An answer counts as wrong if it differs from the answer on the exact result and the result is not exact ($\mathrm{Recall} < 1$ under the tie rule of Section~\ref{sec:setup}); a different answer from an exact result arises from choosing another exact set among ties at $d_k$ and is not counted. Hence, for any operator, wrong answers are a subset of the cases with $\mathrm{Recall} < 1$. The baselines are HNSW32, IVF-Flat, and IVF-RaBitQ with re-ranking (1M only), tuned with the same definition as the price of the guarantee in Section~\ref{sec:cost}, so that the average recall over 1,000 calibration queries disjoint from the test queries reached the target; every baseline reached it. For our method we ran the rule-selected configurations of the main comparison (the exact-search guarantee T1 with $\delta = 0.01$, and the expected-recall guarantee E with $\tau = 0.9$) with 10 independent rotations and three processing paths (batch, 32-thread, and 1-thread), 30,000 trials per dataset and $k$. Time was not measured.

\Needspace{16\baselineskip}
{\centering \textbf{Table \refstepcounter{table}\label{tab:dbop}\thetable{}. Fraction of queries whose tag-aggregate answer over the search results differs from that of exact search (YFCC, 1,000 test queries, target 0.99). Ours: T1 ($\delta = 0.01$), the largest over the three paths of the rate over 10 rotations; baselines: knob (efSearch for HNSW32, $n_{\rm probe}$ for IVF-Flat, $n_{\rm probe}$ and re-ranking factor for IVF-RaBitQ) tuned so that the average recall of calibration queries reached 0.99. $\mathrm{Recall} < 1$ counts inexact results and contains the errors of every operator.}\par}\nopagebreak

{\footnotesize\setstretch{1.1}
\begin{xltabular}{\linewidth}{llYlYlYlYY}
\toprule
\textbf{Data} & \textbf{$k$} & \textbf{Method (knob)} & \textbf{Avg. recall} & \textbf{$\mathrm{Recall} < 1$} & \textbf{Most freq.} & \textbf{Most freq. (inform.)} & \textbf{Top-5} & \textbf{Filtered count} & \textbf{Filter exists} \\
\midrule
\endfirsthead
\multicolumn{10}{l}{\emph{Table~\thetable\ (continued)}} \\
\toprule
\textbf{Data} & \textbf{$k$} & \textbf{Method (knob)} & \textbf{Avg. recall} & \textbf{$\mathrm{Recall} < 1$} & \textbf{Most freq.} & \textbf{Most freq. (inform.)} & \textbf{Top-5} & \textbf{Filtered count} & \textbf{Filter exists} \\
\midrule
\endhead
YFCC-1M & 10 & Ours (T1) & 1.0000 & 0.00\% & 0.00\% & 0.00\% & 0.00\% & 0.00\% & 0.00\% \\
YFCC-1M & 10 & HNSW32 192 & 0.9913 & 3.3\% & 1.0\% & 0.7\% & 3.1\% & 0.2\% & 0.1\% \\
YFCC-1M & 10 & IVF-Flat 64 & 0.9962 & 3.2\% & 1.2\% & 0.5\% & 3.2\% & 0.3\% & 0.1\% \\
YFCC-1M & 10 & IVF-RaBitQ 64, 100 & 0.9960 & 3.4\% & 1.2\% & 0.6\% & 3.4\% & 0.3\% & 0.1\% \\
YFCC-1M & 100 & Ours (T1) & 1.0000 & 0.03\% & 0.00\% & 0.00\% & 0.01\% & 0.00\% & 0.00\% \\
YFCC-1M & 100 & HNSW32 384 & 0.9944 & 21.4\% & 0.5\% & 1.8\% & 19.5\% & 1.3\% & 0.1\% \\
YFCC-1M & 100 & IVF-Flat 64 & 0.9903 & 27.4\% & 1.0\% & 2.4\% & 25.6\% & 2.4\% & 0.2\% \\
YFCC-1M & 100 & IVF-RaBitQ 64, 20 & 0.9899 & 28.5\% & 1.0\% & 2.5\% & 26.6\% & 2.3\% & 0.2\% \\
YFCC-10M & 10 & Ours (T1) & 1.0000 & 0.00\% & 0.00\% & 0.00\% & 0.00\% & 0.00\% & 0.00\% \\
YFCC-10M & 10 & HNSW32 256 & 0.9927 & 5.2\% & 1.9\% & 1.4\% & 5.0\% & 0.3\% & 0.1\% \\
YFCC-10M & 10 & IVF-Flat 128 & 0.9924 & 5.8\% & 1.5\% & 1.5\% & 5.3\% & 0.9\% & 0.1\% \\
YFCC-10M & 100 & Ours (T1) & 1.0000 & 0.04\% & 0.01\% & 0.00\% & 0.03\% & 0.00\% & 0.00\% \\
YFCC-10M & 100 & HNSW32 512 & 0.9929 & 26.8\% & 1.5\% & 2.4\% & 25.4\% & 4.6\% & 0.7\% \\
YFCC-10M & 100 & IVF-Flat 192 & 0.9922 & 21.7\% & 1.1\% & 1.9\% & 19.5\% & 3.6\% & 0.2\% \\
\bottomrule
\end{xltabular}}

As Table~\ref{tab:dbop} shows, indexes tuned to an average recall of 0.99 returned a different most frequent tag for 0.5–1.9\% of queries, a different top-5 list with counts for 3.1–26.6\%, and a different filtered count for up to 4.6\% than exact search. In contrast, our method (T1) changed the answer of any operator in at most 4 of the 10,000 (rotation, query) trials of each path (0.04\%), and never twice for the same query (the most frequent tag changed once, on the 1-thread path of 10M with $k = 100$). Because wrong answers are a subset of the cases with $\mathrm{Recall} < 1$, the exact-search guarantee $\Pr[\mathrm{Recall}(q) < 1] \le \delta$ becomes, for any operator, a per-query guarantee that the answer is wrong with probability at most $\delta$. Ten rotations cannot certify the per-query bound statistically; the bound comes from this argument, and the measurements confirm that they do not contradict it. Among the baselines' queries with recall below 1, the fraction whose answer actually changed depended on how fine the operator is. For the most frequent tag it was 26–41\% at $k = 10$ but 2–9\% at $k = 100$, where the count gap between the first and second tags of the exact answer is 10–12 on average, so the answer survives a changed neighbor (at $k = 10$ the gap is 1.1 on average and 34\% of queries have a tie). For the top-5 tags with counts it was 90–100\%: missing one neighbor almost always made the answer wrong. Larger knobs did not easily remove the errors. Even at efSearch 4,096 (test average recall 0.999), HNSW32 left 4.2–4.8\% top-5 errors at $k = 100$, and on 10M with $k = 100$, IVF-Flat scanning 15\% of the partitions ($n_{\rm probe} = 1{,}536$) still returned inexact results for 0.2\% of queries.

Tuned to an average recall of 0.9, the baselines' wrong answers grew to 2.9–17.2\% for the most frequent tag and 21.6–75.4\% for the top-5 tags. Our expected-recall guarantee (E, $\tau = 0.9$), with a measured average recall of 0.9994–0.9995, gave far fewer wrong answers than the baselines at the same target (most frequent tag 0.08–0.20\%; top-5 0.50–0.54\% at $k = 10$ and 3.95–4.01\% at $k = 100$), but since it does not bound the probability that recall falls below 1, it puts no bound on wrong answers: at $k = 100$, 88 (1M) and 102 (10M) queries had a Clopper–Pearson lower bound on their probability of an inexact result above $\delta = 0.01$. Database operators should therefore use the exact-search guarantee (T1). As limitations of this measurement, YFCC is the only dataset with tag metadata, the operators are tag aggregates, and operators on numeric columns such as averages or sums were not measured; IVF-RaBitQ at 10M was omitted because it required building a new index.

We also measured filtered $k$-NN (Corollary 2) on YFCC with two kinds of predicates: the tag conjunctions of the 1,000 test queries of the Big-ANN filtered track (selectivity below 0.1\% for 498 of them) and random subsets of 0.1\%, 1\%, and 10\% of the vectors, chosen independently of the vectors. In exact search (T1, $\delta = 0.01$, with the index and parameter selection rule used without filters), each of the 22 configurations (predicate, $k$, and execution path) failed in at most 12 of 300,000 (rotation, query) trials on YFCC-1M (300 rotations), and 98.8–99.9\% of the queries were certified individually (Clopper–Pearson upper bound below $\delta$); on YFCC-10M (20 rotations) at most 1 of 20,000 trials failed. Filtered indexes without guarantees lost recall as selectivity fell. On the filtered-track queries with selectivity below 0.1\%, ACORN~\cite{r195} reached a mean recall of only 0.04–0.06, and HNSW32 searching with the predicate did not reach a mean of 0.99 even at efSearch 4,096; at the settings that did reach a mean of 0.99 (chosen on the test queries), 0.2–41\% of the queries still had recall below 1. Our cost does not shrink with selectivity (0.51–0.96 ms at 1 thread on 1M, close to the unfiltered cost), so exact search over the qualifying vectors was faster below 1\% selectivity, whereas above it our method was 6.4–9.7$\times$ faster at 1 thread. Choosing between the two plans by $\lvert S_q\rvert$ alone (exact search if at most 1\% of the vectors qualify), which does not read the rotation, keeps every query exact or $\delta$-guaranteed, and it reduced the mean 1-thread latency of the filtered-track queries to 0.26 ms on 1M (0.67 ms with our method alone and 1.57 ms with exact search alone; 99th percentile 0.99 ms versus 18 ms for exact search) and to 3.65 ms on 10M (8.98 and 22.7 ms).

As the scale grew from 1M to 10M, no setting was slower than FAISS IndexFlatL2 (the 10M scale in Fig.~\ref{fig:speedup}). At $k = 1$, the speedups at 1 thread, 32 threads, and in batch were 16.7$\times$, 81.5$\times$, and 42.5$\times$ on YFCC-10M and 90.4$\times$, 319$\times$, and 61.4$\times$ on msmarco. At 10M scale, the only setting in which the oracle IVF, a post-hoc optimal reference, was faster than our method was 1-thread on YFCC-10M with $k = 1$ (0.48$\times$). IVF with $n_{\rm probe}$ set from calibration queries failed 4 test queries in the same setting. On hotpotqa, which searches 5.23M Wikipedia paragraphs with real questions, our method was faster than FAISS IndexFlatL2 by 17.2–43.9$\times$ at 1 thread, 50.0–122$\times$ at 32 threads, and 10.0–48.6$\times$ in batch.

On DEEP-100M, MSTuring-100M, and SIFT100M, with the scale increased to 100M, our method was again faster than FAISS IndexFlatL2 in all settings (Table~\ref{tab:hundredm}). The speedups were 12.3–20.6$\times$ at 1 thread, 36.3–71.8$\times$ at 32 threads, and 69.0–154$\times$ in batch; in absolute terms, exact search took 57–77 ms at 1 thread and 7.2–12.5 ms at 32 threads and processed 583–1,272 queries/s in batch, versus 873–1,177 ms, 397–521 ms, and 8.3–11.7 queries/s for IndexFlatL2 (Table~\ref{tab:abs}). The batch gain is larger than for exact search at 10M scale (15.8–55.2$\times$) because the batch throughput of IndexFlatL2 drops to 8–12 queries per second at 100M. The failure rate was at most 4\% of $\delta$ in all settings, with no significant violation (Table~\ref{tab:verification}).

\Needspace{16\baselineskip}
{\centering \textbf{Table \refstepcounter{table}\label{tab:hundredm}\thetable{}. Speedup at 100M scale (T1: $\delta = 0.01$; E: $\tau = 0.9$). The oracle is a post-hoc optimal reference chosen with the test ground truth, and values below 1 (bold) are settings where the oracle is faster. Since the $n_{\rm probe}$ grid goes from 8,192 directly to all partitions (100,000), an oracle shown as all partitions may be larger than the actual minimum.}\par}\nopagebreak

{\footnotesize\setstretch{1.1}
\begin{xltabular}{\linewidth}{lllYYYY}
\toprule
\textbf{Data} & \textbf{$k$} & \textbf{Guarantee} & \textbf{$\div$Flat 1-thread · 32-thread · batch} & \textbf{$\div$IVF oracle 1-thread · 32-thread · batch} & \textbf{Oracle $n_{\rm probe}$} & \textbf{Calibrated $n_{\rm probe}$ (failed test queries)} \\
\midrule
\endfirsthead
\multicolumn{7}{l}{\emph{Table~\thetable\ (continued)}} \\
\toprule
\textbf{Data} & \textbf{$k$} & \textbf{Guarantee} & \textbf{$\div$Flat 1-thread · 32-thread · batch} & \textbf{$\div$IVF oracle 1-thread · 32-thread · batch} & \textbf{Oracle $n_{\rm probe}$} & \textbf{Calibrated $n_{\rm probe}$ (failed test queries)} \\
\midrule
\endhead
DEEP-100M & 10 & T1 & 13.1 · 55.2 · 152 & \textbf{0.81} · 1.69 · 13.4 & 6,144 & 6,144 (0) \\
DEEP-100M & 10 & E & 13.2 · 55.1 · 154 & \textbf{0.28} · \textbf{0.64} · 4.9 & 2,048 & 1,536 (1) \\
DEEP-100M & 100 & T1 & 12.3 · 45.8 · 105 & 14.1 · 22.0 · 188 & 100,000 & 8,192 (2) \\
DEEP-100M & 100 & E & 12.5 · 46.5 · 112 & \textbf{0.21} · \textbf{0.44} · 3.3 & 1,536 & 1,024 (4) \\
MSTuring-100M & 10 & T1 & 13.6 · 37.5 · 79.0 & 14.8 · 18.2 · 142 & 100,000 & 100,000 (0) \\
MSTuring-100M & 10 & E & 13.7 · 37.5 · 80.2 & 15.0 · 18.2 · 145 & 100,000 & 100,000 (0) \\
MSTuring-100M & 100 & T1 & 13.2 · 36.3 · 69.0 & 14.5 · 17.6 · 122 & 100,000 & 100,000 (0) \\
MSTuring-100M & 100 & E & 14.0 · 37.3 · 79.6 & 15.4 · 18.1 · 141 & 100,000 & 100,000 (0) \\
SIFT100M & 10 & T1 & 20.5 · 71.8 · 108 & \textbf{0.92} · 1.53 · 12.5 & 4,096 & 1,536 (6) \\
SIFT100M & 10 & E & 20.6 · 70.9 · 110 & \textbf{0.20} · \textbf{0.35} · 2.6 & 768 & 1,536 (0) \\
SIFT100M & 100 & T1 & 17.6 · 60.0 · 91.7 & 19.6 · 28.6 · 257 & 100,000 & 8,192 (2) \\
SIFT100M & 100 & E & 17.5 · 60.0 · 93.9 & \textbf{0.42} · \textbf{0.79} · 5.6 & 2,048 & 1,536 (2) \\
\bottomrule
\end{xltabular}}

The oracle IVF, a post-hoc optimal reference, had shorter latency than our method in some latency settings of the expected-recall guarantee and of exact search ($k = 10$) on DEEP-100M and SIFT100M (1-thread latency in 6 settings, namely the four expected-recall settings and exact search with $k = 10$ on both datasets, with ours at 0.20–0.92$\times$, and 32-thread latency in the four expected-recall settings, with ours at 0.35–0.79$\times$), but in batch our method was 2.6–257$\times$ faster in all settings. This is because IVF scans only some of the partitions, whereas our method also tests the unvisited partitions for the per-query guarantee (Appendix~\ref{app:lowerbound}). With an $n_{\rm probe}$ that allows $\delta$ of the test queries to fail (the $\delta$-oracle), the margins shrink. For exact search with $k = 10$ on DEEP-100M and SIFT100M, the $\delta$-oracle $n_{\rm probe}$ is 1,536; our method was 3.7–5.0$\times$ faster in batch, but IVF had shorter latency (ours 0.22–0.37$\times$ at 1 thread). For exact search with $k = 100$, the $\delta$-oracle needs 6,144 partitions instead of a full scan, so, bounding from the times of measured $n_{\rm probe}$ values, our advantage shrinks to at most 14.6–20.4$\times$ in batch and 1.0–1.5$\times$ at 1 thread (we did not time that $n_{\rm probe}$).

We also measured scalability with the number of threads (SIFT, GIST, laion, $k = 10$, exact search, 1–64 threads). The batch throughput of our method increased up to the number of physical cores (32), reaching 12.7–22.0$\times$ that of 1 thread at 32 threads, and did not increase further at 64 threads, where the two hardware threads of each core are used together. The batch throughput of FAISS IndexFlatL2 at 32 threads was 4.7–12.8$\times$ that at 1 thread, so the gain in batch grew with more threads on SIFT and laion (on SIFT, 15.8$\times$ at 1 thread and 67.4$\times$ at 32 threads) and was similar at 1 and 32 threads on GIST (24.5$\times$, 24.3$\times$). The latency of a single query was 3.0–7.1$\times$ shorter at 32 threads than at 1 thread, with the smallest gain on SIFT, where the work per query is small. Since the single-query latency of FAISS IndexFlatL2 was actually longer at 2–8 threads than at 1 thread, latency is compared only at 1 and 32 threads.

We also measured cost as a function of dimension. On versions of the same data (1M subsets of GIST, arxiv, and msmarco) reduced by PCA to dimensions from 128 up to the original dimension, with the first-stage code length fixed at 128 or 256 bits, the 1-thread speedup over FAISS IndexFlatL2 grew from 19.9–24.6$\times$ at 128 dimensions to 40.4–62.6$\times$ at the original dimension (768–1024). The 1-thread time of FAISS grew 5.9–8.1$\times$ in proportion to the dimension, whereas the time of our method grew only 1.9–3.8$\times$. This is because the testing cost of unvisited vectors is determined by the number of bits, not by the dimension. This trend does not arise because the versions with more dimensions are easier data. As PCA components were added, the relative radius $R_{\rm vis}$ actually grew (arxiv from 0.80 to 0.90, GIST from 0.91 to 0.98, msmarco from 0.93 to 1.05), making the test harder. The failure rate was at most $\delta$ in all cases. In batch, this trend did not appear, because the per-dimension efficiency of FAISS matrix multiplication improves with dimension (Appendix~\ref{app:dimension}). This experiment was measured separately with the kernel prior to the introduction of the adaptive radius, to observe the relative change with dimension.

\bigskip

\section{Conclusion}\label{sec:conclusion}

This paper presented PQR-Search, a method that guarantees a target recall $\tau \in (0, 1]$ for every query in approximate nearest neighbor search in vector databases. The guarantee is defined as a probability over a random orthogonal transformation inside the index, drawn from a secret seed, so it holds for any data and for any queries and updates that do not depend on that transformation, without assumptions on the data or query distribution, and for $\tau < 1$ the same run also satisfies the expected-recall guarantee under a mild condition. If updates are independent of the rotation, this guarantee persists without retuning and without splitting the failure probability $\delta$ between segments. The dynamic index adds only a delta segment, a deletion bitmap, and merges to the static index, so an insertion consists of encoding one vector and re-laying out the delta range of its partition, and a deletion of marking one bit. To obtain this guarantee at low cost, we stored residuals with respect to IVF centroids as 1-bit binary codes after a random orthogonal transformation, and derived a formula that predicts the first-stage pass rate in our index without data-fitted constants and without reading the rotation; without changing the guarantee, we used it to choose parameters as well as the operations that maintain an efficient index in a dynamic environment and their timing. In experiments, indexes tuned for average recall incurred more per-query failures under distribution shift and updates, whereas our method was faster than standard exhaustive search with the same guarantee in all settings of static indexes up to 100M scale, and the failure rate was at most $\delta$ in all search steps of the official streaming runbook and of updates at 100M scale. For database operators that take the search results as input, the exact-search guarantee carries over to a per-query guarantee on the correctness of the answer: indexes tuned for average recall changed the answer of a tag aggregate for up to 27\% of queries, whereas PQR-Search changed it in at most 0.04\% of trials. Under updates, the total cost was higher or lower than that of FAISS depending on the ratio of queries to updates. Future work includes guarantee forms that exploit query logs and a GPU implementation.

\bigskip

\section*{Acknowledgment}

\textit{AI-use disclosure.} The authors conceived the ideas and concepts of this work and planned the research. Anthropic's Claude, used through Claude Code, was used at a substantial level throughout: under the authors' direction, it drafted, edited, and checked the text of all sections, including the abstract and figure captions, and it wrote the code, including the scripts that draw the figures, and carried out the data processing and experiments. The reference list combines entries collected by the authors with additional entries found by the assistant, and all entries were verified. OpenAI's Codex was used to check and review the manuscript and the code; it did not generate content. The authors reviewed all AI-assisted content and take responsibility for its correctness and originality.

\section*{References}

\printbibliography[heading=none]

\clearpage
\appendix
\section{Proofs}\label{app:proofs}

To state the expressions of the main text rigorously, let $\mathcal G$ be the $\sigma$-algebra generated by the data, the queries, and everything determined without looking at $P$ (in the dynamic setting, $\mathcal G_t$, generated by the index state and the queries at time $t$). $P$ follows the Haar distribution on $O(D)$ and is independent of $\mathcal G$. The decisions of Section~\ref{sec:devices} on enabling techniques may depend on $P$, but their failure events are contained in a path-independent union, so the bounds below remain unchanged. In the main text, $\Pr[\cdot]$ and $\mathbb E[\cdot]$ denote $\Pr[\cdot \mid \mathcal G]$ and $\mathbb E[\cdot \mid \mathcal G]$, all guarantees of the main text, being statements about these conditional probabilities, hold almost surely, and “determined independently of $P$” means $\mathcal G$-measurable. Appendix~\ref{app:proofstatic} proves the static guarantees, Appendix~\ref{app:proofdynamic} the dynamic guarantee, and Appendix~\ref{app:proofpass} the pass-rate formula and the soundness of the implementation.

\subsection{Proofs of the Static Guarantees}\label{app:proofstatic}

We first show the distributional fact underlying Lemma 1. If $\boldsymbol{Y}$ is uniformly distributed on $S^{n-1}$, then $Y_1^2 \sim \mathrm{Beta}(\tfrac12, \tfrac{n-1}2)$ and its sign is independent of $\lvert Y_1\rvert$. This follows directly from the facts that $\boldsymbol{Y}$ has the same distribution as a normalized standard Gaussian vector $\boldsymbol{g}$ and that $g_1^2$ and $\sum_{i \ge 2}g_i^2$ are independent chi-square variables. Next, for fixed unit vectors $\boldsymbol{o}$, $\boldsymbol{q}$, let $\rho = \langle \boldsymbol{o}, \boldsymbol{q}\rangle$, $\tilde{\boldsymbol{x}} = P^{-1}\boldsymbol{o}$, and $\tilde{\boldsymbol{y}} = P^{-1}\boldsymbol{q}$, and, when $\lvert\rho\rvert < 1$, define $\boldsymbol{e} = (\tilde{\boldsymbol{y}} - \rho\tilde{\boldsymbol{x}})/\sqrt{1-\rho^2}$. For a fixed orthogonal matrix $H$, $PH^\top$ is also Haar, so $(H\tilde{\boldsymbol{x}}, H\tilde{\boldsymbol{y}})$ and $(\tilde{\boldsymbol{x}}, \tilde{\boldsymbol{y}})$ have the same distribution. This distribution lies on the set of pairs of unit vectors with inner product $\rho$, and $O(D)$ acts transitively on this set, so the invariant probability measure on it is unique. On the other hand, the pair constructed by drawing $\tilde{\boldsymbol{x}}$ uniformly and taking as $\boldsymbol{e}$ the normalized projection of an independent Gaussian vector onto $\tilde{\boldsymbol{x}}^\perp$ is also invariant, so the two distributions are equal. Hence, given $\tilde{\boldsymbol{x}}$, $\boldsymbol{e}$ is uniformly distributed on the unit sphere of $\tilde{\boldsymbol{x}}^\perp$, and if $D \ge 3$, then for any unit vector $\boldsymbol{u} \in \tilde{\boldsymbol{x}}^\perp$ that is a measurable function of $\tilde{\boldsymbol{x}}$, $Z = \langle \boldsymbol{u}, \boldsymbol{e}\rangle$ is independent of $\tilde{\boldsymbol{x}}$ and $Z^2 \sim \mathrm{Beta}(\tfrac12, \tfrac{D-2}2)$. RaBitQ~\cite{r62} states that $Z$ is independent of $\langle\bar{\boldsymbol{o}}, \boldsymbol{o}\rangle$ and follows the same distribution, and the proof in its technical report (arXiv:2405.12497, Appendix B.3) conditions on $\tilde{\boldsymbol{x}}$; the conclusion above states this independence from all of $\tilde{\boldsymbol{x}}$ for any $\boldsymbol{u}$ that is a function of $\tilde{\boldsymbol{x}}$, which covers prefix codes and 2-bit codes. Since the distribution of $Z$ is symmetric about 0, the cumulative distribution function of $\zeta = \sqrt{D-1}\,Z$ is, for $s \ge 0$, $G_D(-s) = \Pr[Z < -s/\sqrt{D-1}] = \tfrac12\Pr[Z^2 > s^2/(D-1)]$, which is Eq.~\eqref{eq:gd}.

Lemma 1 follows immediately from this. Decompose the code vector $\bar{\boldsymbol{x}}$ in the rotated coordinates as $\bar{\boldsymbol{x}} = A\tilde{\boldsymbol{x}} + \boldsymbol{v}_\perp$ ($A = \langle\bar{\boldsymbol{x}}, \tilde{\boldsymbol{x}}\rangle$, $\boldsymbol{v}_\perp \perp \tilde{\boldsymbol{x}}$) and substitute $\tilde{\boldsymbol{y}} = \rho\tilde{\boldsymbol{x}} + \sqrt{1-\rho^2}\,\boldsymbol{e}$; then $\langle\bar{\boldsymbol{x}}, \tilde{\boldsymbol{y}}\rangle = \rho A + \sqrt{1-\rho^2}\langle \boldsymbol{v}_\perp, \boldsymbol{e}\rangle$. Applying the above fact to $\boldsymbol{u} = \boldsymbol{v}_\perp/\lVert \boldsymbol{v}_\perp\rVert$ gives $\langle \boldsymbol{v}_\perp, \boldsymbol{e}\rangle = \lVert \boldsymbol{v}_\perp\rVert Z$. Let $\nu_\ell = \lVert\bar{\boldsymbol{x}}^{(\ell)}\rVert^2$ be the squared norm of the code vector (a 1-bit prefix-truncated code keeps only $m_\ell$ coordinates of magnitude $1/\sqrt D$, so $\nu_\ell = m_\ell/D$); since the orthogonal transformation preserves inner products, $A = \langle\bar{\boldsymbol{x}}^{(\ell)}, P^{-1}\boldsymbol{o}\rangle = \langle\bar{\boldsymbol{o}}^{(\ell)}, \boldsymbol{o}\rangle$. Hence $\lVert\boldsymbol{v}_\perp\rVert = \sqrt{\nu_\ell - A^2}$, and letting $\zeta = \sqrt{D-1}\,Z$ and dividing both sides by $A$ yields $\hat\rho^{(\ell)} - \rho = \sqrt{1-\rho^2}\,\kappa^{(\ell)}\zeta^{(\ell)}$, i.e., Eq.~\eqref{eq:err}. This argument holds as long as the code is a function only of $\tilde{\boldsymbol{x}}$ and of randomness independent of $P$, so it applies equally to 2-bit codes.

Part (a) of Theorem 1 is shown by cases. If the residual is 0 or the normalization constant of the code is at most 0, the rule passes the vector, so the left-hand side is the empty set. If $\Phi = 0$, the rule prunes a vector only when $x > r$, and since the distance is then $x$, a pruned vector lies outside the radius. If $\rho = \pm1$, the estimate is exact, so a pruned vector has $\rho < \theta(r) - f\kappa\varepsilon_0 \le \theta(r)$ (since $\varepsilon_0 \ge 0$), i.e., lies outside the radius. In the remaining case ($\lvert\rho\rvert < 1$), if a vector is pruned and lies within the radius, then $\rho \ge \theta(r)$ together with Eq.~\eqref{eq:rule} and Eq.~\eqref{eq:err} gives $\sqrt{1-\rho^2}\,\kappa\,\zeta < \theta(r) - \rho - f\kappa\varepsilon_0 \le -f\kappa\varepsilon_0$. If $\kappa = 0$ or $\theta(r) \ge 1$, this is a contradiction; otherwise $\zeta < 0$ (the right-hand side is at most 0 because $\varepsilon_0 \ge 0$) and $\sqrt{1-\rho^2} \le f$ (if $\theta \le 0$, then $\sqrt{1-\rho^2} \le 1 \le f$; if $\theta > 0$, then $\sqrt{1-\rho^2} \le \sqrt{1-\theta^2} \le f$). Therefore $f\zeta \le \sqrt{1-\rho^2}\,\zeta < -f\varepsilon_0$, and $\zeta < -\varepsilon_0$. Equivalently, if $\zeta \ge -\varepsilon_0$, then $\rho = \hat\rho - \sqrt{1-\rho^2}\,\kappa\zeta \le \hat\rho + \sqrt{1-\rho^2}\,\kappa\varepsilon_0 \le \hat\rho + f\kappa\varepsilon_0$, where the last step is $(f - \sqrt{1-\rho^2})\,\kappa\varepsilon_0 \ge 0$ and uses $\varepsilon_0 \ge 0$. This assumption ($\delta_\ell \le 1/2$) is needed: if $\varepsilon_0 < 0$, a vector on the boundary ($\rho = \theta(r)$, $\kappa > 0$) with $f > \sqrt{1-\rho^2}$ is pruned whenever $\zeta < f\lvert\varepsilon_0\rvert/\sqrt{1-\rho^2}$, a threshold above $-\varepsilon_0 = \lvert\varepsilon_0\rvert$, so (a) and (b) fail. Part (b) follows from the facts that the conditional distribution of $\zeta$ is fixed and that $\varepsilon_0$ is $\mathcal G$-measurable, and part (c) from the union bound obtained by applying (a) at each stage.

We first fix the treatment of ties. Let $\mathcal K$ be the family of valid $k$-nearest sets obtainable by interchanging vectors at equal distance; then the recall in the main text is, precisely, $\mathrm{Recall}(q) = \max_{S \in \mathcal K}\lvert\mathcal N_k(q) \cap S\rvert/k$, and without ties $\mathcal K = \{\mathcal N_k^*(q)\}$, so it coincides with Eq.~\eqref{eq:recall}. Lemma 2 is shown as follows. All candidates closer than $d^* = d_k$ enter the result, and the remaining slots are filled with candidates at distance $d^*$. Take a valid set $S \in \mathcal K$ that contains as many of the tied vectors of the result as possible. Then every true neighbor at distance less than $d^*$ enters the result if it is a candidate, and as many slots at distance $d^*$ can be put into $S$ as the result fills, so $\lvert\mathcal N_k(q) \cap S\rvert \ge \lvert\mathcal N_k^*(q) \cap C\rvert$; dividing this by $k$ gives Lemma 2. In Theorem 2, the visited vectors are a subset of all vectors, so $r \ge d_k$, and if no unvisited true neighbor is pruned, these neighbors pass the triangle-inequality range and all stages and enter the candidates, so by Lemma 2 the result is exact. Therefore a failure is contained in the event that some unvisited true neighbor is pruned at some stage, and Theorem 1 and the union bound give Eq.~\eqref{eq:union}.

Theorem 5 is proved as follows. Suppose $v \in V$ is pruned at stage $\ell$ at the radius $r_t$ of some point in time. By condition (i), $r_t$ is the maximum of the distances of $k$ distinct actual vectors, so $r_t \ge d_k \ge d_v$, and $v$ lies within that radius. Theorem 1 (a) holds whatever the radius is, i.e., even if $r_t$ is a function of $P$, so $\zeta_v^{(\ell)} < -\varepsilon_{0,\ell}$. The triangle-inequality range does not exclude $v$ either, since $\lvert x_v - \Phi\rvert \le d_v \le r_t$. The event $\{\zeta_v^{(\ell)} < -\varepsilon_{0,\ell}\}$ is defined only by $(\boldsymbol{o}_v, \boldsymbol{q}, P)$ and $\varepsilon_{0,\ell}$ and does not use the radius, the test order, or the choice of vectors whose exact distances are computed first; so even if the same vector is tested twice at two radii, the event is a single one, and at every sample point the failure event is contained in $\bigcup_{v \in V}\bigcup_{\ell}\{\zeta_v^{(\ell)} < -\varepsilon_{0,\ell}\}$. $V$ is determined by the visited vectors that set the initial radius and is therefore $\mathcal G$-measurable, and by C1 $\varepsilon_{0,\ell}$ is also $\mathcal G$-measurable, so by Theorem 1 (b) and the union bound the conditional probability of this union is at most $\lvert V\rvert\sum_\ell G_D(-\varepsilon_{0,\ell})$. If no vector of $V$ is pruned, exact distances are computed for all of them and they enter the candidates by C4; the visited true neighbors are also among the candidates, so by Lemma 2 the result is exact. This is Theorem 2. Theorems 3 and 4 and Corollary 1 use only the bound on the expectation of $M \le \sum_{v \in V}\mathbb 1[\bigcup_\ell\{\zeta_v^{(\ell)} < -\varepsilon_{0,\ell}\}]$ and Lemma 2, so they hold as they are. Regarding re-testing, when the same vector is tested again with the same code at a smaller radius $r' \le r$, $\theta(r) - f\kappa\varepsilon_0$ is decreasing in $r$ ($\theta$ decreases in $r$, the $f$ of the implementation increases in $r$, and $\varepsilon_0 \ge 0$), so a vector pruned at $r$ is also pruned at $r'$. The word “distinct” in condition (i) is necessary, because if the same vector enters the candidates twice, the $k$-th element of the multiset of distances can be smaller than $d_k$.

Theorem 3 follows from $\mathbb E[M] \le \lvert V\rvert\beta \le k\beta = \delta(m+1)$ and Markov's inequality. When all that is known is that each neighbor fails with probability $p$, the coupling that selects one subset of size $m+1$ with probability $kp/(m+1)$ and makes only its elements fail attains this bound, so the bound cannot be improved using the marginal probabilities alone. Proposition 1 below shows that an actual data configuration also comes close to this bound. The tightness is meant only in this sense: an analysis that uses only the per-neighbor probabilities cannot give a smaller bound, and actual data attain the bound exactly only with $k = m + 1$ ($G = 1$, $m + 1$ duplicates in one unvisited partition) and otherwise within a second-order term. It does not exclude a smaller failure probability for the algorithm on other data or a sharper analysis that uses the dependence among neighbors or stages, and it is separate from the cost lower bounds of Appendix~\ref{app:lowerbound}. Theorem 4 applies linearity of expectation to $\mathrm{Recall} \ge 1 - M/k$. Corollary 1 follows from the fact that the proof of Theorem 3 bounds $\mathbb E[M] \le \lvert V\rvert\beta \le \delta(m+1)$ before applying Markov's inequality. By Lemma 2, $\mathbb E[\mathrm{Recall}] \ge 1 - \delta(m+1)/k$, so if this value is at least $\tau$, i.e., if $\delta(m+1) \le (1-\tau)k$, then $\mathbb E[\mathrm{Recall}] \ge \tau$. The proofs of Theorems 5 and 6 reproduce this expectation-bound step as it is, so the corollary also holds under the adaptive radius and under updates.

\begin{thmblock}{\textbf{Proposition 1} (Tightness of Theorem 3):}Suppose there is a single test stage ($L = 1$), the per-vector value $\sqrt{1-\max(\theta(r),0)^2}$ is used as the slack factor, $\rho > 0$, $k = G(m+1)$, and $D \ge \max(2G, 8)$. In each of $G$ unvisited partitions, place $m+1$ identical vectors at distance $r - \eta$ from the query, such that in partition $g$ the unit vectors of the data and the query normalized with respect to the centroid are $\boldsymbol{o}_g$ and $\boldsymbol{q}_g = \rho\,\boldsymbol{o}_g + \sqrt{1-\rho^2}\,\boldsymbol{e}_g$, with the $2G$ vectors $\boldsymbol{o}_g$, $\boldsymbol{e}_g$ orthonormal, and place $k$ vectors at distance $r$ in the visited partitions. Let $p = G_D(-\varepsilon_0)$ (equal to the failure probability budget $\beta$, since there is a single stage). Then, as $\eta \to 0$, the limit of $\Pr[\mathrm{Recall}(q) < \tau]$ is at least $Gp - \binom G2\lambda^*p^2$, where $\lambda^*$ is a constant determined only by $D$ and $\varepsilon_0$. In particular, if $G = 1$ ($k = m + 1$), this limit equals the bound $kp/(m+1) = p$ of Theorem 3.\end{thmblock}

Such a configuration can be constructed, for example, by placing the query at the origin and the centroid of partition $g$ at $-\Phi\,\boldsymbol{q}_g$. The proof is as follows. Identical vectors in the same partition have the same $\boldsymbol{o}$ and $\boldsymbol{q}$, and hence the same error $\zeta$ and the same test outcome; so if the event $D_g$ that group $g$ is pruned occurs, $m+1$ true neighbors are missed at once and recall falls below $\tau$. Hence the failure event contains $\bigcup_g D_g$. As $\eta \to 0$, the margin tends to 0, $\theta(r) \to \rho$, and the slack factor tends to $\sqrt{1-\rho^2}$, so $D_g$ converges monotonically to $\{\zeta_g < -\varepsilon_0\}$ and $\Pr[D_g] \to p$. If $G = 1$, this completes the proof. If $G \ge 2$, the probability of the union is at least the sum of the probabilities of the individual events minus the sum of the probabilities of the pairwise intersections ($\Pr[\bigcup_g D_g] \ge \sum_g\Pr[D_g] - \sum_{g<h}\Pr[D_g \cap D_h]$), so it suffices to show $\Pr[D_g \cap D_h] \le \lambda^*p^2$. Since $\boldsymbol{o}_g$, $\boldsymbol{e}_g$, $\boldsymbol{o}_h$, $\boldsymbol{e}_h$ are orthonormal, given $P^{-1}\boldsymbol{o}$ of the two groups, the two error directions form an orthonormal pair drawn uniformly in the remaining $D-2$ dimensions, and $\zeta_g$ and $\zeta_h$ are functions of the $2 \times 2$ block of this pair. Applying $\det(I - B^\top B) \le e^{-\lVert B\rVert_F^2}$ to the formula of Diaconis, Eaton, and Lauritzen for the block density of a uniform orthogonal matrix (summarized in Jiang~\cite{r178}), the density of this block is at most $\kappa_J$ times the density of a Gaussian block whose entries are independent with variance $1/(D-7)$. In the Gaussian block the two errors are independent and each has tail probability at most $\bar\Phi(\varepsilon_0\sqrt{(D-7)/(D-1)})$ (where $\bar\Phi$ is the tail probability of the standard normal distribution), so one may take $\lambda^* = \kappa_J\,[\bar\Phi(\varepsilon_0\sqrt{(D-7)/(D-1)})/p]^2$. $\kappa_J$ is determined by the normalizing constant of that density formula and equals 1.042, 1.020, and 1.007 for $D = 128, 256, 768$. In a Monte Carlo experiment ($k = 4$, $m = 1$, $G = 2$), the failure rate $\div$ the bound of Theorem 3 was 1.006 at $D = 128$ and 0.990 at $D = 768$, i.e., 1 within sampling error.

\subsection{Proof of the Dynamic Guarantee}\label{app:proofdynamic}

In part (a) of Theorem 6, the update sequence is given, and by C5 the state at time $t$ (live vectors, centroids, assignments, segments) is independent of the rotation; since the rotation was drawn independently of this state, it is Haar conditionally on $\mathcal G_t$. The visited set and the error-bound parameters are functions of exact distances and residual norms and are therefore $\mathcal G_t$-measurable, and by C7 the radius is the $k$-th of the exact distances of distinct live vectors, so it satisfies condition (i) of Theorem 5 on $L_t$. Unvisited true neighbors are live, so by C7 they are not caught by the bitmap and are processed by the test path of their own segment, and by C6 the premise of Lemma 1 holds. The pruning event of each vector is defined by the code of its own segment and the current centroid, so the proof of Theorem 5 is repeated verbatim on $L_t$. The terms of the union and of Markov's inequality are only the $k$ vectors of $\mathcal N_{k,t}^*(q)$, so the failure probability $\delta$ need not be split. In (b), $\zeta$ is a function only of $(\boldsymbol{o}, \boldsymbol{q}, P)$, so if the centroid and the deterministic slack are unchanged between two times and $r_t \le r_s$, then $\theta(r) - f\kappa\varepsilon_0$ at time $t$ is greater than or equal to that at time $s$ ($\varepsilon_0 \ge 0$ is constant, $\theta$ decreases in $r$, and $f$ increases in $r$). Hence a neighbor pruned at time $s$ is also pruned at time $t$. In (c), the rotation schedule is part of the update sequence and a fresh rotation is independent of the past, so it is Haar conditionally on the past, and events at two times that use different rotations are conditionally independent.

\subsection{Pass-Rate Formula and Soundness of the Implementation}\label{app:proofpass}

Substituting Eq.~\eqref{eq:gap} into Eq.~\eqref{eq:passprob}, when the numerator is nonnegative, $\sqrt{1-\rho^2} \le 1$ gives, for fixed $f$ and $\sigma_{\det}$, a decreasing upper bound that depends only on the margin.

The soundness of the implementation reduces to the argument of Section~\ref{sec:process}. If the accumulated value used in the test is at least the exact value, and the approximation of the right-hand side together with the deterministic slack is at most the exact right-hand side for every rotation, then every vector pruned by the implementation is also pruned by the exact rule. The slack may depend on the rotation. This argument applies to the 8-bit quantization of nibble LUTs, the LUTs and accumulation of later stages, the reconstruction of 8-bit metadata (minimum rounded down, step rounded up), the fp32 error of per-vector slack factors, the fp32 addition in the 2-bit final stage, passing all vectors when the radius is infinite, and passing vectors with zero residual without testing. The implementation used in the experiments also covers, in the same way, the errors of the query rotation, of the distance $\Phi$ to the centroid, of the exact distances entering the radius, and of the floating-point computation of the right-hand side. For the rotation and data-side values, the values computed in fp64 at index construction serve as the reference. The query rotation is computed with an fp32 matrix and fp64 accumulation, and per-coordinate bounds determined from the fp32 rounding of the matrix and the accumulation error are subtracted from the query term of the test. For $\Phi$, an upper bound on its relative error is added to the slack, and partitions for which this bound exceeds $4 \times 10^{-6}$ are recomputed with fp64 centroids. Distances entering the radius are enlarged by an upper bound on their fp32 computation error. The remaining premises are these fp64 reference computations and the given fp32 query coordinates. The time added by this treatment was within 2\% in geometric mean on six 1M datasets for 1-thread and 32-thread latency and for batch throughput alike (up to 17\% at 32 threads).

\bigskip

\section{Lower Bounds for Per-Query Guarantees}\label{app:lowerbound}

This appendix proves three lower bounds on the cost of per-query guarantees. Theorems 7 and 8 are lower bounds within the probe model below, and Theorem 9 is a lower bound in the word-RAM model under SETH. Throughout, $\tau \in (0, 1]$, $\delta \in [0, 1)$, and $m = \lfloor(1-\tau)k\rfloor$. Since $\tau > 0$, we have $m + 1 \le k$, and, as in Section~\ref{sec:problem}, missing $m+1$ or more true neighbors implies $\mathrm{Recall}(q) \le (k-m-1)/k < \tau$.

\textbf{Probe model.} Consider a probe model in which the algorithm knows the public information (the query, the candidates, and the public structure of the index) at no cost, and learns the content of a hidden cell only by probing that cell. The cost is the number of probed cells. Precisely, an instance consists of public information $J$ and the contents $y_1, \dots, y_{n'}$ of $n'$ hidden cells, and $J$ contains the query and the vectors and ids of the candidate set $C$. The algorithm uses internal randomness $\omega$ that is independent of the instance (the orthogonal transformation $P$ for the index of the main text); the record of cell $i$ is a function of $y_i$ and $\omega$ only, and reading this record is a probe of cell $i$. The next cell to probe may be chosen adaptively from $J$, $\omega$, and the records read so far. The output consists of $k$ items chosen from $C$ and the probed cells; a cell that is output without having been probed is counted as probed. Let $Q$ be the set of probed hidden cells and $T = \lvert Q\rvert$. The IVF index of the main text belongs to this model as long as its public structure (centroids, assignments, residual norms) is fixed independently of the contents of the hidden cells; the hidden content is then the direction of the residual, and the record is its code. Structures whose access paths change with the contents, such as graphs, LSH, and trees, lie outside this model.

\textbf{Hard family of instances.} Fix the public information $J$, and let $C$ consist of $k$ vectors whose distances from the query are distinct and lie in $[1, 2]$. For each cell, fix a near content at distance less than $1/2$ from the query and a far content at distance greater than 3, with distinct distances for the near contents of different cells. For a set of cells $B \subseteq \{1, \dots, n'\}$, let $I_B$ be the instance in which the cells in $B$ have their near contents and the other cells their far contents. All $I_B$ share the same $J$, and in $I_\varnothing$ the candidates are the correct answer. If $\lvert B\rvert \le k$, all distances are distinct, so the true $k$-NN set is uniquely $B$ together with the $k - \lvert B\rvert$ nearest vectors of $C$. Since the output lies in $C \cup Q$, we have $k(1 - \mathrm{Recall}(q)) \ge \lvert B \setminus Q\rvert$. This family can be realized within an IVF index. For example, for a vector with residual norm 2 in a partition whose centroid is at distance 2 from the query, changing only the direction of the residual, with the centroid and the residual norm unchanged, makes its distance to the query any value between 0 and 4.

\textbf{Coupling.} Fix $\omega$, and let $Q_\omega$ be the set of cells probed by the run on $I_\varnothing$ and $T_\varnothing = \lvert Q_\omega\rvert$. If $Q_\omega \cap B = \varnothing$, the run on $I_B$ is identical to the run on $I_\varnothing$. Indeed, the two instances differ only in the cells of $B$, so two runs that start from the same $J$ and $\omega$ read the same records and choose the same cells until a cell of $B$ is probed for the first time (by induction on the number of probes); since the run on $I_\varnothing$ probes no cell of $B$, the two runs coincide to the end and have the same output. Hence, in this case, the output on $I_B$ contains no cell of $B$, and $k(1 - \mathrm{Recall}(q)) \ge \lvert B\rvert$ if $\lvert B\rvert \le k$.

\begin{thmblock}{\textbf{Lemma 3} (Hypergeometric Avoidance Probability):}For integers $N \ge 1$ and $1 \le u \le N$, the probability that $u$ cells chosen uniformly at random among $N$ cells all avoid a fixed set of $t$ cells is, for $t = 0, 1, \dots, N$, \[\mathrm{Hg}_N(u, t) = \binom{N-t}{u}\Big/\binom{N}{u} = \prod_{j=0}^{u-1}\Bigl(1 - \frac{t}{N-j}\Bigr).\] The function obtained by extending this product to real $t \in [0, N-u+1]$ and setting it to 0 for $t \in [N-u+1, N]$ (also written $\mathrm{Hg}_N(u, t)$) is continuous, nonincreasing, and convex on $[0, N]$. Moreover, $\mathrm{Hg}_N(u, t) \ge \bigl(1 - t/(N-u+1)\bigr)^u$ for $0 \le t \le N-u+1$.\end{thmblock}

\textit{Proof.} Draw the $u$ cells one at a time. Given that the first $j$ cells avoided the fixed $t$ cells, the $(j+1)$-th cell also avoids them with conditional probability $(N-t-j)/(N-j)$, and the product of these is the product form. For an integer $t \ge N-u+1$, the factor with $j = N-t \le u-1$ is 0, so the extended function agrees with the probability at integer points. On $[0, N-u+1]$, every factor $f_j(t) = 1 - t/(N-j)$ is linear, decreasing, and nonnegative, so the product is nonincreasing, and it is convex because $(\prod_j f_j)'' = \sum_{i \ne j}f_i'f_j'\prod_{l \ne i, j}f_l \ge 0$. At $t_0 = N-u+1$ we have $f_{u-1}(t_0) = 0$, so the function is continuous; its left derivative is at most 0 and its right derivative is 0, so the function obtained by joining the two intervals is also convex. The lower bound follows from $f_j(t) \ge 1 - t/(N-u+1) \ge 0$ for $j \le u-1$. $\blacksquare$

\begin{thmblock}{\textbf{Theorem 7} (Cost Lower Bound in the Probe Model):}In the probe model, let $n' \ge m+1$, and suppose that an algorithm gives the per-query $(\tau, \delta)$-recall guarantee $\Pr_\omega[\mathrm{Recall}(q) < \tau] \le \delta$ on $I_B$ for every $B$ of size $m+1$. Then, even on the instance $I_\varnothing$ in which the candidates are the correct answer, the expected number of probes satisfies $\mathrm{Hg}_{n'}(m+1, \mathbb E\,T_\varnothing) \le \delta$, and hence $\mathbb E\,T_\varnothing \ge (n'-m)\bigl(1 - \delta^{1/(m+1)}\bigr)$.\end{thmblock}

\textit{Proof.} We apply Yao's principle~\cite{r174} to an adversarial distribution that chooses the defect configuration at random: draw $B$ uniformly from the sets of $m+1$ cells, independently of $\omega$. Since an average is at most the maximum, $\delta \ge \mathbb E_B\Pr_\omega[\mathrm{Recall}(q) < \tau \text{ on } I_B]$. By the coupling, the failure event on $I_B$ contains $\{Q_\omega \cap B = \varnothing\}$. Given $\omega$, $Q_\omega$ is a fixed set of size $T_\varnothing$, so $\Pr_B[Q_\omega \cap B = \varnothing] = \mathrm{Hg}_{n'}(m+1, T_\varnothing)$. Exchanging the order of averaging and applying Jensen's inequality with the convexity in Lemma 3, \[\delta \ge \mathbb E_\omega\bigl[\mathrm{Hg}_{n'}(m+1, T_\varnothing)\bigr] \ge \mathrm{Hg}_{n'}(m+1, \mathbb E\,T_\varnothing).\] If $\mathbb E\,T_\varnothing \ge n'-m$, the final inequality is trivial. Otherwise, the lower bound of Lemma 3 gives $\bigl(1 - \mathbb E\,T_\varnothing/(n'-m)\bigr)^{m+1} \le \delta$, which yields the claim. $\blacksquare$

For example, for $k = 100$, $\tau = 0.9$, and $\delta = 0.01$, we have $m = 10$ and $1 - 0.01^{1/11} = 0.3421$, so even on the instance in which the candidates are correct, $\mathbb E\,T_\varnothing \ge 0.3421(n'-10)$. That is, when there are many hidden cells, about 34.2\% of them must be probed (the fraction is at least 0.342 once $n' \ge 51{,}227$). Thus, in this model, $\tau$, $\delta$, and $k$ enter the cost as the probed fraction $1 - \delta^{1/(m+1)} = 1 - e^{-\ln(1/\delta)/(m+1)}$, not through the logarithm in the error-bound parameter. On this family, the condition of Theorem 7 cannot be improved: an algorithm that probes $t$ cells chosen uniformly at random, with $\mathrm{Hg}_{n'}(m+1, t) \le \delta$, and outputs the $k$ nearest among $C$ and the probed cells fails on $I_B$ with $\lvert B\rvert = m+1$ with probability exactly $\mathrm{Hg}_{n'}(m+1, t)$, because once it probes any cell of $B$, all probed cells of $B$ enter the output and at most $m$ neighbors are missed.

\begin{thmblock}{\textbf{Theorem 8} (Deterministic Skipping):}Under the assumptions of Theorem 7, for every set $Z$ of $m+1$ cells, $\Pr_\omega[Q_\omega \cap Z \ne \varnothing] \ge 1-\delta$ on $I_\varnothing$. In particular, a deterministic algorithm can skip at most $m$ hidden cells, even on the instance $I_\varnothing$ in which the candidates are correct.\end{thmblock}

\textit{Proof.} Let $B = Z$. By the coupling, if $Q_\omega \cap Z = \varnothing$, then $m+1$ or more neighbors are missed on $I_Z$ and $\mathrm{Recall}(q) < \tau$, so $\Pr_\omega[Q_\omega \cap Z = \varnothing] \le \delta$. For a deterministic algorithm, $Q_\omega$ is a single fixed set; if $m+1$ or more hidden cells were left unprobed, taking $m+1$ of them as $Z$ would give $\Pr_\omega[Q_\omega \cap Z = \varnothing] = 1 > \delta$, a contradiction. $\blacksquare$

If $m = 0$ (e.g., $\tau = 1$), a deterministic rule must probe all hidden cells. Hence the statement “certification without assumptions must read all vectors” is true only for deterministic rules. A randomized rule only needs to hit every set of $m+1$ cells with probability at least $1-\delta$, and the cost of doing so is the lower bound of Theorem 7.

\begin{thmblock}{\textbf{Theorem 9} (Lower Bound Conditional on SETH):}Fix $k \ge 1$, $\tau \in (0, 1]$, and $\delta \in [0, 1)$, and assume randomized SETH~\cite{r106,r108}. Then for every $c \ge 1$ and $\varepsilon \in (0, 1]$ there is a constant $\gamma$, determined by $c$ and $\varepsilon$ alone, such that for any dimension $D(n)$ (a function computable in time polynomial in $n$) with $\gamma\log_2 n \le D(n) \le n^c$ for all sufficiently large $n$, there is no word-RAM data structure that preprocesses $n$ vectors with distinct ids (duplicates of the same vector allowed) in time $O(n^c)$, returns $k$ ids per query in time $O(n^{1-\varepsilon})$, and, without assumptions on the data or queries, gives for every query (i) the $(\tau, \delta)$-recall guarantee $\Pr[\mathrm{Recall}(q) < \tau] \le \delta$ or (ii) the expected-recall guarantee $\mathbb E[\mathrm{Recall}(q)] \ge \tau$. Here the probabilities are over the internal randomness of the data structure, which is independent of the data and queries. Consequently, when $D = \omega(\log n)$ and $D$ is at most polynomial in $n$, no data structure with polynomial-time preprocessing that gives such guarantees can achieve query time $O(n^{1-\varepsilon})$.\end{thmblock}

Randomized SETH is the following hypothesis: for every $\varepsilon' > 0$ there is an integer $w$ such that no randomized algorithm (with error probability at most 1/3) decides the satisfiability of $w$-CNF formulas with $n_v$ variables in time $O(2^{(1-\varepsilon')n_v})$~\cite{r106,r108}. The proof also uses the sparsification lemma~\cite{r107}: for every $w$ and $\varepsilon'' > 0$ there is a constant $K$ such that a $w$-CNF $\varphi$ with $n_v$ variables can be written, in time $2^{\varepsilon'' n_v}\mathrm{poly}(n_v)$, as a disjunction of $w$-CNFs $\psi_1, \dots, \psi_s$ ($s \le 2^{\varepsilon'' n_v}$) over the same variables, each with at most $Kn_v$ clauses. The proof is the split-and-list reduction to the orthogonal vectors problem, whose data-structure version is given in~\cite{r111}.

\textit{Proof of Theorem 9.} Let $c' = c+1$, $\alpha = 1/(2c')$, and $\varepsilon_3 = \varepsilon\alpha$. Apply SETH with $\varepsilon' = \varepsilon_3/4$ to obtain $w$, apply the sparsification lemma with $(w, \varepsilon_3/2)$ to obtain $K$, and set $\gamma = 6c'K$. Suppose such a data structure $\mathcal A$ exists. We decide the satisfiability of a $w$-CNF $\varphi$ as follows.

\begin{enumerate}
\item Using the sparsification lemma, rewrite $\varphi$ as $\psi_1, \dots, \psi_s$. The formula $\varphi$ is satisfiable if and only if some $\psi_i$ is. If some $\psi_i$ has no clauses, answer “satisfiable” immediately.
\item For $\psi = \psi_i$ with $\mu$ clauses ($1 \le \mu \le Kn_v$), split its variables into a set $V_1$ of $n_1 = \lceil\alpha n_v\rceil$ variables and the remaining set $V_2$. For each assignment $a$ of $V_1$, define $\boldsymbol{a} \in \{0, 1\}^\mu$ by $a_j = 1$ if no $V_1$-literal of clause $j$ is true under $a$, and $a_j = 0$ otherwise. Define $\boldsymbol{b}$ in the same way for each assignment $b$ of $V_2$. Clause $j$ is satisfied by $(a, b)$ if and only if $a_jb_j = 0$, so $\psi$ is satisfiable if and only if there is a pair with $\langle\boldsymbol{a}, \boldsymbol{b}\rangle = 0$.
\item Per coordinate, let $G(0) = (1, 0, 0)$, $G(1) = (0, 1, 0)$, $H(0) = (1, 1, 0)$, and $H(1) = (1, 0, 1)$. Then $\lVert G(a_j) - H(b_j)\rVert^2 = 1 + 2a_jb_j$, so the coordinatewise concatenations $G(\boldsymbol{a}), H(\boldsymbol{b}) \in \{0, 1\}^{3\mu}$ satisfy $\lVert G(\boldsymbol{a}) - H(\boldsymbol{b})\rVert^2 = \mu + 2\langle\boldsymbol{a}, \boldsymbol{b}\rangle$. That is, an orthogonal pair has the smallest possible distance $\sqrt\mu$, and every other pair has distance at least $\sqrt{\mu+2}$. Padding with zero coordinates to dimension $D(n)$ does not change distances.
\item The data consist of $n = k2^{n_1}$ vectors, $k$ copies of each $G(\boldsymbol{a})$ with distinct ids, and the queries are the $2^{n_v - n_1}$ vectors $H(\boldsymbol{b})$. Since $n \ge 2^{\alpha n_v}$, we have $3\mu \le 3Kn_v \le (3K/\alpha)\log_2 n = \gamma\log_2 n \le D(n)$. If some $\boldsymbol{a}$ is orthogonal to $\boldsymbol{b}$, there are at least $k$ vectors at distance $\sqrt\mu$, so $d_k = \sqrt\mu$ and every valid true $k$-NN set, obtained by exchanging vectors at equal distance, consists only of vectors at distance $\sqrt\mu$. Hence, if $\mathrm{Recall}(q) > 0$, the output contains a vector of an orthogonal pair. We check the $k$ outputs directly in $O(k\mu)$ time and answer “satisfiable” only when an orthogonal pair is found, so there are no false positives.
\item Build $r$ copies of $\mathcal A$ with mutually independent internal randomness and issue all queries to each. For a query that has an orthogonal pair, the probability that one data structure misses the pair is at most $\delta$ in case (i), and at most $1-\tau$ in case (ii), because $\mathrm{Recall}(q) \le 1$ gives $\Pr[\mathrm{Recall}(q) > 0] \ge \mathbb E[\mathrm{Recall}(q)] \ge \tau$. The queries are determined by $\varphi$ in advance and are independent of the internal randomness, so the per-query guarantee applies as is, and the $r$ outcomes are independent. With $\rho = \max(\delta, 1-\tau) < 1$ and $r = \lceil\ln 3/\ln(1/\rho)\rceil$ ($r = 1$ if $\rho = 0$), the algorithm answers “satisfiable” with probability at least $1 - \rho^r \ge 2/3$ when $\varphi$ is satisfiable, and always answers “unsatisfiable” otherwise. Here $r$ is a constant depending only on $\tau$ and $\delta$.
\item We count the time. For one $\psi$, building the vectors takes $(2^{n_1} + 2^{n_v - n_1})\mathrm{poly}(n_v)$, and writing the zero-padded input and building the $r$ data structures takes $O(r(nD(n) + n^c)) = O(n^{c'}) = O(k^{c'}2^{c'(\alpha n_v + 1)}) = O(2^{n_v/2})$. Each query vector is formed in $O(\mu)$ time by overwriting the first $3\mu$ coordinates of a zero-filled buffer, and the query phase takes $r2^{n_v - n_1}\bigl(O(n^{1-\varepsilon}) + O(k\mu)\bigr) \le 2^{n_v - \varepsilon n_1}\mathrm{poly}(n_v) \le 2^{(1-\varepsilon_3)n_v}\mathrm{poly}(n_v)$ (using $n^{1-\varepsilon} \le k2^{(1-\varepsilon)n_1}$). Since $\varepsilon_3 \le 1/4$, one $\psi$ takes $2^{(1-\varepsilon_3)n_v}\mathrm{poly}(n_v)$; adding up the $s \le 2^{\varepsilon_3 n_v/2}$ formulas and the sparsification time, the total is $2^{(1-\varepsilon_3/2)n_v}\mathrm{poly}(n_v) = O(2^{(1-\varepsilon')n_v})$.
\end{enumerate}

This is a randomized algorithm with one-sided error that solves $w$-CNF satisfiability in time $O(2^{(1-\varepsilon')n_v})$, contradicting randomized SETH. The last sentence of the theorem follows by taking $c$ larger than both the exponent of the preprocessing time and the degree of $D(n)$, so that the preprocessing is $O(n^c)$ and $D(n) \le n^c$, and noting that $D(n) = \omega(\log n)$ implies $D(n) \ge \gamma\log_2 n$ for all sufficiently large $n$. $\blacksquare$

Because the distances of the $k$ outputs can be checked in $O(kD)$ time, there are no false positives, so Theorem 9 holds even when $\delta$ is close to 1. This conclusion is in the word-RAM model, so unlike Theorems 7 and 8 it covers graphs, LSH, and trees alike. It is, however, an asymptotic statement about worst-case instances of binary vectors with $D = \Omega(\log n)$, and it says nothing about whether sublinear cost is possible on real data. Lower bounds of the same form are known for approximate nearest neighbor search~\cite{r112}; what we add is only the $k$-fold replication (step 4) and the one-sided amplification (step 5) that carry them over to recall guarantees.

\section{Supplementary Experiments and Discussion}\label{app:extra}

\subsection{Comparison with Other Baselines}\label{app:baselines}

Our own GEMM baseline, an exhaustive search based on matrix multiplication that we optimized ourselves, was 1.5–8.4$\times$ faster than FAISS IndexFlatL2 (default setting) in batch. Against this baseline, our method was 1.03–37$\times$ faster in batch in all settings except laion ($k = 100$, T1) (the smallest margin was 1.03$\times$ on SSNPP ($k = 100$, T1)), and on laion ($k = 100$, T1) this baseline was 1.2$\times$ faster. In the main text we excluded this baseline, following the principle of using well-known public software with standard settings. For the comparison with GPUs we used GpuIndexFlatL2 of FAISS~\cite{r23} (RTX 4090) (GPU times were measured in early October 2026). In single-query latency, our method with 32 threads was faster than the GPU in all 44 settings in which the GPU was measured (eight 1M datasets and three 10M datasets) (3.5–31.7$\times$ at 1M scale and at least 7.4$\times$ at 10M scale). With 1 thread, our latency was longer than the GPU's in four exact-search settings: SSNPP and SSNPP-10M ($k = 100$) and laion ($k = 10, 100$). Batch throughput was higher on the GPU when the data fit in GPU memory. The GPU led in 31 of the 32 settings at 1M scale, and the throughput ratio of our method to the GPU had a geometric mean of 0.44 and a minimum of 0.11. In contrast, on msmarco (fp32, 36 GB), which exceeds GPU memory, our method was 5.4–10.3$\times$ faster in batch. Since the 1-bit codes take only about 1.6 GB even for 100M vectors, we leave a GPU implementation of our method to future work.

\subsection{Numerical Results of the Main Speed Comparison}\label{app:speedup}

Table~\ref{tab:speedup} lists the values shown in Fig.~\ref{fig:speedup} of Section~\ref{sec:cost} together with the $n_{\rm probe}$ of IVF. The 32-thread speedups over IVF were measured with the setting that parallelizes over the partitions to be searched (Section~\ref{sec:setup}). The oracle $n_{\rm probe}$ is the minimum value at which, for every test query, the result equals that of FAISS IndexFlatL2 or is exact when ties are judged strictly, and the test-failure column is also counted with ties judged strictly.

\Needspace{16\baselineskip}
{\centering \textbf{Table \refstepcounter{table}\label{tab:speedup}\thetable{}. Speedup for exact search (T1, $\delta = 0.01$).}\par}\nopagebreak

{\footnotesize\setstretch{1.1}
\begin{xltabular}{\linewidth}{llYYlYYY}
\toprule
\textbf{Data} & \textbf{$k$} & \textbf{$\div$Flat 1-thread} & \textbf{$\div$Flat 32-thread} & \textbf{$\div$Flat batch} & \textbf{$\div$IVF oracle 1-thread · 32-thread · batch} & \textbf{IVF $n_{\rm probe}$ oracle / calibrated} & \textbf{Test failures of calibrated $n_{\rm probe}$} \\
\midrule
\endfirsthead
\multicolumn{8}{l}{\emph{Table~\thetable\ (continued)}} \\
\toprule
\textbf{Data} & \textbf{$k$} & \textbf{$\div$Flat 1-thread} & \textbf{$\div$Flat 32-thread} & \textbf{$\div$Flat batch} & \textbf{$\div$IVF oracle 1-thread · 32-thread · batch} & \textbf{IVF $n_{\rm probe}$ oracle / calibrated} & \textbf{Test failures of calibrated $n_{\rm probe}$} \\
\midrule
\endhead
SIFT & 10 & 38.1 & 39.8 & 59.1 & 8.2 · 4.8 · 35.0 & 192 / 192 & 0 \\
SIFT & 100 & 29.5 & 32.9 & 47.0 & 12.5 · 7.2 · 46.9 & 384 / 384 & 0 \\
YFCC & 10 & 29.9 & 62.5 & 39.8 & 6.1 · 6.0 · 28.7 & 192 / 128 & 2 \\
YFCC & 100 & 21.0 & 49.4 & 25.9 & 11.3 · 12.6 · 49.6 & 512 / 512 & 0 \\
agnews & 10 & 54.0 & 134 & 24.5 & 41.5 · 45.9 · 165 & 768 / 1000 & 5* \\
agnews & 100 & 37.1 & 62.8 & 14.7 & 27.8 · 22.3 · 109 & 768 / 1000 & 0 \\
arxiv & 10 & 59.5 & 162 & 34.9 & 17.8 · 22.0 · 77.8 & 384 / 128 & 17 \\
arxiv & 100 & 34.7 & 75.6 & 19.6 & 27.3 · 26.0 · 136 & 1024 / 1024 & 0 \\
GIST & 10 & 48.0 & 120 & 22.0 & 24.0 · 27.5 · 97.2 & 384 / 1000 & 0 \\
GIST & 100 & 32.4 & 54.7 & 13.8 & 30.3 · 21.6 · 115 & 768 / 1000 & 0 \\
yandex & 10 & 12.7 & 48.2 & 18.2 & 13.7 · 22.9 · 77.5 & 1000 / 768 & 1 \\
yandex & 100 & 8.5 & 26.3 & 11.5 & 9.3 · 13.2 · 49.0 & 1000 / 1000 & 1* \\
SSNPP & 10 & 10.2 & 34.9 & 18.2 & 10.3 · 15.7 · 76.7 & 1000 / 1000 & 0 \\
SSNPP & 100 & 3.3 & 13.8 & 8.5 & 3.4 · 6.5 · 36.1 & 1000 / 1000 & 0 \\
laion & 10 & 13.8 & 41.4 & 11.7 & 12.7 · 19.4 · 88.5 & 1000 / 1000 & 0 \\
laion & 100 & 7.1 & 20.5 & 5.8 & 6.5 · 9.5 · 36.9 & 1000 / 1000 & 0 \\
DEEP1M & 10 & 20.0 & 26.1 & 34.0 & 7.8 · 5.4 · 29.6 & 384 / 192 & 2 \\
DEEP1M & 100 & 14.0 & 19.2 & 19.4 & 11.3 · 7.8 · 36.9 & 768 / 512 & 1 \\
GloVe & 10 & 14.9 & 28.3 & 23.0 & 14.0 · 13.4 · 50.2 & 1024 / 768 & 8 \\
GloVe & 100 & 3.7 & 17.4 & 12.3 & 4.0 · 8.0 · 31.6 & 1183 / 1024 & 8 \\
YFCC-10M & 10 & 18.9 & 71.6 & 55.2 & 2.8 · 5.3 · 39.7 & 1536 / 256 & 12 \\
YFCC-10M & 100 & 16.4 & 64.8 & 37.9 & 5.1 · 9.3 · 53.5 & 3072 / 4096 & 0 \\
SSNPP-10M & 10 & 12.3 & 37.9 & 29.6 & 12.2 · 17.8 · 129 & 10000 / 10000 & 0 \\
SSNPP-10M & 100 & 6.3 & 21.9 & 15.8 & 6.6 · 10.2 · 70.4 & 10000 / 10000 & 0 \\
msmarco & 10 & 73.3 & 283 & 53.4 & 53.1 · 90.8 · 354 & 6144 / 1536 & 7 \\
msmarco & 100 & 50.6 & 165 & 40.9 & 46.9 · 69.2 · 328 & 8192 / 8192 & 0 \\
\bottomrule
\end{xltabular}}

* Test failures occurred even though the calibrated $n_{\rm probe}$ covered all partitions (agnews) or equaled the oracle (yandex). Since scanning all partitions is equivalent to exhaustive search, these failures arise from ties in fp32 distances, on which FAISS IndexFlatL2 also errs (9 queries on agnews and 1 on yandex).

\subsection{Per-Neighbor Guarantees and PUFFINN}\label{app:puffinn}

PUFFINN~\cite{r179} solves with an LSH forest the $(k, \delta)$-NN problem, which asks to find each true $k$-nearest neighbor with probability at least $1-\delta$. It scans multiple LSH tries starting from the longest hash matches and stops after examining $\ln(1/\delta)/p^{i}$ tries ($i$ is the length of the matching hash), a number determined by the collision probability $p$ between the query and the current $k$-th candidate; by the monotonicity that the collision probability of the true $k$-th neighbor is at least $p$, it shows that the probability of missing each neighbor is at most $\delta$. This is a guarantee that each query's expected recall is at least $1-\delta$, of the same form as our expected-recall guarantee, but since it is a per-neighbor probability, the probability of finding all $k$ neighbors is only at least $1-k\delta$ by the union bound. Still, since the expected number $M$ of missed neighbors is at most $k\delta$, Markov's inequality gives the tail bound $\Pr[M \ge m+1] \le k\delta/(m+1)$, which equals the $(\tau, \delta)$-recall guarantee for a per-neighbor probability of $\delta(m+1)/k$. The implementation supports only cosine and Jaccard similarity, filters candidates before exact distance computation with a sketch filter whose threshold is placed at the expected value, selects hash functions from a shared pool, and approximates collision probabilities by a Monte Carlo table. The proof does not cover these steps. Insertions require a rebuild, deletions are not supported, and the original paper verified the guarantee by average recall. Our guarantee subsumes this problem as follows.

\begin{thmblock}{\textbf{Corollary 3} (Per-Neighbor Guarantee):}In each mode of Theorems 2–4 (and likewise with the adaptive radius of Theorem 5), for every query and every true $k$-nearest neighbor $v$ of it, the probability that $v$ is missed is at most $\beta$. Hence the expected-recall guarantee mode ($\beta = 1-\tau$) solves the $(k, \delta_{\rm P})$-NN problem of PUFFINN~\cite{r179}, which asks to find each true neighbor with probability at least $1-\delta_{\rm P}$, with $\delta_{\rm P} = 1-\tau$ ($\delta_{\rm P}$ is PUFFINN's per-neighbor failure probability). The exact-search guarantee ($\beta = \delta/k$) solves the same problem with $\delta/k$ and in addition ensures that all $k$ neighbors are found with probability at least $1-\delta$, and the $(\tau, \delta)$-recall guarantee ($\beta = \delta(m+1)/k$) solves the same problem with $\delta(m+1)/k$ while further bounding the tail of the number of missed neighbors.\end{thmblock}

Corollary 3 follows directly from Theorem 1. True neighbors in visited partitions are tested with exact distances and are therefore never missed. An unvisited true neighbor, under the condition that the test radius is at least the true $k$-th distance (C2 or Theorem 5), is pruned at each test stage $\ell$ with probability at most $G_D(-\varepsilon_{0,\ell})$, and the error-bound parameters are set so that the sum of these probabilities is at most $\beta$ (Theorem 1(c)); hence, by the union bound, it is missed with probability at most $\beta$. What the exact-search guarantee and the $(\tau, \delta)$-recall guarantee add is given by Theorem 2 and Theorem 3, respectively. The per-neighbor guarantee alone only ensures, by the union bound, that all $k$ neighbors are found with probability at least $1-k\beta$.

We also verified the per-neighbor bound experimentally. On the eight 1M datasets and imagenet-align with $k \in \{10, 100\}$, we measured the miss frequency of each (query, true neighbor) pair over 20 independent rotations not shared with the other audits and three processing paths (batch, 32-thread, and 1-thread). Under the expected-recall guarantee ($\tau = 0.9$, $\beta = 0.1$), the miss rate pooled over all pairs was 0.0002–0.017, and 52–90 pairs per path had a Clopper–Pearson 95\% lower bound exceeding $\beta$. Because the pairs are very numerous, such pairs also arise by chance, and their number falls far short of the number expected by chance if the true probability of every pair were exactly $\beta$ (37,994), consistent with the per-neighbor bound. The minimum per-query mean recall was 0.935. For the $(0.9, 0.01)$-recall guarantee and the exact-search guarantee, whose $\beta$ is small, pair-level tests lack resolution, so we examined the pooled rate instead, which was at most 9\% and 7\% of $\beta$, respectively.

We also compared with the public implementation of PUFFINN (commit fd86b0d) (Table~\ref{tab:puffinn}). The datasets are yandex, laion, and arxiv, which are normalized to unit length so that L2 and cosine rankings coincide; $k = 10$, and the target recalls are 0.9 and 0.95 ($\delta = 0.1, 0.05$). Following the GloVe 1M 4 GB setting of the original paper, we set the index size to 381 hash tables, and we rebuilt the index 10 times with different hash seeds (5 times for the version with the sketch filter disabled). For hash functions we used independent hashing, the default of the Python interface, a choice closer to the assumptions of the proof than the shared pool used in the experiments of the original paper. The officially recommended default setting (with the sketch filter) fell short of the target even in average recall; the per-neighbor miss rate was 1.2–4.1$\times$ $\delta$, and 15–93\% of queries had a repetition-averaged recall significantly below $1-\delta$. Disabling the sketch filter brought the per-neighbor miss rate within $\delta$ but made queries 4–9$\times$ slower, and at $\delta = 0.05$, 4–11\% of queries were still significantly below the target. The shortfalls of the filter-disabled version occurred almost entirely at the $k$-th neighbor, in cases where the relative distance gap to the $(k+1)$-th neighbor was $10^{-4}$–$10^{-3}$, i.e., the order was swapped in 16-bit fixed-point distances. Our method, configured for the expected-recall guarantee at the same level ($\tau = 1-\delta$), had no query significantly below the target and was 3.1–17.5$\times$ faster in 1-thread latency than the default setting and 27–76$\times$ faster than the filter-disabled version. PUFFINN processes one query with one thread and has no interface for processing multiple queries together. Results were measured with the same tie handling as in the other experiments (Section~\ref{sec:setup}). The build time of PUFFINN was 31, 60, and 149 s on yandex, laion, and arxiv, respectively (ours: 11.3 s on laion and 27.3 s on arxiv — Table~\ref{tab:build}).

\Needspace{16\baselineskip}
{\centering \textbf{Table \refstepcounter{table}\label{tab:puffinn}\thetable{}. Comparison of PUFFINN with our method configured for the expected-recall guarantee at the same level ($\tau = 1-\delta$) ($k = 10$; PUFFINN indexes were built 10 times for the default and 5 times with the filter disabled). Queries below target are those whose repetition-averaged recall is significantly lower than the target.}\par}\nopagebreak

{\small\setstretch{1.1}
\begin{xltabular}{\linewidth}{llYYY}
\toprule
\textbf{Data} & \textbf{Target} & \textbf{PUFFINN default: mean recall · neighbor miss $\div$ $\delta$ · queries below target · 1-thread ms} & \textbf{PUFFINN filter off: mean recall · neighbor miss $\div$ $\delta$ · queries below target · 1-thread ms} & \textbf{Ours: queries below target · 1-thread ms} \\
\midrule
\endfirsthead
\multicolumn{5}{l}{\emph{Table~\thetable\ (continued)}} \\
\toprule
\textbf{Data} & \textbf{Target} & \textbf{PUFFINN default: mean recall · neighbor miss $\div$ $\delta$ · queries below target · 1-thread ms} & \textbf{PUFFINN filter off: mean recall · neighbor miss $\div$ $\delta$ · queries below target · 1-thread ms} & \textbf{Ours: queries below target · 1-thread ms} \\
\midrule
\endhead
yandex & 0.9 & 0.876 · 1.24 · 14.8\% · 4.39 & 0.950 · 0.51 · 0.1\% · 23.1 & 0 · 0.75 \\
yandex & 0.95 & 0.921 · 1.58 · 18.6\% · 5.86 & 0.976 · 0.49 · 4.0\% · 41.2 & 0 · 0.84 \\
laion & 0.9 & 0.824 · 1.77 · 41.3\% · 33.9 & 0.951 · 0.49 · 0.4\% · 132.7 & 0 · 2.35 \\
laion & 0.95 & 0.886 · 2.29 · 42.2\% · 43.0 & 0.971 · 0.58 · 8.7\% · 186.5 & 0 · 2.46 \\
arxiv & 0.9 & 0.728 · 2.72 · 90.7\% · 3.82 & 0.977 · 0.23 · 0.5\% · 33.6 & 0 · 1.23 \\
arxiv & 0.95 & 0.794 · 4.13 · 93.4\% · 4.19 & 0.978 · 0.45 · 11.3\% · 33.8 & 0 · 1.25 \\
\bottomrule
\end{xltabular}}

\subsection{Dimension and Relative Cost}\label{app:dimension}

The dimension experiment of Section~\ref{sec:scale} was performed on several versions of the same data reduced by PCA to dimensions ranging from 128 up to the original dimension (GIST 960, arxiv 768, and a 1M subset of msmarco 1024; $k = 10$, one rotation; a run separate from Table~\ref{tab:speedup}). In batch, the speedup over FAISS IndexFlatL2 decreased from 37.9–56.1$\times$ at 128 dimensions to 29.2–42.4$\times$ at the original dimension, because the per-dimension efficiency of FAISS's matrix multiplication improves with dimension. With the first-stage code length set equal to the dimension, the test error alone reduced the pass rate by 9–80$\times$, but the first-stage cost also grew in proportion to dimension, so the 1-thread speedup was 21.5–24.4$\times$, independent of dimension. Hence the growth of the gain with dimension is a property that holds when the first-stage code length is fixed.

\subsection{Unsuccessful Attempts, Design History, and Settings}\label{app:negative}

Several attempts to reduce the cost of testing unvisited vectors were ineffective. Examples include methods that shrink the radius or build partitions with uniform radii, deterministic certification using PQ~\cite{r19} centroids, enlarging the set of visited partitions, pruning with partition summaries (spheres, boxes, PCA, and cones), the pigeonhole principle based on subspace partitioning~\cite{r20,r101}, prediction of the maximum cosine in the query direction, and residual whitening. Subspace partitioning produced few candidates, but finding those candidates cost as much as exhaustive search, and even when the maximum cosine in the query direction was known by an oracle, the surviving vectors decreased by only 27–58\%. The common cause is that the maximum cosine in the query direction, the information that separates the answers, exists only at the level of individual vectors, so partition summaries cannot provide a deterministic upper bound. Methods that introduce a data model and prune at the partition level could filter 65–89\% of the vectors on text data without reading them, but the surviving vectors that dominate the cost are concentrated in a few partitions near the query, so these decreased by only 1–11\%. The throughput gain was 0–20\%, and guarantee violations of up to 1.7\% occurred. These negative results are the grounds for our choice of an assumption-free per-query guarantee.

We disclose the following regarding the measurements. Early exact computation and estimate-ordered processing, which assist the adaptive radius, their enabling rule, global centering, $m_1 = D$, and the cost model of the parameter selection rule were designed or revised after seeing the first measurements on laion, SSNPP, and yandex. Early exact computation selects the $2k + 20$ vectors with the smallest distances in the first-stage estimates of the first 64 or 192 partitions in $\Phi$ order for 1-thread, and of the first 64 or 128 partitions in partition order for batch. The 1-thread scan visits the first 132 partitions in $\Phi$ order first and the rest in their stored order. The enabling rule times the candidate paths over 3 rounds on calibration queries (200 for latency and 1,000 for batch) and chooses the simpler option (off) if its median is within 1.05$\times$ of the fastest candidate; this threshold was initially set to 1.02$\times$ and raised after we observed settings in which the calibration and test results disagreed. Estimate-ordered processing is a candidate only for 32-thread latency, and a variant that removes synchronization barriers in batch was not used because it gave no gain in 32-thread batch. The batch thread count of 32 was chosen by measuring the batch throughput of our method and FAISS IndexFlatL2 with 24–64 threads on laion ($k = 100$), SIFT ($k = 10$), and SSNPP ($k = 100$) and taking the value that maximizes the minimum, over all (dataset, method) pairs, of the ratio to each pair's own peak throughput (0.94; FAISS was 6\% faster with 63–64 threads on SIFT and SSNPP, our method peaked at 32 threads or was within 2\% of its peak there, and slow runs increased sharply at 60 or more threads). In addition, the calibration data used to set the machine constants of the rule overlap with the evaluation data. The evidence for generalization that does not overlap with calibration comes from arxiv, the three 10M datasets, imagenet-align, and DEEP1M and GloVe, which were added after the rule was fixed. The oracle $n_{\rm probe}$ of IVF and the oracles of some configurations were chosen with the test queries. Batch measurements varied by $\pm$5–10\% between runs, and FAISS at 10M scale was measured in the first round only. Multi-bit codes for prefix stages, a dual first-stage index with 128 and 256 bits, block skipping, block sharing among multiple queries in batch, and a prefilter on a 4-bit copy were tested at the implementation level but gave no gain (0.88–1.03$\times$), because the first stage is compute-bound and different queries make few visits per partition, leaving nothing to share. To the gains of split and dissolution in policy K, we add a term equal to the change in the number of partitions multiplied by the fixed cost of testing one partition (a machine constant set to 73 ns if the first-stage data fit in the L3 cache and 397 ns otherwise) and by the probability that the partition becomes a test target at the initial radius of the calibration queries. Unlike the official evaluation, the streaming measurements used 1,000 queries, fp64 ground truth, and zero-padding to 112 dimensions, with no time limit, and IVF was trained on the first batch and never retrained. In one instance, the first attempt at the streaming measurement violated C5 by measuring the constants of the maintenance policy with the rotation used in the measurement. Those results were discarded and the measurement was repeated, and the code now rejects such use.

\subsection{An Empirical Formula for Relative Radius and Pass Rate}\label{app:relradius}

Summarizing the inputs of Eq.~\eqref{eq:passprob} in a single variable yields the \textit{relative radius} $R_{\mathrm{vis}} = r_{\rm vis}/\sqrt{x\Phi}$, the initial radius (C2; hereafter $r_{\rm vis}$) divided by the residual scale. For each query, $x$ and $\Phi$ are taken as medians over the unvisited vectors within the triangle-inequality range, and the value for a dataset is the median over queries. This variable shares the idea of relative contrast~\cite{r118} but differs in that its denominator is the residual scale and its numerator is the test radius actually used in search. Regressing the first-stage pass rate $p_1$ on $R_{\mathrm{vis}}$ over the 26 combinations of 13 datasets and $k \in \{10, 100\}$ gives \begin{equation}\mathrm{logit}\,p_1 \approx -17.3 + 14.4\,R_{\mathrm{vis}}\qquad(R^2 = 0.908)\label{eq:rvis}\end{equation} On the 12 settings measured at the initial radius with the adaptive radius disabled, $R^2 = 0.95$. In contrast, the residual norm $x$ divided by the mean query–data distance alone gave $R^2$ of at most 0.08. Hence what determines the cost difference between datasets is not the residual norm itself but the relative distance of the neighbors with respect to the residual scale.

This formula is a summary across datasets, and its direction can be reversed between partitions within the same dataset or between time points before and after updates. For example, the dynamic index without maintenance (Section~\ref{sec:dynamic}) had a smaller $R_{\mathrm{vis}}$ than the maintained index (0.88 vs. 1.01) but a higher pass rate (0.018 vs. 0.011). When centroids drift away, $x$ and $\Phi$ grow together so that $R_{\mathrm{vis}}$ decreases, but the margin in Eq.~\eqref{eq:gap} also shrinks and more vectors outside the radius pass. Therefore, when comparing time points or deciding on maintenance, we use Eq.~\eqref{eq:passprob} directly (Section~\ref{sec:maintenance}).

\subsection{Details of the Optimization Techniques}\label{app:techniques}

The operation and implementation of the optimization techniques of Section~\ref{sec:devices} are as follows.

Early exact computation scans some partitions at the front in $\Phi$ order using only first-stage codes and computes in advance the exact distances of a few vectors with the smallest estimated distances. This first-stage reading serves to select vectors and is not a test; after the radius is reduced with the selected vectors, all unvisited partitions, including those scanned, are tested as usual with the reduced radius. The selected vectors enter the candidates with their exact distances, and if they pass the test again during the scan, they are filtered by id so that they are not added to the candidates twice (Section~\ref{sec:process}). Since batch queries are processed grouped by partition, early exact computation is also done in partition order. Estimate-ordered processing is used when one query is processed by multiple threads: among the vectors that passed the first stage in a partition, a few with small estimated distances are first processed through the later stages and their exact distances to reduce the radius, and the remaining passing vectors are then processed with the reduced radius. If too few vectors are computed first, the radius does not come down enough; if too many, exact distance computations increase.

The first-stage code length $m_1$ is chosen from $m_1 \in \{128, 256, D\}$. On data with a low first-stage pass rate, a short code discards most vectors, whereas on data with a high pass rate, reading the entire code at once reduces the cost of the later stages.

The 2-bit final stage adds one more test stage after the last 1-bit stage with $m_\ell = D$. The 2-bit code quantizes each coordinate of $P^{-1}\boldsymbol{o}$ on a 4-level uniform grid symmetric about 0, and, as in Extended RaBitQ~\cite{r63}, the grid scale is determined from that vector alone so as to maximize the cosine between the code vector and $P^{-1}\boldsymbol{o}$. The error scale is the value of Eq.~\eqref{eq:kappa} with $m_\ell/D$ replaced by the squared norm of the code vector, typically about half that of the 1-bit code ($0.365/\sqrt D$ vs. $0.755/\sqrt D$). This stage reduces the number of vectors whose exact distances are computed from the original vectors, but it uses an additional $16 + D/4$ bytes per vector, so it pays off only when the fraction of vectors passing the final 1-bit stage is large.

Global centering reduces the deterministic slack. The slack of 8-bit LUT quantization is proportional to the norm of the query vector, so on data whose vector mean lies far from the origin, such as data with integer coordinates in 0–255 and no mean subtraction, $\lVert\boldsymbol{q}_r\rVert/\Phi$ is much larger than 1 and the deterministic slack $\sigma_{\det}$ grows large enough to reach the error scale $\kappa^{(1)}$. Building the LUT from $P^{-1}(\boldsymbol{q}_r - \boldsymbol{c}_0)$, with the mean $\boldsymbol{c}_0$ of the whole data subtracted, and subtracting the contribution of $P^{-1}(\boldsymbol{c} - \boldsymbol{c}_0)$ from the per-vector coefficients leaves the test values unchanged and reduces only the slack. The number of front partitions, the number of vectors computed first, the pass-rate threshold, and the number of queries used for calibration change only the cost; their values are given in Appendix~\ref{app:negative}.

\subsection{Comparison of Partitioning Methods}\label{app:partition}

The comparison of the ten partitioning methods summarized in Section~\ref{sec:ablation} is as follows. At the same radius, k-means~\cite{r159} and its variants k-means++~\cite{r160}, hierarchical k-means~\cite{r161}, spherical k-means~\cite{r162}, and anisotropic quantization~\cite{r25} differed little, at 0.95–1.03$\times$ of balanced k-means (except spherical k-means on GIST, at 1.12–1.21$\times$), and within the same dataset the order of pass rates was determined by the relative residual size (rank correlation 0.88–0.96 among the ten methods). The picture changes, however, when each method uses the radius it obtains itself. On 1M datasets with in-distribution queries (text, GIST, and SIFT), with the radius fixed by the visited vectors, a global centroid was 4.8–18.3$\times$ slower and the random projection tree~\cite{r163} 1.6–5.5$\times$ slower, because poor centroids make the visited partitions contain fewer neighbors and thus worsen the radius as well. Shrinking the radius with the adaptive radius while scanning the unvisited region reduced this loss to 1.3–2.5$\times$ for the global centroid and 1.2–1.6$\times$ for the random projection tree, and the same trend appeared at 10M scale (from 2.8–25.5$\times$ to 1.18–1.95$\times$). These results support the explanation in Section~\ref{sec:costlaw} that partitions play two roles, providing centroids and acquiring the radius. On laion and SSNPP, however, where the visited partitions contain few true neighbors, the difference between the two partitionings was small (1.0–1.7$\times$ with a fixed radius and 1.1–1.6$\times$ with the adaptive radius). Graph partitioning, which splits the $k$-NN graph with METIS~\cite{r164} to gather neighbors into the same partition, and subdividing only partitions with large residuals gave no gain.

\subsection{Effects of the Techniques and Data Characteristics}\label{app:devices}

On which datasets the techniques of Section~\ref{sec:devices} are enabled is determined by which term of Eq.~\eqref{eq:passprob} is large on each dataset. In terms of cost, our datasets fall into three groups. Data with in-distribution queries and low intrinsic dimension (SIFT, YFCC, arxiv, agnews, GIST, and msmarco) have a relative radius $R_{\mathrm{vis}}$ of 0.81–0.97 and an initial radius within 1.03$\times$ of the true $k$-th distance, so the first stage already filters out almost everything, and little remains to be gained by adjusting the radius, bits, or slack. Data with OOD queries (laion, yandex, hotpotqa, and imagenet-align) have an $R_{\mathrm{vis}}$ of 1.16–1.54; the partitions visited in $\Phi$ order contain few true neighbors, so the visited radius exceeds the true $k$-th distance (1.07$\times$ on laion), and many vectors pass the first stage. Integer data with high intrinsic dimension and coordinates far from the origin (SSNPP) have an $R_{\mathrm{vis}}$ of 1.22–1.24 and a large LUT quantization slack. Setting $k = 100$ raises $R_{\mathrm{vis}}$ on all datasets, shifting them toward the second group. Table~\ref{tab:devices} summarizes, for each technique, the term it targets, the condition for enabling it, and its effect in the settings where the rule enabled it in our measurements. Because the defaults are set without regard to the guarantee level, $m_1 = 256$, which benefits exact search on text and GIST at $k = 100$, increased the 1-thread time by 1.07–1.34$\times$ under the expected-recall guarantee. This is because the error-bound parameters of the expected-recall guarantee are small, so the pass rate is already low with 128 bits; choosing $m_1$ per mode resolves this.

\Needspace{16\baselineskip}
{\centering \textbf{Table \refstepcounter{table}\label{tab:devices}\thetable{}. Terms targeted by the techniques and conditions for enabling them (effects are speedups, in the settings where the rule enabled the technique, over the configuration with it disabled).}\par}\nopagebreak

{\small\setstretch{1.1}
\begin{xltabular}{\linewidth}{YYYY}
\toprule
\textbf{Technique} & \textbf{Term of Eq.~\eqref{eq:passprob}} & \textbf{Enabling condition} & \textbf{Effect (our measurements)} \\
\midrule
\endfirsthead
\multicolumn{4}{l}{\emph{Table~\thetable\ (continued)}} \\
\toprule
\textbf{Technique} & \textbf{Term of Eq.~\eqref{eq:passprob}} & \textbf{Enabling condition} & \textbf{Effect (our measurements)} \\
\midrule
\endhead
Adaptive radius & Radius $r$ in margin $g$ & Always (effective when the visited radius exceeds the true $k$-th distance) & laion 1-thread 5.6–10.6·batch 3.8–5.8, SSNPP 1.45–2.09·1.23–1.59, yandex 1.12–1.36·1.02–1.17 \\
Early exact computation·estimate-ordered processing & Same term (when the radius comes down) & When calibration queries are at least 5\% faster (laion, yandex, GIST $k = 100$, SSNPP-10M $k = 100$, etc.) & Only in settings chosen by the enabling rule \\
$m_1 \in \{128, 256, D\}$ & Error scale $\kappa^{(1)}$ & 256 if the calibration pass rate > 5\%; $D$ if the pass rate is high even at $r \approx d_k$ & $k = 100$ text·GIST T1 1-thread 1.19–1.80$\times$ (expected-recall guarantee: see the paragraph above) \\
2-bit final stage & Error scale $\kappa^{(L)}$ & Pass fraction of the final 1-bit stage at least a few percent, and latency metrics & laion·yandex T1 1-thread 1.16–1.59$\times$ \\
Global centering & Deterministic slack $\sigma_{\det}$ & $\lVert\boldsymbol{q}_r\rVert/\Phi \gg 1$ (SSNPP 5.7 $\rightarrow$ 0.97) & SSNPP 1-thread 1.36–1.83·32-thread 1.37–1.62·batch 1.22–1.46 \\
\bottomrule
\end{xltabular}}

\end{document}